\documentclass[useAMS,usenatbib,twocolumn]{mn2e}
\pdfoutput=1
\usepackage{amssymb}
\usepackage{amsmath}
\usepackage{natbib}
\usepackage[pdftex]{graphicx}
\usepackage{subfigure}
\usepackage{booktabs}
\usepackage{tabularx}
\usepackage{xcolor}

\begin{document}

\title[GSF instability]
{Angular momentum transport by the GSF instability: nonlinear simulations at the equator}
  \author[A. J. Barker, C. A. Jones \& S. M. Tobias]{A. J. Barker\thanks{Email address: A.J.Barker@leeds.ac.uk}, C. A. Jones and S. M. Tobias\\ Department of Applied Mathematics, School of Mathematics, University of Leeds, Leeds, LS2 9JT, UK
}

\pagerange{\pageref{firstpage}--\pageref{lastpage}} \pubyear{2019}
\maketitle
\label{firstpage}

\begin{abstract}
We present an investigation into the nonlinear evolution of the Goldreich-Schubert-Fricke (GSF) instability using axisymmetric and three-dimensional simulations near the equator of a differentially rotating radiation zone. This instability may provide an important contribution to angular momentum transport in stars and planets. We adopt a local Boussinesq Cartesian shearing box model, which represents a small patch of a differentially rotating stellar radiation zone. Complementary simulations are also performed with stress-free, impenetrable boundaries in the local radial direction. The linear and nonlinear evolution of the equatorial axisymmetric instability is formally equivalent to the salt fingering instability. This is no longer the case in 3D, but we find that the instability behaves nonlinearly in a similar way to salt fingering. Axisymmetric simulations -- and those in 3D with short dimensions along the local azimuthal direction -- quickly develop strong jets along the rotation axis, which inhibit the instability and lead to predator-prey-like temporal dynamics. In 3D, the instability initially produces homogeneous turbulence and enhanced momentum transport, though in some cases jets form on a much longer timescale. We propose and validate numerically a simple theory for nonlinear saturation of the GSF instability and its resulting angular momentum transport. This theory is straightforward to implement in stellar evolution codes incorporating rotation. We estimate that the GSF instability could contribute towards explaining the missing angular momentum transport required in red giant stars, and play a role in the long-term evolution of the solar tachocline. 
\end{abstract}


\section{Introduction}

The additional mixing and angular momentum transport caused by various hydrodynamic (or magnetohydrodynamic) instabilities of differential rotation can significantly modify the global properties and internal structure of stars (e.g.~\citealt{ARM2018}). The changes induced by these effects are found to be very sensitive to how they are modelled in stellar evolution codes (e.g.~\citealt{Meynetetal2013}), but the underlying physics behind these processes is at present poorly understood. 

Recent observational advances in helio- and astero-seismology have highlighted our poor understanding of the mechanisms of angular momentum transport in the radiation zones of the Sun \citep{Thompson2003,Tachocline2007} and solar-type stars, as well as intermediate-mass stars at various stages of evolution \citep{ARM2018}, particularly during the red giant phase (e.g.~\citealt{Cantielloetal2014,Eggenberger2017}). To interpret these (and future) observations, it is now essential to understand better the mechanisms of angular momentum transport and the resulting mixing of chemical elements in stars.

A potential key player in angular momentum transport in stellar radiation zones is the Goldreich-Schubert-Fricke (GSF) instability \citep{GS1967,Fricke1968}. This is an axisymmetric hydrodynamic instability of differential rotation. It is essentially a centrifugal instability enabled by the action of thermal diffusion, which neutralises the otherwise stabilising effects of buoyancy in a stably-stratified region. The instability grows if the differential rotation is sufficiently strong (e.g.~\citealt{KnoblochSpruit1982,Rashid2008,Caleo2016b}). Stellar evolution codes usually incorporate the transport due to the GSF, and various other hydrodynamical instabilities, as a diffusion of angular momentum with a prescribed diffusivity. However, current models are inadequate; they are, for example, unable to reproduce the observed rotational evolution of sub-giants and early red giant stars \citep{Cantielloetal2014,Eggenberger2017}. In addition, a diffusive approximation for the angular momentum transport is not always appropriate. For example, in stably-stratified turbulent flows, momentum transport can be anti-frictional rather than frictional (e.g.~\citealt{McIntyre2002,TDH2007}).

The GSF instability may also occur in astrophysical discs, where it has been proposed as a mechanism to drive turbulence and to stir solids in the regions of protoplanetary discs that are stable to the magneto-rotational instability. In this context, it has been referred to as the Vertical Shear Instability (VSI) (e.g.~\citealt{Urpin1998,Nelsonetal2013,BL2015,LinYoudin2015,LatterPap2018}). Global simulations of the VSI in protoplanetary discs have been performed by e.g.~\cite{Nelsonetal2013} and \cite{StollKley2014}, which demonstrate that it produces wave activity but weak levels of angular momentum transport.

In the context of stellar and planetary interiors, the nonlinear evolution of the GSF instability has only been studied previously using axisymmetric simulations by \cite{Kory1991} and \cite{Rashid2010}. However, no previous work has studied its three-dimensional nonlinear evolution. In this series of papers, we will present the results of three-dimensional (and some axisymmetric) simulations of the nonlinear evolution of the GSF instability in a local Cartesian model. This is the first paper in the series, and herein we will focus on the properties of the instability near the equator, since this is the simplest case.

The equatorial regions in a local model represent a special case for the GSF instability. The rotation profile is locally barotropic (invariant along the rotation axis), which means very strong differential rotation is required to drive the instability. In particular, we require the differential rotation to be centrifugally unstable according to Rayleigh's criterion for the instability to operate, i.e. the angular momentum must decrease radially. At the equator, GSF is also formally equivalent to the salt fingering instability in both its linear and axisymmetric nonlinear evolution \citep{Knobloch1982}, even if the three-dimensional nonlinear problems are strictly not equivalent. The formal analogy is between salinity (or heavy elements) and angular momentum, with an unstable angular momentum gradient behaving just as an unstable salinity gradient in driving the instability on short enough length-scales that thermal diffusion can operate efficiently. This analogy means that we already have some idea about the nonlinear evolution of the equatorial GSF instability based on extensive prior work on the salt fingering instability by e.g.~\citet{Den2010,Den2011,Traxler2011,Brownetal2013,GaraudBrummell2015,Garaud2018}. However, a study such as ours is required because nonlinear equatorial GSF fundamentally differs from salt fingering in three dimensions, and its three-dimensional evolution has not been explored previously. 

Our goal is to understand the nonlinear evolution of the GSF instability and also to derive physically-motivated prescriptions for the transport of angular momentum that can be straightforwardly implemented in stellar evolution codes. The structure of this paper is as follows. In \S \ref{model} we describe our model and the numerical methods. We then review the key properties of the linear axisymmetric equatorial GSF instability in \S \ref{axilin}, and its formal equivalence with salt fingering in \S \ref{saltfinger}, before proceeding to discuss the results of our nonlinear simulations in \S \ref{nonlinear}. We propose and validate numerically a simple theory for the saturation of the GSF instability, and its consequent rates of angular momentum transport, in \S \ref{theoryvalidation}. Finally, we discuss the astrophysical implications of our results and present our conclusions in \S \ref{implications} and \ref{conclusions}.

\section{Local Cartesian model: small patch of a radiation zone}
\label{model}

\begin{figure}
  \begin{center}
    \subfigure{\includegraphics[trim=0cm 0cm 0cm 0cm, clip=true,width=0.45\textwidth]{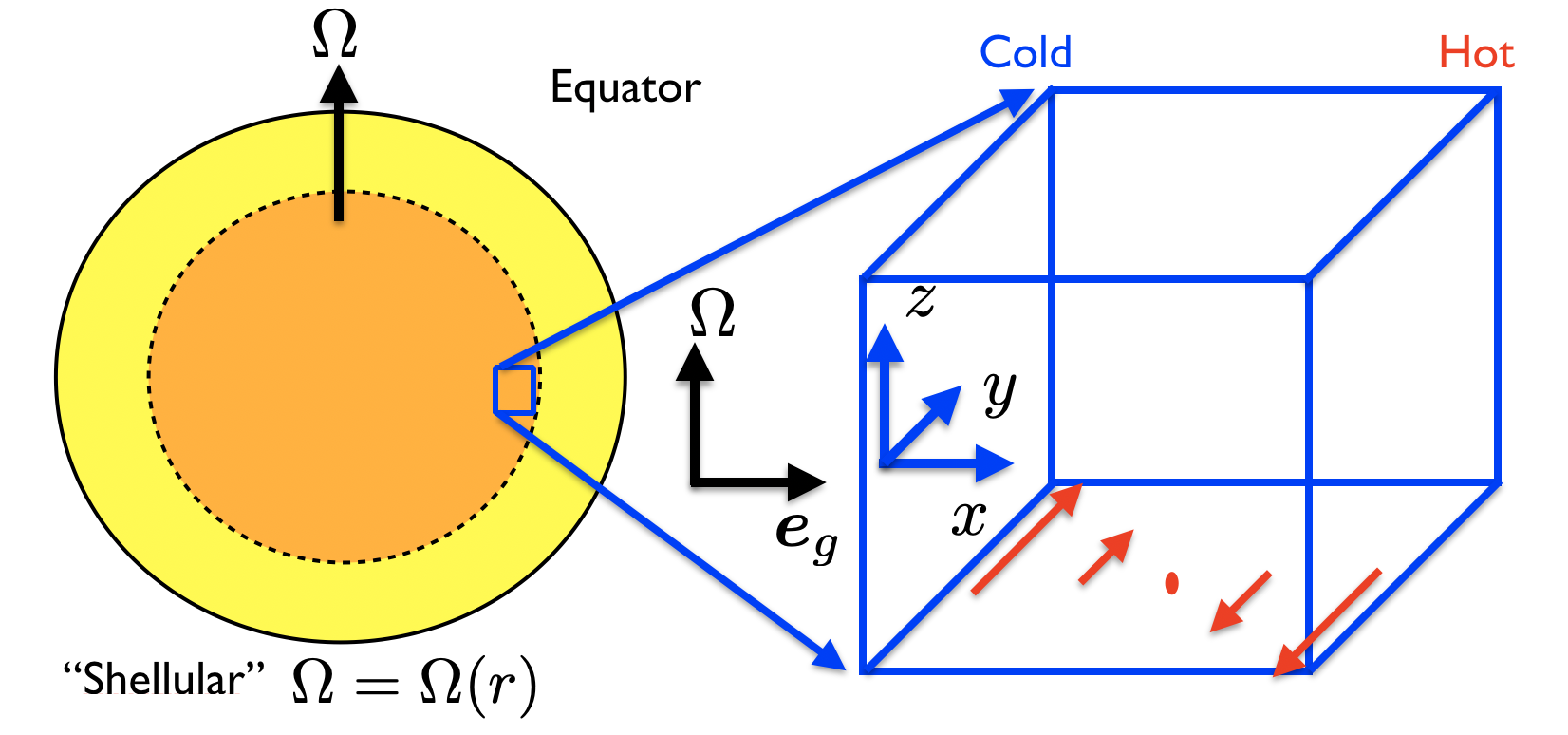}}
    \end{center}
  \caption{Local Cartesian model to study the GSF instability at the equator. For illustration, the dark orange region may represent a radiation zone and the yellow region an overlying convection zone. The Cartesian domain would therefore represent a small patch of the radiation zone, such as in the solar tachocline, for example. At the equator, the local gravity vector is normal to the stratification surfaces, and $\boldsymbol{e}_g=\boldsymbol{e}_x$.}
  \label{modelfig}
\end{figure}

We consider a local Cartesian representation of a small patch of a stably-stratified radiation zone of a differentially rotating star (or planet). Our focus here is on dynamics near the equator, and we will adopt coordinate axes $(x,y,z)$ defined such that $x$ is the local radial, $y$ is the local azimuthal, and $z$ points along the rotation axis (see Fig.~\ref{modelfig}), which is the local latitudinal direction. Note that our choice of coordinates differs from \citet{Rashid2008}. We consider a domain of size $L_x\times L_y \times L_z$. The star is assumed to possess a ``shellular" differential rotation profile (e.g.~\citealt{Zahn1992}), such that the angular velocity $\Omega(r)$ depends only on spherical radius\footnote{More complex differential rotation profiles can be considered, if desired.}, $r$. However, at the equator, this is equivalent to considering a rotation profile that instead varies with cylindrical radius. The differential rotation can be locally decomposed into a uniform rotation $\boldsymbol{\Omega}=\Omega \boldsymbol{e}_z$ and a linear (radial) shear flow $\boldsymbol{U}_0=-\mathcal{S}x\boldsymbol{e}_y$, where $\mathcal{S}$ is the local value of $\mathrm{d}\Omega/\mathrm{d}\ln r$. 

Since the instability operates on lengthscales that are much shorter than a pressure scale height, we will adopt the Boussinesq approximation. In this case, perturbations to the shear flow $\boldsymbol{U}_{0}$ in a frame rotating with an angular velocity $\boldsymbol{\Omega}$, are governed by
\begin{eqnarray}
\label{Eq1}
&&D\boldsymbol{u} + 2\boldsymbol{\Omega}\times \boldsymbol{u} +\boldsymbol{u}\cdot \nabla \boldsymbol{U}_0 = -\nabla p +  \theta\boldsymbol{e}_{x}+ \nu\nabla^{2}\boldsymbol{u}, \\
\label{Eq2}
&& D\theta +\mathcal{N}^2 \boldsymbol{u}\cdot\boldsymbol{e}_{x}=  \kappa\nabla^{2}\theta, \\
\label{Eq3}
&& \nabla \cdot \boldsymbol{u} = 0, \\
\label{Eq4}
&& D \equiv \partial_{t} + \boldsymbol{u}\cdot \nabla + \boldsymbol{U}_0 \cdot \nabla,
\end{eqnarray}
where $\boldsymbol{u}$ is the velocity perturbation and $p$ is a pressure. We define our ``temperature perturbation" $\theta=\alpha g T$, where $\alpha$ is the thermal expansion coefficient, $g$ is the acceleration due to gravity and $T$ is the usual temperature perturbation, so that $\theta$ has the units of an acceleration. We adopt a background temperature profile $T(\boldsymbol{x})$, with uniform gradient $\alpha g\nabla T = \mathcal{N}^2 \boldsymbol{e}_{x}$, where $\mathcal{N}^2>0$ is the square of the buoyancy frequency in a stably-stratified radiation zone. We also adopt a constant kinematic viscosity $\nu$ and thermal diffusivity $\kappa$. Here the background reference density has also been set to unity. 

At the equator, the rotation is constant on cylinders and surfaces of constant density and pressure are aligned. The model here is therefore equivalent to the shearing box model of an accretion disc with both radial stratification and shear, and with rotation being locally constant on cylinders. If we consider a shellular profile of differential rotation, this is no longer true at other latitudes, and the normal to stratification surfaces must be determined by the thermal wind equation if $\mathcal{S}, \mathcal{N}^2$ and $\Omega$ are prescribed. Alternatively, we can impose a temperature gradient and use the thermal wind equation to constrain the (baroclinic) shear.

In our simulations we adopt $\Omega^{-1}$ as our unit of time and the lengthscale \textbf{$d$} to define our unit of length, where 
\begin{eqnarray}
d=\left(\frac{\nu\kappa}{\mathcal{N}^2}\right)^{\frac{1}{4}}.
\end{eqnarray}
The reason for this choice, by analogy with other double-diffusive problems (e.g.~\citealt{Garaud2018}), is that the fastest growing mode typically has a wavelength $O(d)$. This choice permits us to select the box size conveniently relative to the wavelength of the fastest growing linear modes. We define $N=\mathcal{N}/\Omega$ to be our dimensionless buoyancy frequency and $S=\mathcal{S}/\Omega$ to denote our dimensionless shear rate, which can be thought of as a Rossby number. 
We also define the Prandtl number
\begin{eqnarray}
\mathrm{Pr} = \frac{\nu}{\kappa}.
\end{eqnarray}
This problem then has 3 independent non-dimensional parameters: $S$, $\mathrm{Pr}$ and $N^2$, in addition to the dimensions of the box, $L_x$, $L_y$ and $L_z$ in units of $d$, and the numerical resolution. These parameters define the simulations performed, which are listed in table 1.
We also define the Ekman number
\begin{eqnarray}
\mathrm{E} = \frac{\nu}{\Omega d^2} = \mathrm{Pr}^{1/2} N,
\end{eqnarray}
which can be used as an alternative independent parameter replacing $N$,
and the Richardson number
\begin{eqnarray}
\mathrm{Ri} = \frac{\mathcal{N}^2}{\mathcal{S}^2}=\mathrm{E}^2\mathrm{Pr}^{-1}S^{-2},
\end{eqnarray}
which is not an independent quantity here. These non-dimensional numbers allow results to be compared with those of \cite{Rashid2008}.

The non-dimensional momentum and heat equations can be written in the form
\begin{eqnarray}
\label{Eq1ND}
&&D\boldsymbol{u} + 2\boldsymbol{e}_z\times \boldsymbol{u} -Su_x\boldsymbol{e}_y = -\nabla p +  \theta\boldsymbol{e}_{x}+ \mathrm{E}\nabla^{2}\boldsymbol{u}, \\
\label{Eq2ND}
&& D\theta +N^2u_x=  \frac{\mathrm{E}}{\mathrm{Pr}}\nabla^{2}\theta,
\end{eqnarray}
where we have scaled the time by $\Omega^{-1}$, lengths by $d$, velocities by $\Omega d$ and the temperature $
T=\theta/g \alpha$  by $\Omega^2 d / g \alpha$. We have not added hats to denote non-dimensional quantities (i.e.~$u_x, u_y, u_z$ and $\theta$) to simplify the presentation. All formulae below are written using dimensionless quantities unless otherwise specified.

A modified version of the Cartesian pseudo-spectral code SNOOPY is used for most of the simulations \citep{Lesur2005}. This uses a basis of shearing waves to deal with the linear spatial variation of $\boldsymbol{U}_0$, which is equivalent to using shearing-periodic boundary conditions in $x$. In real space, these specify that
\begin{eqnarray}
u_x\left(-\frac{L_x}{2},y,z,t\right)=u_x\left(\frac{L_x}{2},(y-S L_{x} t)\textrm{mod}(L_y),z,t\right),
\end{eqnarray}
and similarly for the other variables. We adopt periodic boundary conditions in $y$ and $z$. The code uses a 3rd order Runga-Kutta time-stepping scheme and deals with the diffusion terms using an integrating factor. Further details regarding the code can be found in e.g.~\citet{Lesur2005}. The parameters and numerical resolutions (i.e.~the number of Fourier modes in $x$, $y$ and $z$) that we adopt are listed in Table \ref{Table}, and we note that the nonlinear terms are fully de-aliased using the 3/2 rule. 

We have thoroughly tested the code to ensure that it correctly captures the linear growth of the GSF instability (according to the predictions of \S~\ref{axilin} below). We have also tested a few axisymmetric simulations against \citet{GaraudBrummell2015} to ensure that the instability correctly behaves in the same manner as the nonlinear evolution of the salt fingering instability in the relevant parameter regime (see \S~\ref{saltfinger} for an explanation). We ensure that each simulation is adequately resolved by either running selected simulations at higher resolution to ensure convergence of the bulk statistics, or by ensuring that the relative spectral kinetic energy in the modes at the de-aliasing wavenumber is no larger than $10^{-3}$ of the maximum.

We also enforce the box-averaged velocity components (i.e. the zero wavenumber mode) to be zero periodically (with a typical period of between 1 and 20 timesteps) to avoid unphysical growth of these quantities. This was found to be necessary when the flow is centrifugally unstable, since this component can grow owing to small numerical errors even though it is not coupled nonlinearly to the other modes (and so should not grow if it is zero initially). 

A number of three-dimensional simulations were also performed using the spectral element code Nek5000 \citep{Nek5000}. These simulations solve Eqs.~\ref{Eq1}--\ref{Eq4} for the same linear shear flow as above. In a star, the shear will slowly evolve in time, and this imposed shear corresponds to the value of the shear at a particular moment in its evolution. This allows us to consider different boundary conditions to shearing-periodic conditions in $x$, and these simulations are presented only in \S\ref{nekcomp}. In particular, we adopt impenetrable, stress-free, fixed temperature conditions at the boundaries in $x$ for these simulations. These specify that
\begin{eqnarray}
\theta=u_x=\partial_x u_y=\partial_x u_z=0 \;\;\;\; \text{on} \;\;\;\; x=\pm\frac{L_x}{2},
\end{eqnarray}
which corresponds with the setup considered by \cite{Rashid2008}, albeit using a different definition of the coordinate axes. Nek5000 partitions the domain into a set of $\mathcal{E}$ non-overlapping elements, and within each element the velocity components and the pressure are represented as tensor product Legendre polynomials of order $\mathcal{N}_p$ and $\mathcal{N}_p-2$, respectively, defined at the Gauss-Lobatto-Legendre and Gauss-Legendre points. The total number of grid points is $\mathcal{E}\mathcal{N}_p^3$. We use a 3rd order implicit-explicit scheme with a variable time-step determined by a target CFL number (typically chosen to be 0.3). Our typical resolution is $\mathcal{E}=20^3$ and $\mathcal{N}_p=10$ (15 for the nonlinear terms), unless otherwise specified. The nonlinear terms are fully de-aliased by using a polynomial order that is 3/2 larger for their evaluation than the resolutions that are specified in Table \ref{Table}. We have tested our setup of the GSF instability in Nek5000 by validating the code against the linear growth rates discussed in \S~\ref{axilin}).

\section{Axisymmetric linear instability}
\label{axilin}

The fastest growing modes in linear theory are axisymmetric (i.e.~have an azimuthal wavenumber $k_y=0$), and in our local model all variables vary as $\mathrm{Re}[\exp \left(\mathrm{i}k_x + \mathrm{i}k_z + st\right)]$, where $k_x$ and $k_z$ are the radial and latitudinal (along $\boldsymbol{\Omega}$) wavenumbers. For clarity, we use the dimensional form of Eqs.~\ref{Eq1}--\ref{Eq4} for the formulae in this section. The growth rate $s$ can be shown to satisfy (e.g.~\citealt{GS1967,Acheson1978,KnoblochSpruit1982,LatterPap2018})
\begin{eqnarray}
\label{disprel}
s_\nu^2 s_\kappa + a s_\kappa + b s_\nu=0,
\end{eqnarray}
where $s_\nu=s+\nu k^2$, $s_\kappa=s+\kappa k^2$, and
\begin{eqnarray}
a &=&\kappa_{ep}^2 \frac{k_z^2}{k^2},  \\
b &=& \mathcal{N}^2 \frac{k_z^2}{k^2},
\end{eqnarray}
and $k^2=k_x^2+k_z^2$. We also define the squared epicyclic frequency $\kappa_{ep}^2=2\Omega(2\Omega-\mathcal{S})$. In the GSF instability we consider cases which are stable in the absence of diffusion, but which are unstable when diffusion is added and the Prandtl number is small. 
The criterion for the non-diffusive (dynamical) $\nu = \kappa =0$ problem to give stability is the Solberg-H{\o}iland criterion for axisymmetric adiabatic, inviscid, perturbations, which are stable if
\begin{eqnarray}
\label{Solberg}
\kappa_{ep}^2+\mathcal{N}^2>0.
\end{eqnarray}
When diffusion is restored, thermal diffusion can allow instability to occur even when  Eq.~\ref{Solberg} is satisfied.  
The criterion for instability is now
\begin{eqnarray}
\label{growtheq}
\kappa_{ep}^2+\mathrm{Pr} \mathcal{N}^2 < 0.
\end{eqnarray}
Note that Pr $<1$, $\mathcal{N}^2 > 0$ and $\kappa_{ep}^2<0$ are required for the GSF instability to operate.
With shearing-periodic boundary conditions in $x$, the fastest growing modes are ``elevator modes" that do not vary along $x$ ($k_x=0$), with $k_z\ne 0$. On the other hand, only modes with $k_x\ne 0$ are permitted when impermeable boundaries are considered in $x$. The restoring action of stratification is maximal on the modes excited at the equator, and the instability requires strongly differentially-rotating flows that are centrifugally unstable with $\kappa_{ep}^2<0$. The fastest growing $k_z$ may be determined by maximising Eq.~\ref{disprel} with respect to $k_z$, i.e.~by solving
\begin{eqnarray}
2\mathrm{Pr} s_\nu s_{\kappa}+s_\nu^2+a+\mathrm{Pr} b=0.
\label{maxgrowthkz}
\end{eqnarray}
The maximum growth rate and the corresponding wavenumber $k_z$ are then determined by solving Eqs.~\ref{disprel} and \ref{maxgrowthkz}. 
\begin{figure}
  \begin{center} 
    \subfigure{\includegraphics[trim=3.5cm 0cm 6cm 0cm, clip=true,width=0.4\textwidth]{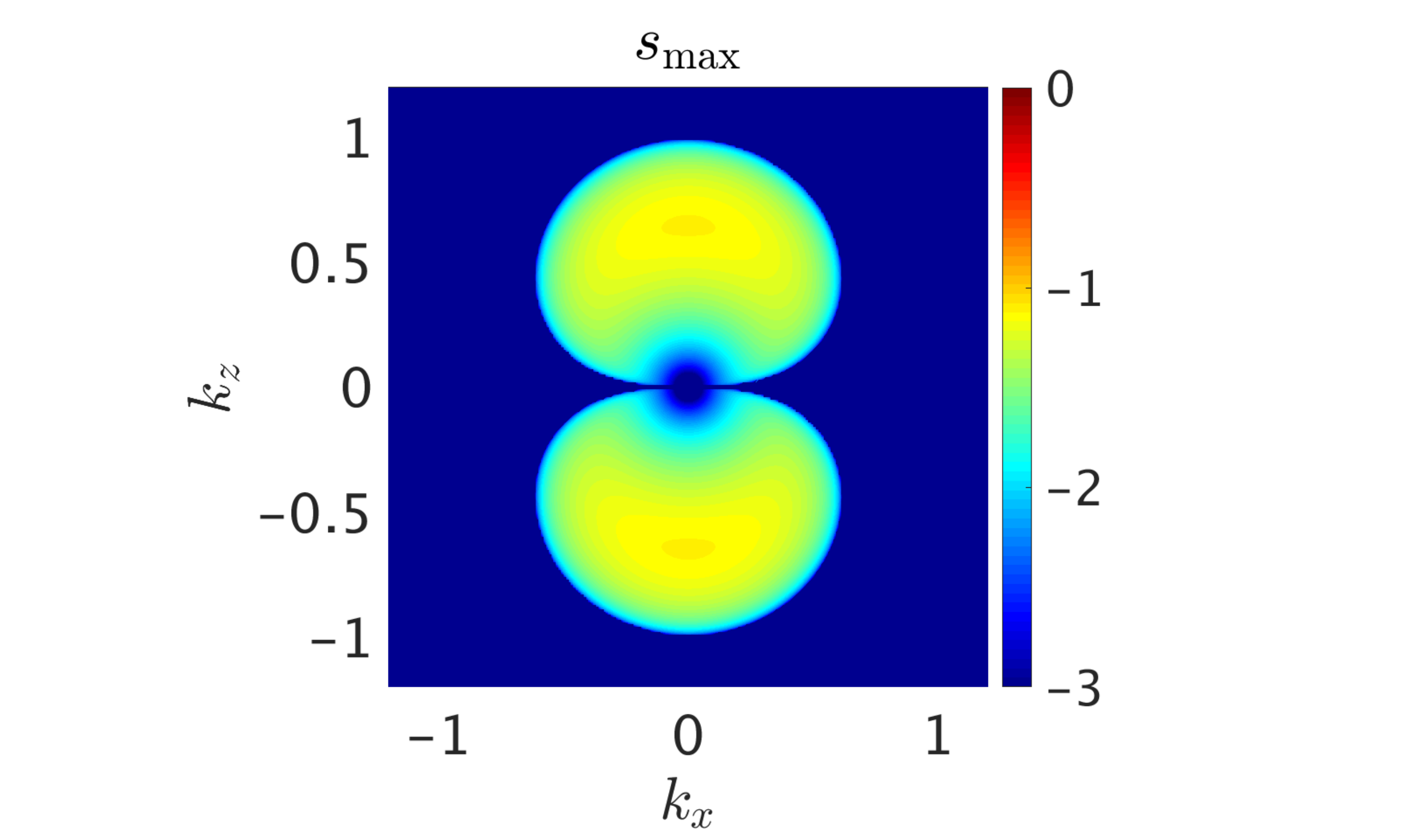}}
    \subfigure{\includegraphics[trim=5cm 0cm 6.5cm 0cm, clip=true,width=0.4\textwidth]{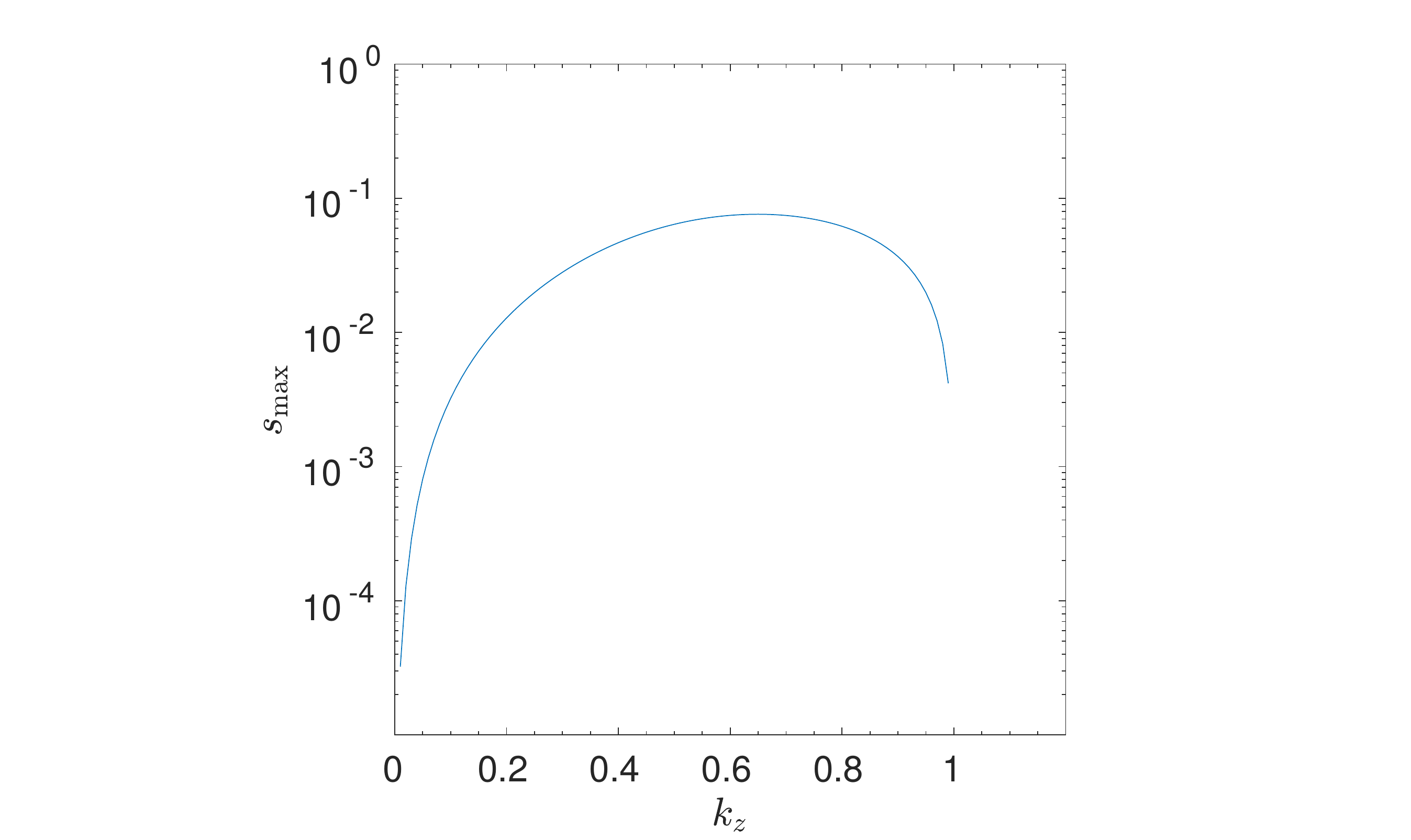}}
    \end{center}
  \caption{Top: base 10 logarithm of the growth rate of the axisymmetric GSF instability on the $(k_x,k_z)$-plane with $S=2.1, N^2=10, \mathrm{Pr}=10^{-2}$. Bottom: growth rate of elevator modes on a log-scale with $k_x=0$ as a function of $k_z$. The top panel shows that the fastest growing modes at the equator have $k_x=0$, and that these modes have $k_z=O(d^{-1})$.}
  \label{growthrate}
\end{figure}

In Fig.~\ref{growthrate} we illustrate the growth rate on the $(k_x,k_z)$ plane (top panel), and for elevator modes as a function of $k_z$ (bottom panel), for an example with \textbf{$S=\mathcal{S}/\Omega=2.1, N^2=\mathcal{N}^2/\Omega^2=10$} and Pr$=10^{-2}$. The top panel shows that the fastest growing modes have $k_x=0$, and both panels demonstrate that the wavelength of the fastest growing mode (and that of the unstable modes in general, for these parameters) is $O(d)$.

\section{Analogy with salt fingering for the axisymmetric equatorial GSF}
\label{saltfinger}

In this section, we briefly re-iterate the equivalence of the axisymmetric equatorial GSF instability at the equator with the well-studied salt fingering instability. This is helpful to understand the nonlinear evolution of the instability, and allows us to check our axisymmetric simulations against the two-dimensional simulations of salt fingering by e.g. \citet{GaraudBrummell2015}. This analogy was first discussed by \cite{GS1967} and demonstrated formally by \cite{Knobloch1982}. The nonlinear equations governing the salt fingering instability for axisymmetric ($y$-invariant) flows are (e.g.~\citealt{Garaud2018}), in dimensional form,
\begin{eqnarray} 
\label{SF1}
D u_x &=& -\partial_x p + (\theta-\mu) + \nu \nabla^2 u_x, \\
D u_z &=& -\partial_z p + \nu \nabla^2 u_z, \\
D \theta &=& -\mathcal{N}^2 u_x + \kappa \nabla^2 \theta, \\
D \mu &=& -\mathcal{N}^2_{\mu} u_x + \kappa_\mu \nabla^2 \mu, \\
D u_y &=& \nu \nabla^2 u_y, \\
\label{SF5}
D &=& \partial_t + (u_x\partial_x + u_z\partial_z),
\end{eqnarray}
where we have taken the local gravity direction to be along $x$ to be consistent with our setup in \S~\ref{model}, $\mu$ is the salinity (or heavy element content), $\mathcal{N}^2_{\mu}$ is the background salinity (or heavy element) gradient and $\kappa_\mu$ is the saline diffusivity (or diffusivity of heavy elements). For $y$-invariant solutions, $u_y$ is passively advected and plays no role in the evolution of any instabilities.

For comparison, the evolution of the axisymmetric ($y$-invariant) equatorial GSF instability is governed by Eqs.~\ref{Eq1}--\ref{Eq4} (with $\partial_y=0$), i.e., by
\begin{eqnarray} 
D u_x &=& -\partial_x p + (\theta+2\Omega u_y) + \nu \nabla^2 u_x, \\
D u_z &=& -\partial_z p + \nu \nabla^2 u_z, \\
D \theta &=& -\mathcal{N}^2 u_x + \kappa \nabla^2 \theta, \\
D u_y &=& - (2\Omega -\mathcal{S})u_x + \nu \nabla^2 u_y, \\
D &=& \partial_t + (u_x\partial_x + u_z\partial_z),
\end{eqnarray}
where these are written in dimensional form.
 This system is formally equivalent to Eqs.~\ref{SF1}-\ref{SF5} for fully nonlinear axisymmetric solutions as long as $\kappa_{\mu}=\nu$, and we also identify $\mu=-2\Omega u_y$, and $\mathcal{N}^2_{\mu}=2\Omega(2\Omega-\mathcal{S})=\kappa_{ep}^2$, so that $\mu$ represents an angular momentum stratification. 
The linear GSF instability is therefore also equivalent to salt fingering. 
However, it should be realised that for fully three-dimensional (non-$y$-invariant) solutions, the nonlinear equations describing salt fingering and GSF are no longer equivalent owing to the presence of $u_y$ in the advection term. This means that while the axisymmetric evolution of both instabilities is formally identical (so we should obtain similar results to e.g.~\citealt{GaraudBrummell2015} and \citealt{Xie2019}), the three-dimensional evolution of the equatorial GSF instability differs from that of salt fingering.

\section{Nonlinear results}
\label{nonlinear}

We wish to explore the nonlinear evolution of the instability and how this differs between axisymmetric and 3D simulations as a function of $S, N^2, \mathrm{Pr}$ and $L_y$ (which can be used to probe the importance of three-dimensional effects). We also wish to analyse the efficiency of angular momentum transport, which is quantified by the Reynolds stress $\langle u_x u_y \rangle$, where $\langle \cdot \rangle$ represents a volume average. An ``effective viscosity" can also be defined on dimensional grounds by the sum of the kinematic viscosity with $\nu_E$, where $\nu_{E} = \frac{1}{S}\langle u_x u_y\rangle$, whether or not the turbulence acts in the manner of an eddy diffusion for angular momentum.

We initialise the flow using solenoidal random noise of amplitude $10^{-3}$ for all wavenumbers in the range ${\hat{i},\hat{j},\hat{k}}\in[1,21]$, where $k_x=\frac{2\pi}{L_x}\hat{i}$, $k_y=\frac{2\pi}{L_y}\hat{j}$, and $k_z=\frac{2\pi}{L_z}\hat{k}$. The domain size is chosen so that $L_x=L_z=100 d$, which is sufficient to contain several wavelengths of the fastest growing modes in each of our simulations. $L_y$ is varied separately in the 3D simulations.

\subsection{Illustrative axisymmetric simulations}
\label{2Dresults}

\begin{figure}
  \begin{center}
   \subfigure[$K$]{\includegraphics[trim=0.4cm 0cm 0cm 0cm, clip=true,width=0.45\textwidth]{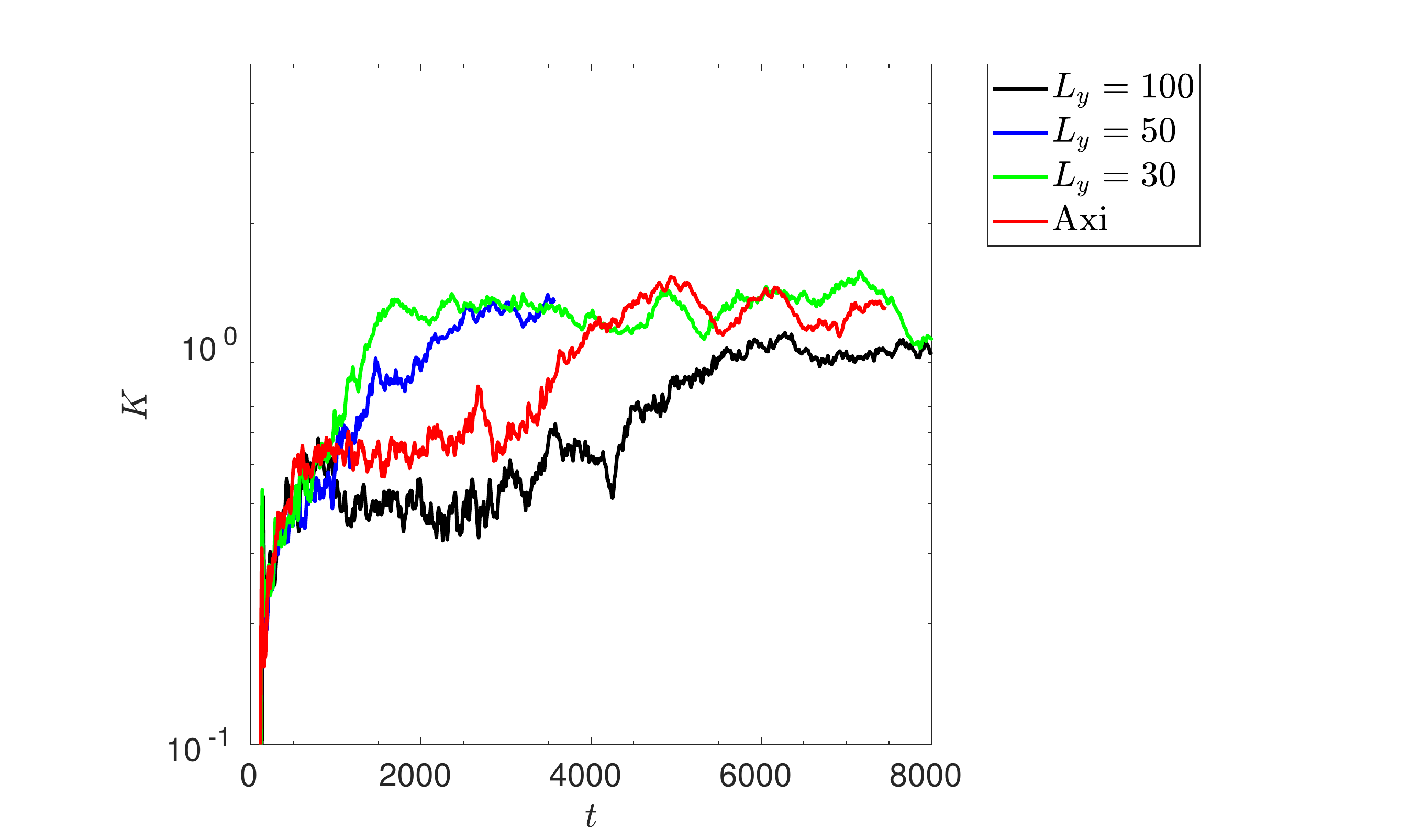}}
    \subfigure[$\langle u_xu_y\rangle$]{\includegraphics[trim=0.4cm 0cm 0cm 0cm, clip=true,width=0.45\textwidth]{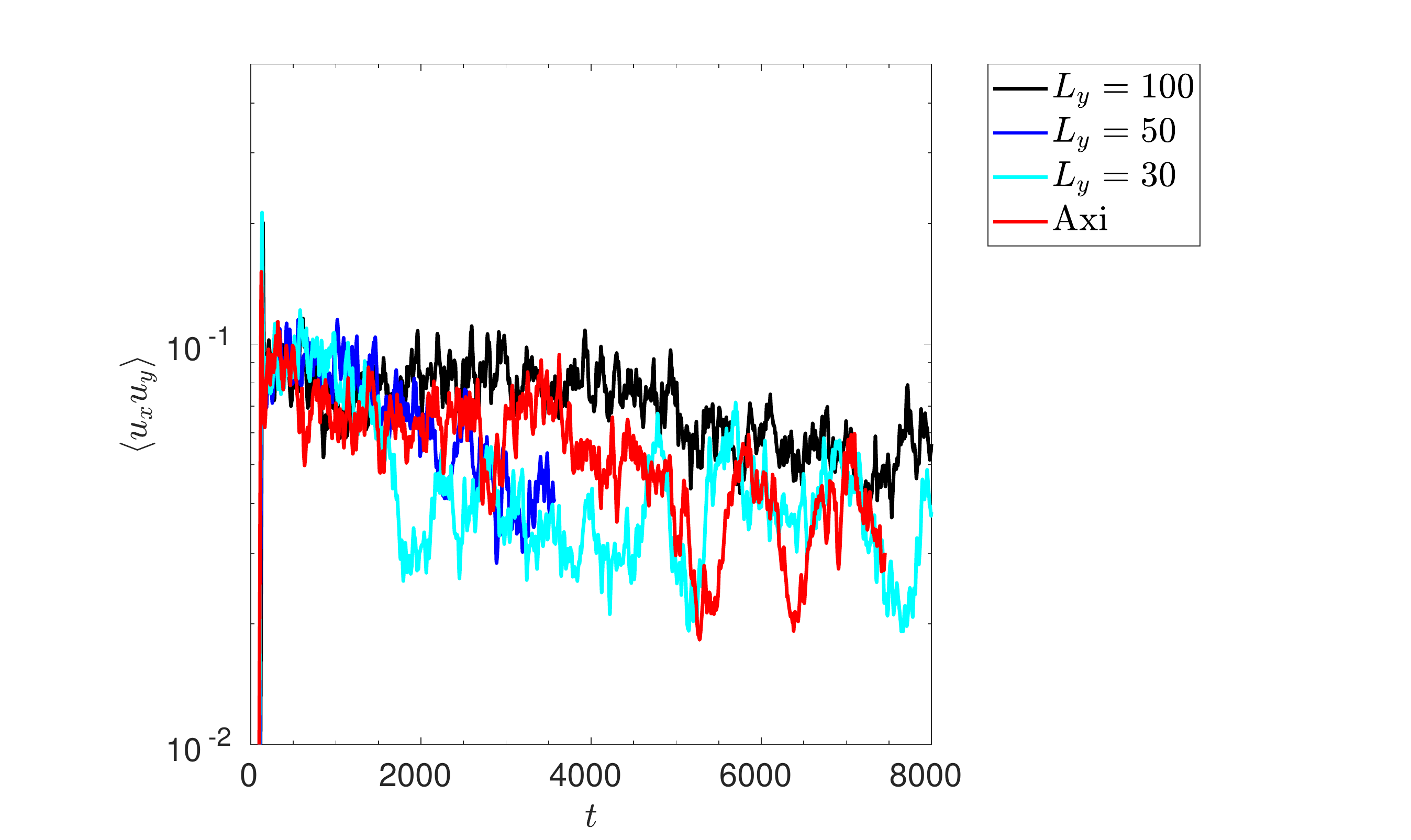}}
    \subfigure[$\sqrt{\langle u_z^2\rangle}$]{\includegraphics[trim=0.4cm 0cm 0cm 0cm, clip=true,width=0.45\textwidth]{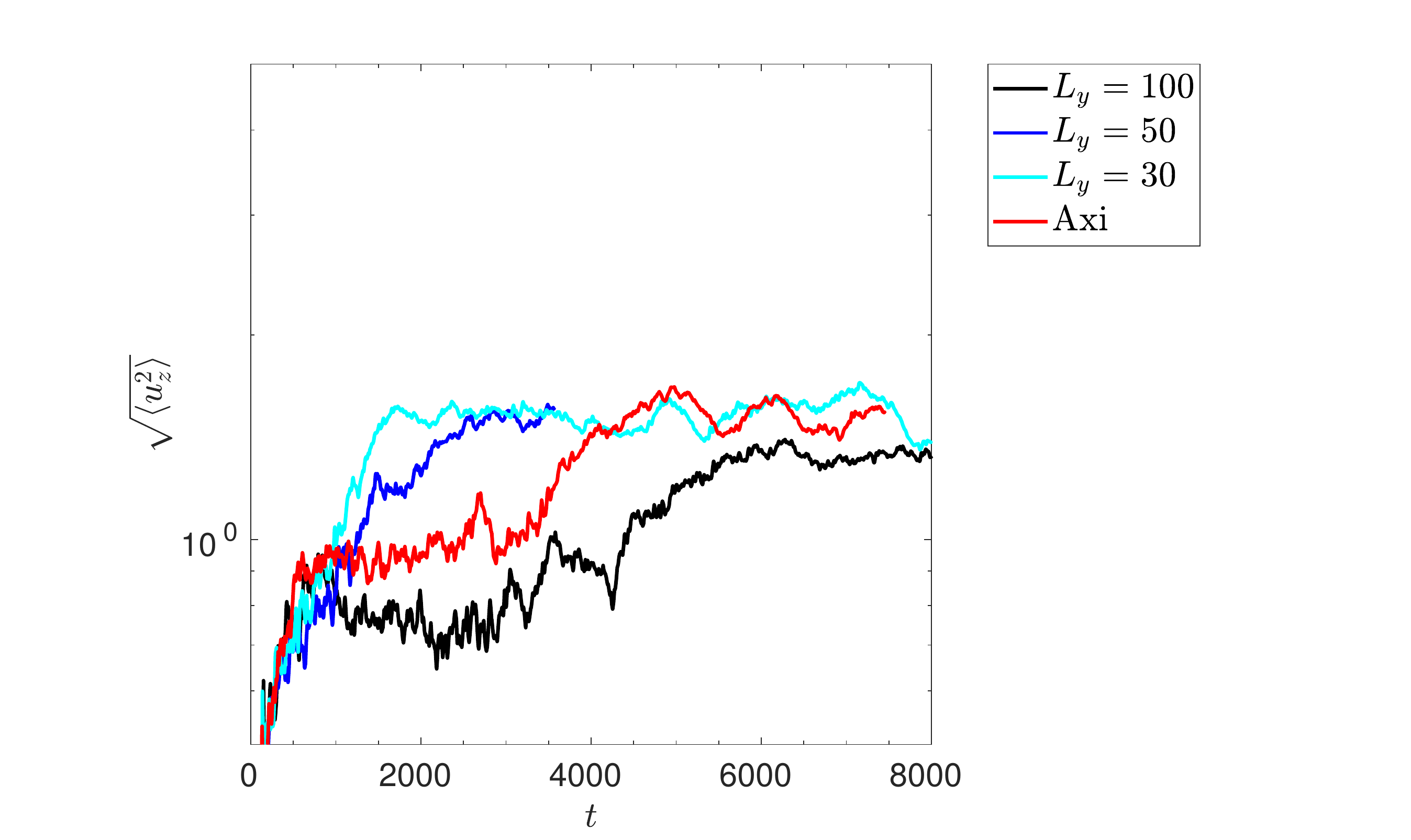}}
     \subfigure[$\langle u_x\theta\rangle$]{\includegraphics[trim=0.4cm 0cm 0cm 0cm, clip=true,width=0.45\textwidth]{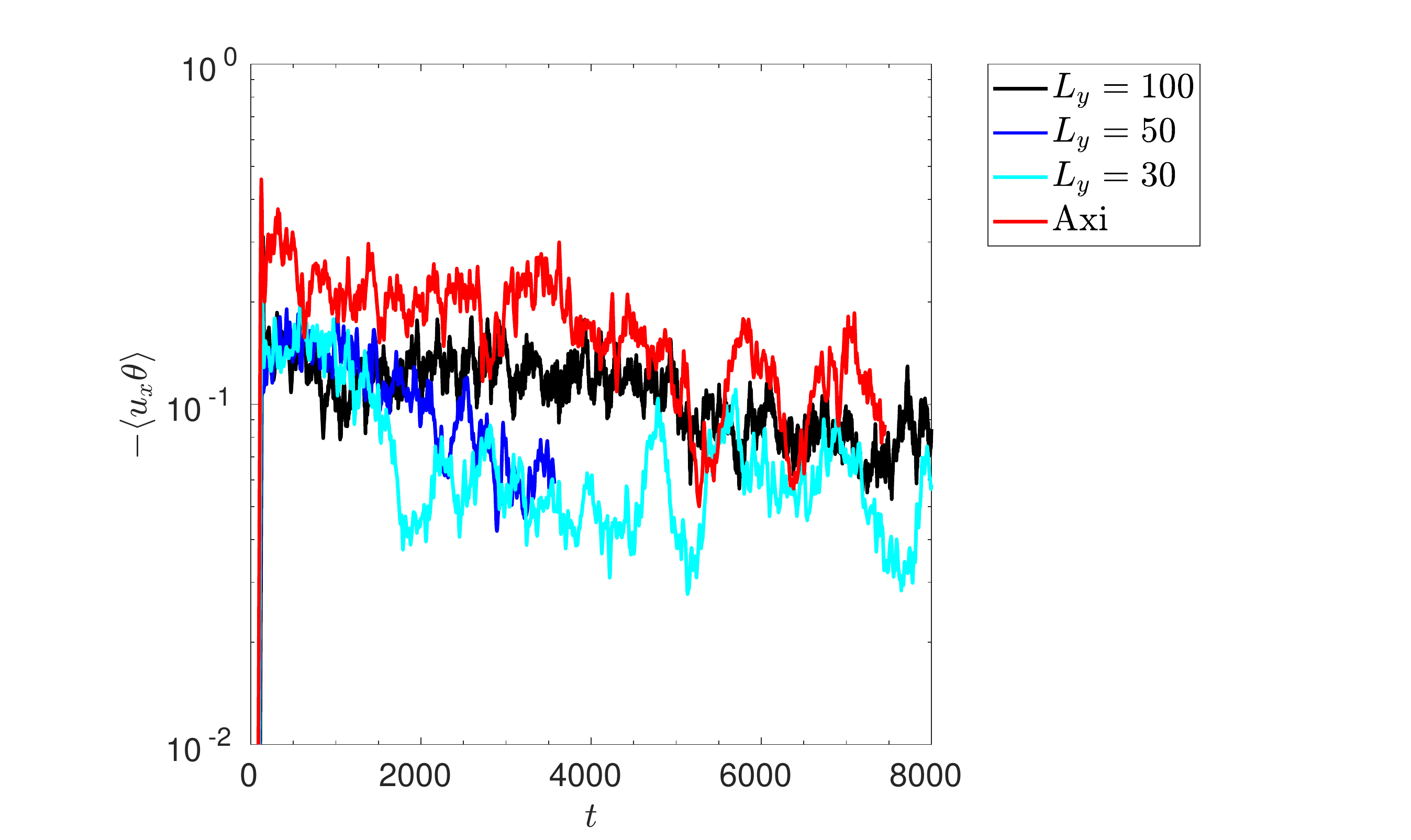}}
    \end{center}
  \caption{Temporal evolution of $K$, $\langle u_xu_y\rangle$, $v_z$, and $-\langle u_x\theta\rangle$, in a set of simulations with $S=2.1, N^2=10, \mathrm{Pr}=10^{-2}$, with various different $L_y$. The strong dependence on $L_y$ illustrates the importance of three-dimensional effects on the nonlinear evolution of the GSF instability.}
  \label{S2p1_tplots}
\end{figure}

\begin{figure*}
  \begin{center}
     \subfigure[$t=100$]{\includegraphics[trim=1cm 7cm 0.5cm 7cm, clip=true,width=0.34\textwidth]{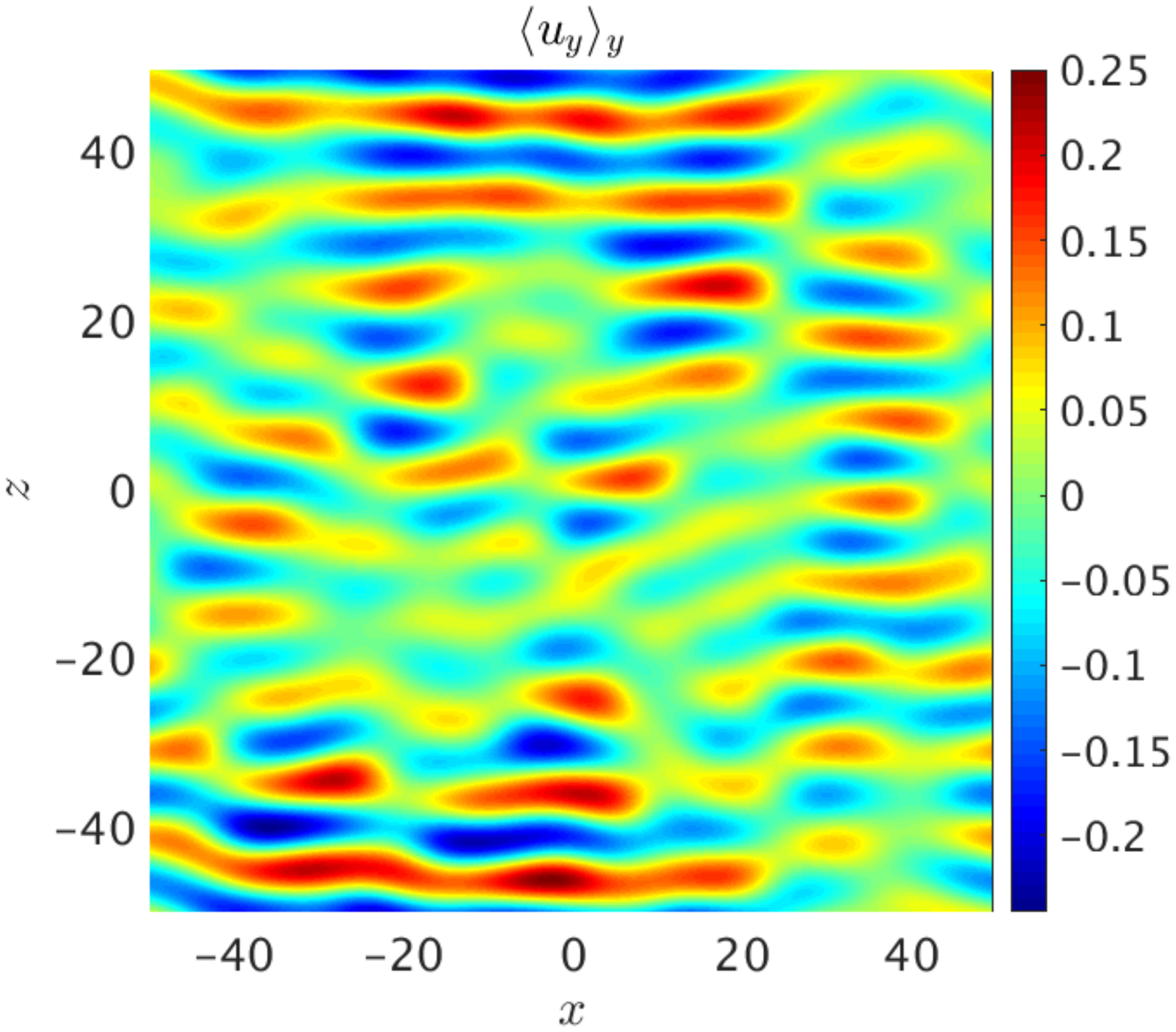}}
    \subfigure[$t=100$]{\includegraphics[trim=1cm 7cm 0.5cm 7cm, clip=true,width=0.34\textwidth]{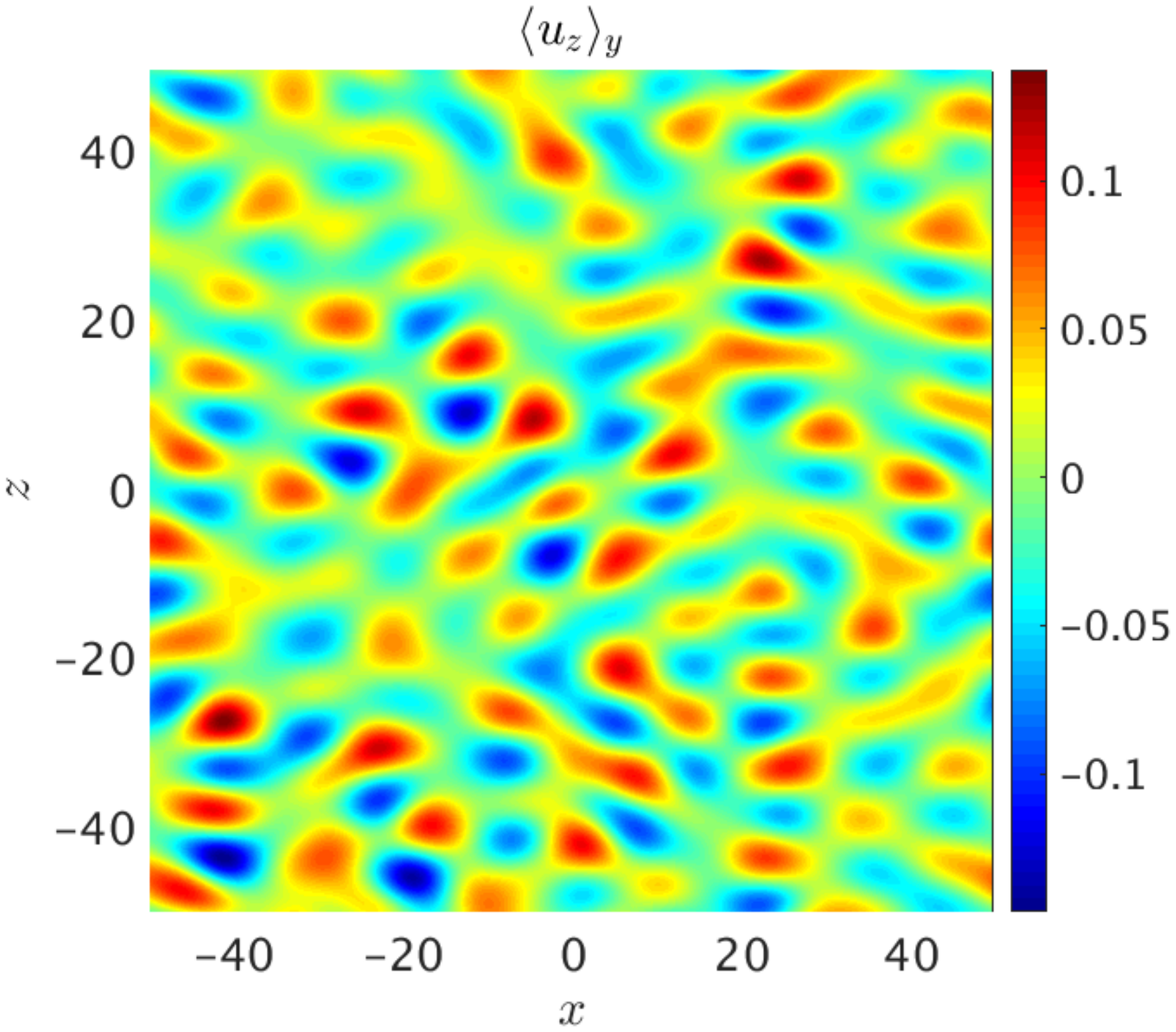}} \\
    \subfigure[$t=130$]{\includegraphics[trim=1cm 7cm 0.5cm 7cm, clip=true,width=0.34\textwidth]{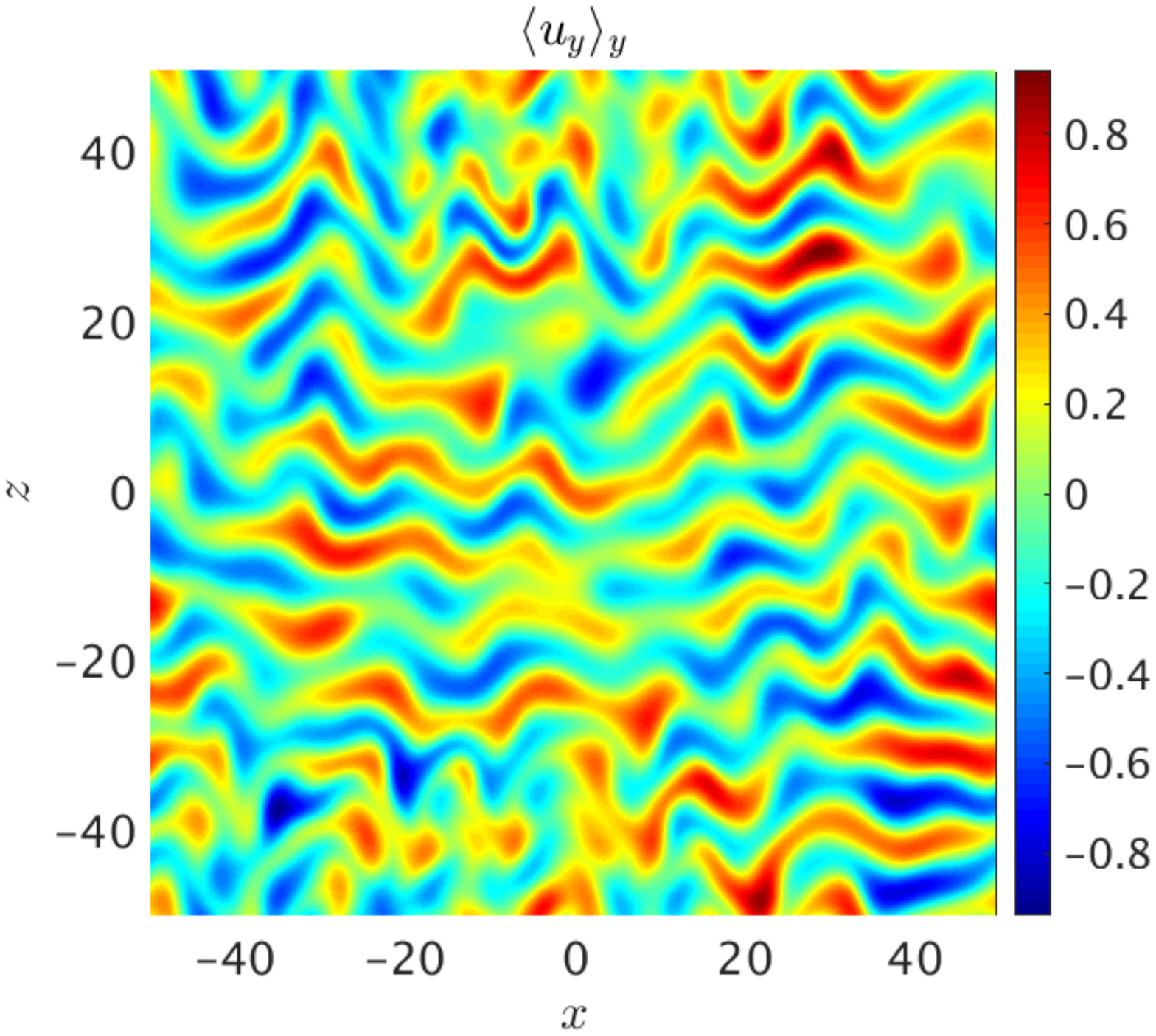}}
    \subfigure[$t=130$]{\includegraphics[trim=1cm 7cm 0.5cm 7cm, clip=true,width=0.34\textwidth]{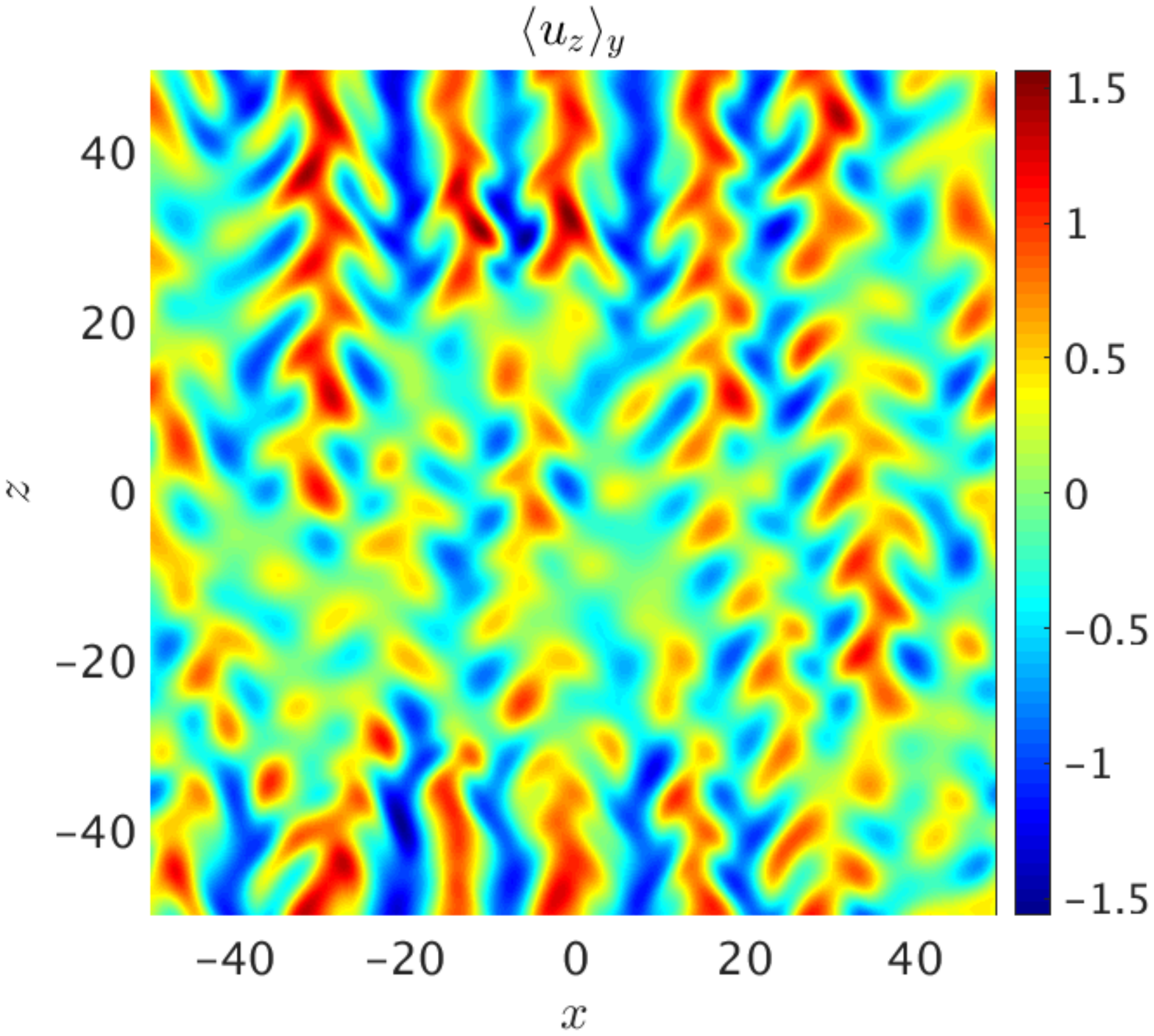}} \\
     \subfigure[$t=2000$]{\includegraphics[trim=1cm 7cm 0.5cm 7cm, clip=true,width=0.34\textwidth]{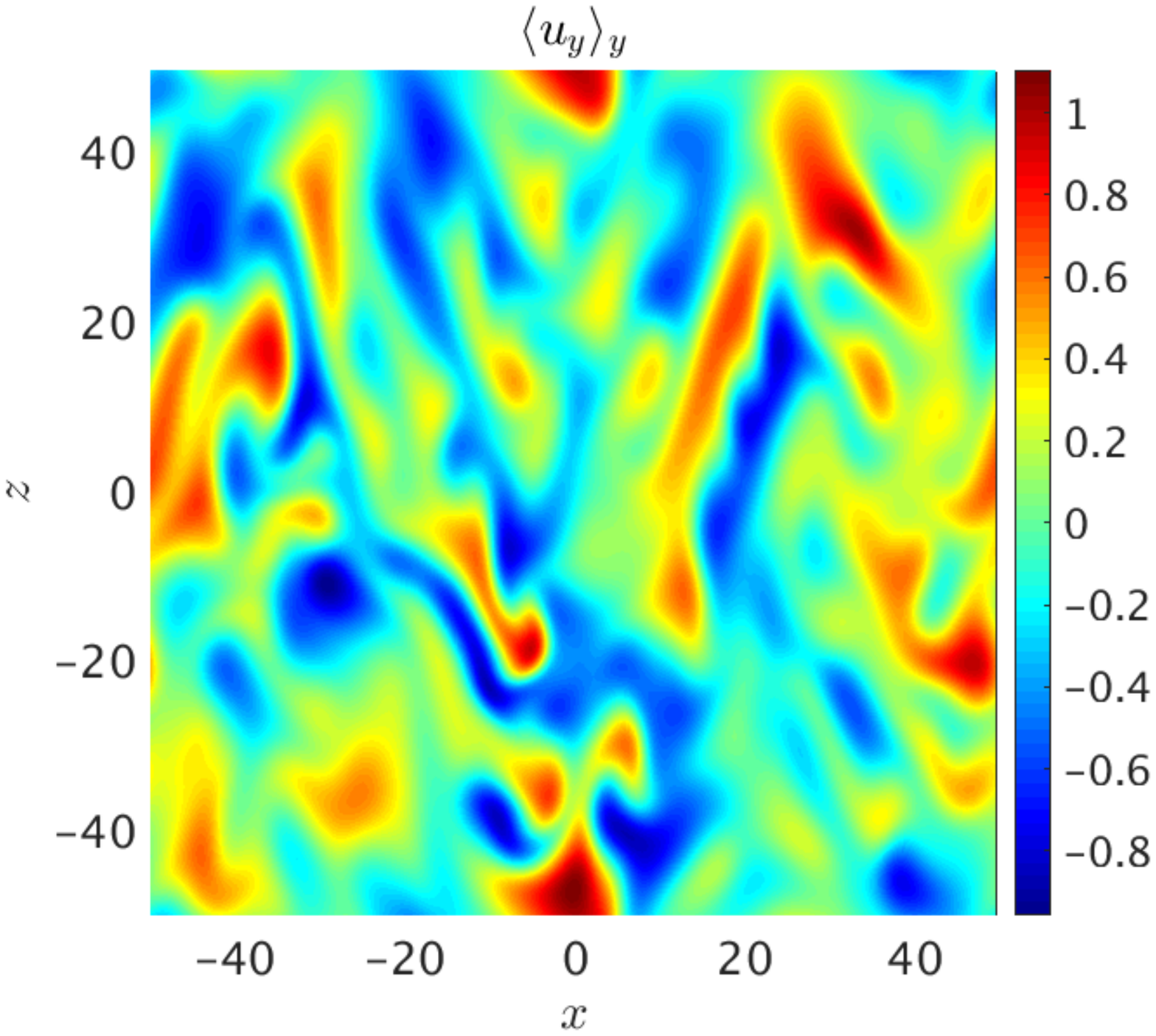}}
    \subfigure[$t=2000$]{\includegraphics[trim=1cm 7cm 0.5cm 7cm, clip=true,width=0.34\textwidth]{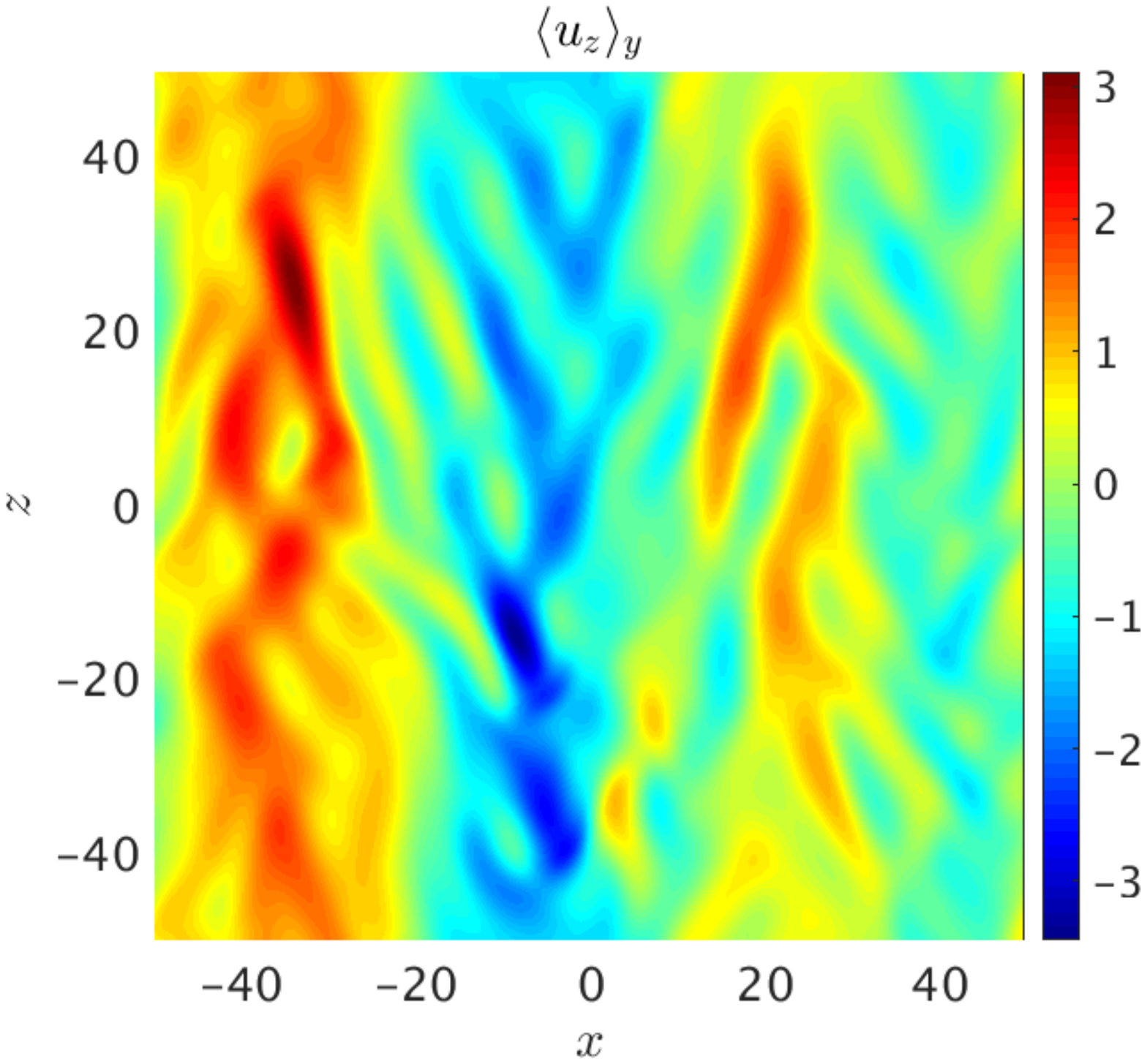}} \\
     \subfigure[$t=5000$]{\includegraphics[trim=1cm 7cm 0.5cm 7cm, clip=true,width=0.34\textwidth]{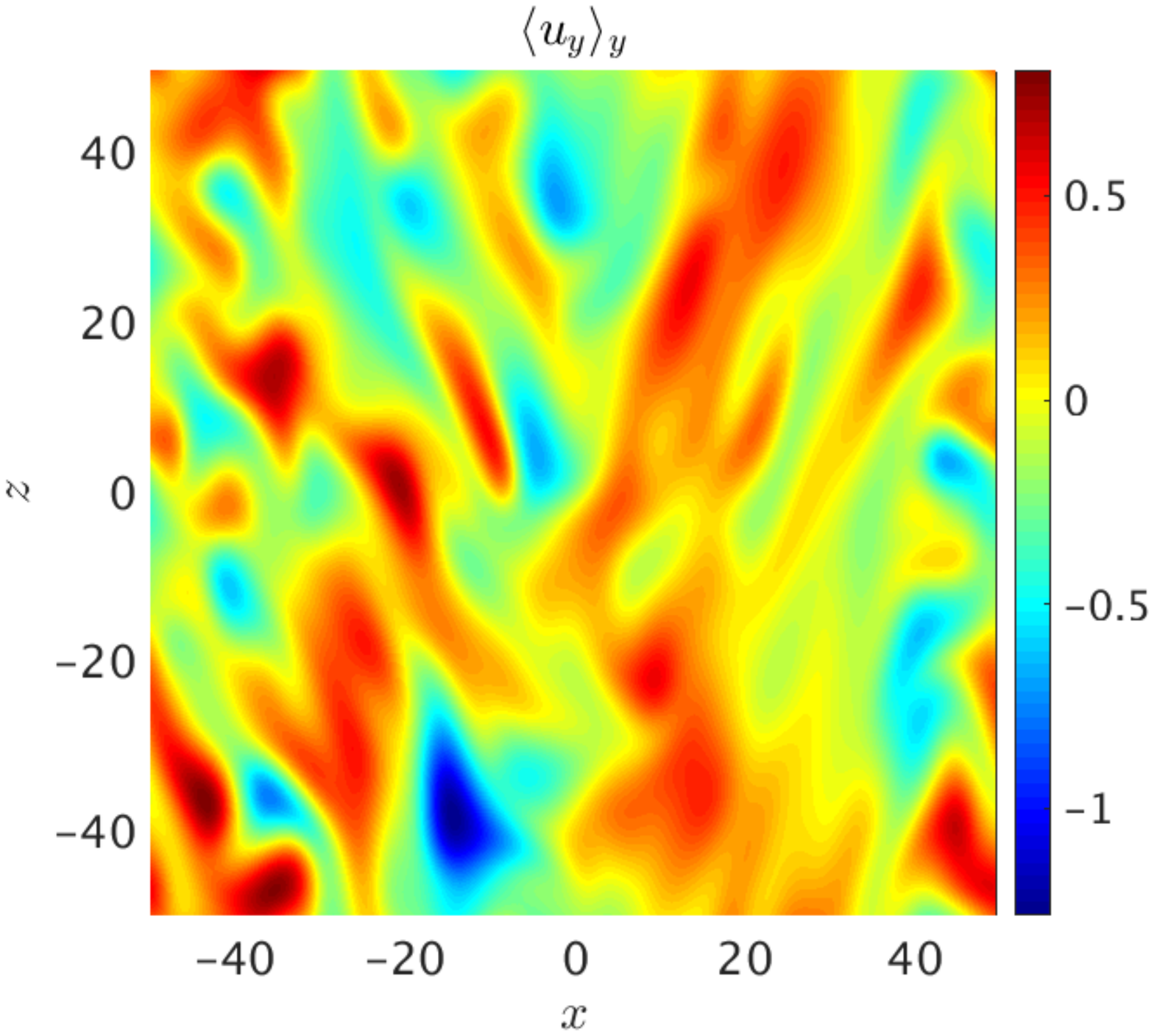}}
    \subfigure[$t=5000$]{\includegraphics[trim=1cm 7cm 0.5cm 7cm, clip=true,width=0.34\textwidth]{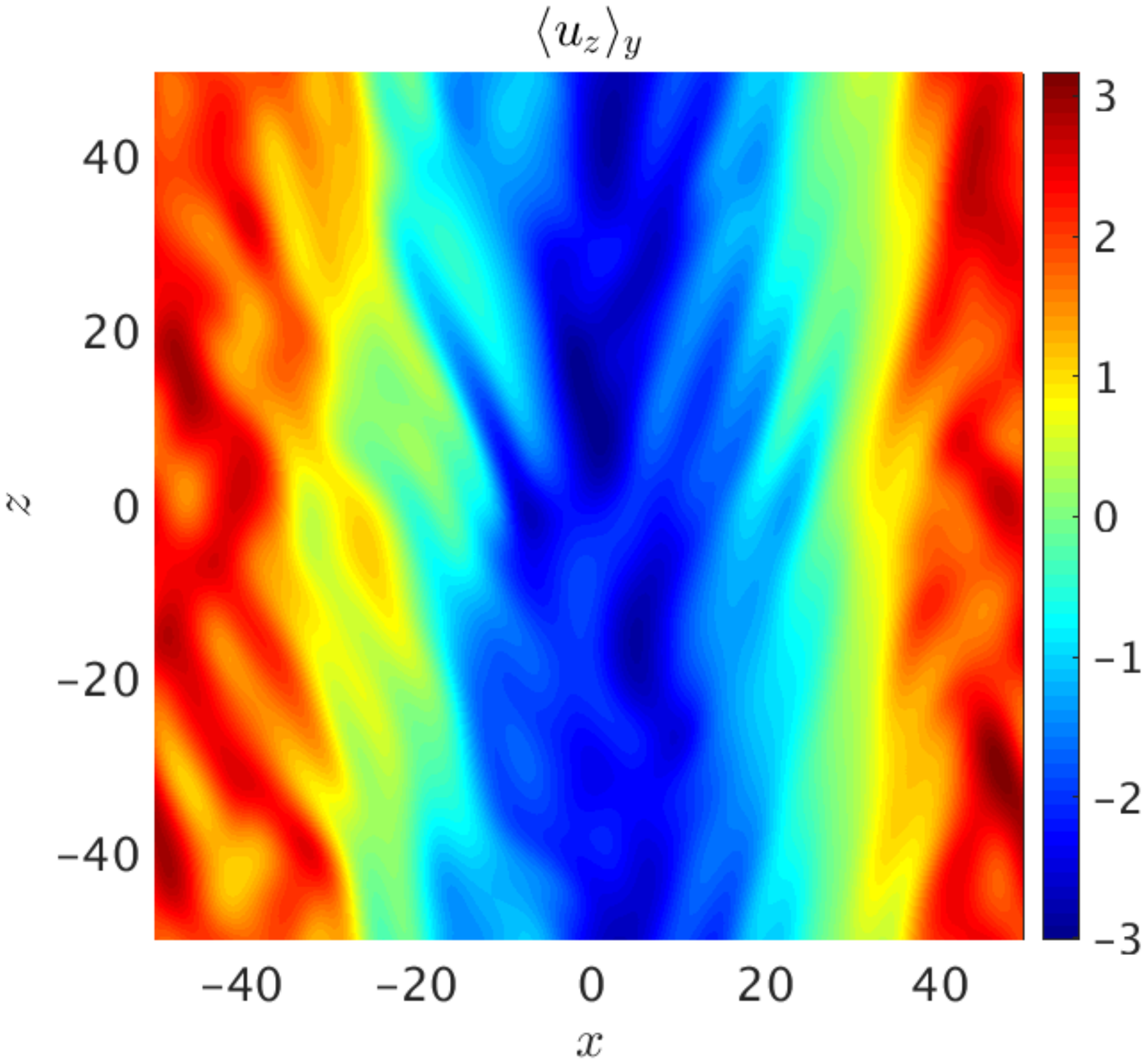}}
    \end{center}
  \caption{Snapshots of $y$-averaged $u_y$ and $u_z$ in the $(x,z)$-plane for an axisymmetric simulation with $S=2.1, N^2=10, \mathrm{Pr}=10^{-2}$ at various times. The top panels show the linear growing modes, which have $k_x\sim 0$. The middle panels show the initial nonlinear saturation, followed by the formation of latitudinal (along $z$) jets. The bottom panels shows the strong latitudinal shear flows that have developed in the later stages, and their effects on the propagation of unstable finger-like motions in $x$.}
  \label{S2p1_2Duyplots}
\end{figure*}

\begin{figure}
  \begin{center}
     \subfigure{\includegraphics[trim=1cm 7cm 0.5cm 7cm, clip=true,width=0.45\textwidth]{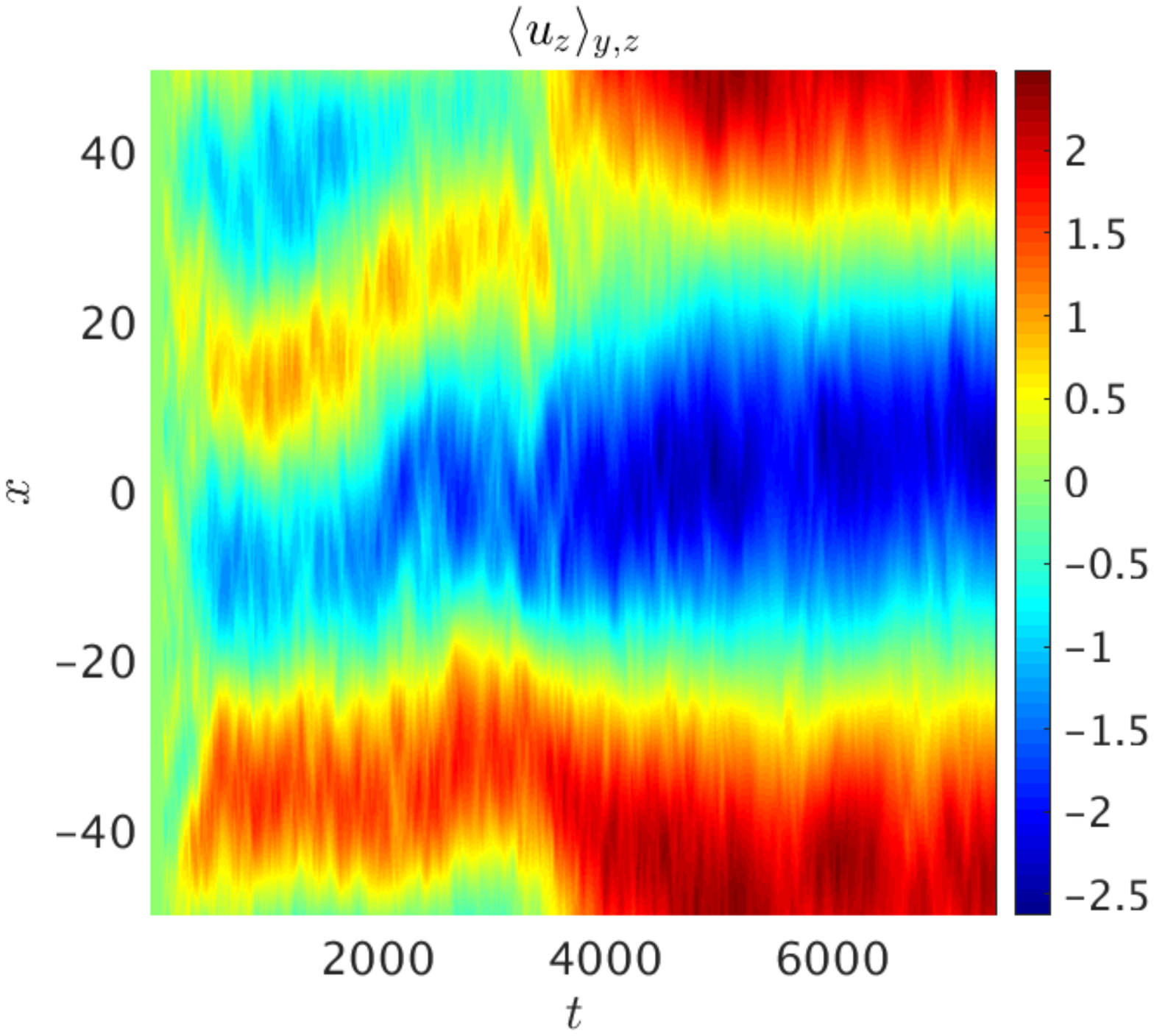}}
\end{center}
  \caption{Hovm\"{o}ller diagram showing the $x$-averaged $u_z$ velocity component as a function of $x$ and $t$. This shows the formation and the merging of latitudinal jets in the axisymmetric simulation with $S=2.1, N^2=10, \mathrm{Pr}=10^{-2}$, as also shown in Fig.~\ref{S2p1_2Duyplots}.}
  \label{Hovmoller}
\end{figure}

We begin with a set of illustrative axisymmetric simulations with Pr$=10^{-2}$, $N^2=10$, with $S=2.1$ and $S=2.5$. Note that $S>2$ is required for the shear flow to be centrifugally unstable in the absence of stratification, but that the flow is stable in the absence of thermal diffusion according to Eq.~\ref{Solberg} (which would require $S>7$ for adiabatic axisymmetric instability).

Fig.~\ref{S2p1_tplots} shows the evolution of the volume-averaged kinetic energy $K=\frac{1}{2}\langle |\boldsymbol{u}|^2\rangle$ (top panel), $\langle u_xu_y\rangle$ (top middle), the RMS latitudinal velocity $v_z=\sqrt{\langle u_z^2\rangle}$ (bottom middle), and minus the radial buoyancy flux $-\langle u_x\theta\rangle$, in simulations with $S=2.1$. The axisymmetric simulation is shown as the red line. This figure also shows the results from 3D simulations with various $L_y$, which will be discussed in \S~\ref{results}. Snapshots of the $y$-averaged $u_y$ and $u_z$ velocity components in the $(x,z)$-plane at several times are shown in Fig.~\ref{S2p1_2Duyplots} for the axisymmetric simulation with $S=2.1$.

After the initial saturation at $t\sim130$, there is a secondary growth of strong latitudinal shear flows (along $z$), which quickly dominate the energy, as shown in the third panel of Fig.~\ref{S2p1_tplots}. The spatial structure of the flow is shown in the bottom two right panels of Fig.~\ref{S2p1_2Duyplots}, which illustrates that these latitudinal flows possess significant radial (along $x$) shear. Following the development of these strong latitudinal shear flows, the angular momentum transport is significantly inhibited, and undergoes chaotic bursty dynamics, as is shown in the second panel of Fig.~\ref{S2p1_tplots}.

The development of these strong latitudinal shear flows in an axisymmetric simulation is shown in snapshots at various times in Fig.~\ref{S2p1_2Duyplots}, and as a Hovm\"{o}ller diagram ($y$ and $z$-averaged $u_z$ as a function of $x$ and $t$) in Fig.~\ref{Hovmoller}. During the linear growth phase, at $t=100$, $u_y\sim u_z$, but shortly after the initial saturation, $u_z$ jets grow, and these jets merge until this component dominates. By $t\sim 2000$, the latitudinal shear flow persists in a configuration with two wavelengths in $x$, and strongly affects the radial propagation of finger-like motions, as can be seen from $u_y$ in the bottom two left panels. The latitudinal shear flow then merges to form a single wavelength in $x$ by $t\sim 4000$, which further enhances the shear and reduces the angular momentum transport by shearing the radial fingers and reducing their radial extent. The buoyancy flux is also reduced when strong shears develop, which indicates that these jets act as barriers to transport, much like zonal flows (e.g.~\citealt{Diamond2005}).

\begin{figure}
  \begin{center}
     \subfigure[$K$]{\includegraphics[trim=0cm 0cm 0cm 0cm, clip=true,width=0.45\textwidth]{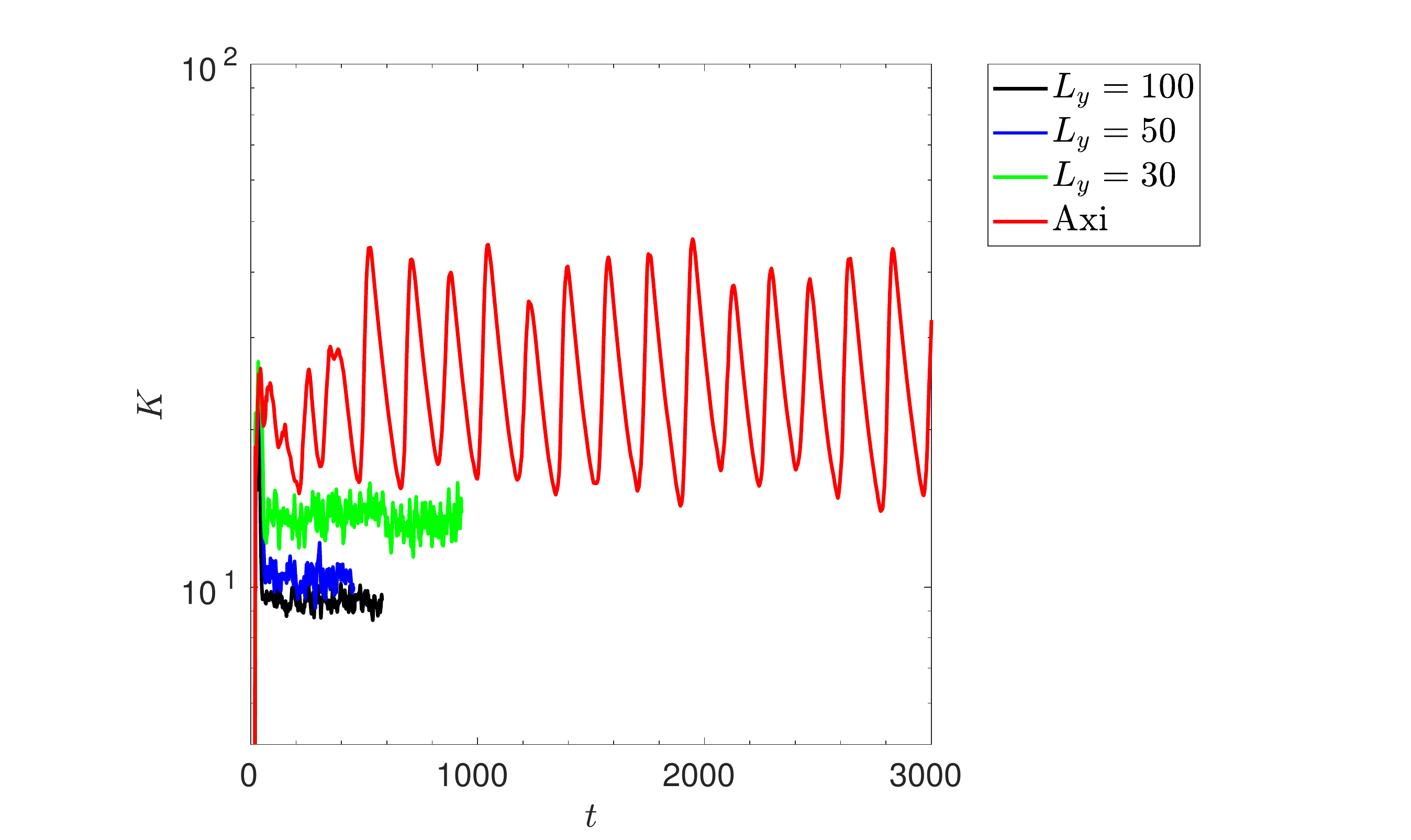}}
    \subfigure[$\langle u_xu_y\rangle$]{\includegraphics[trim=0.4cm 0cm 0cm 0cm, clip=true,width=0.45\textwidth]{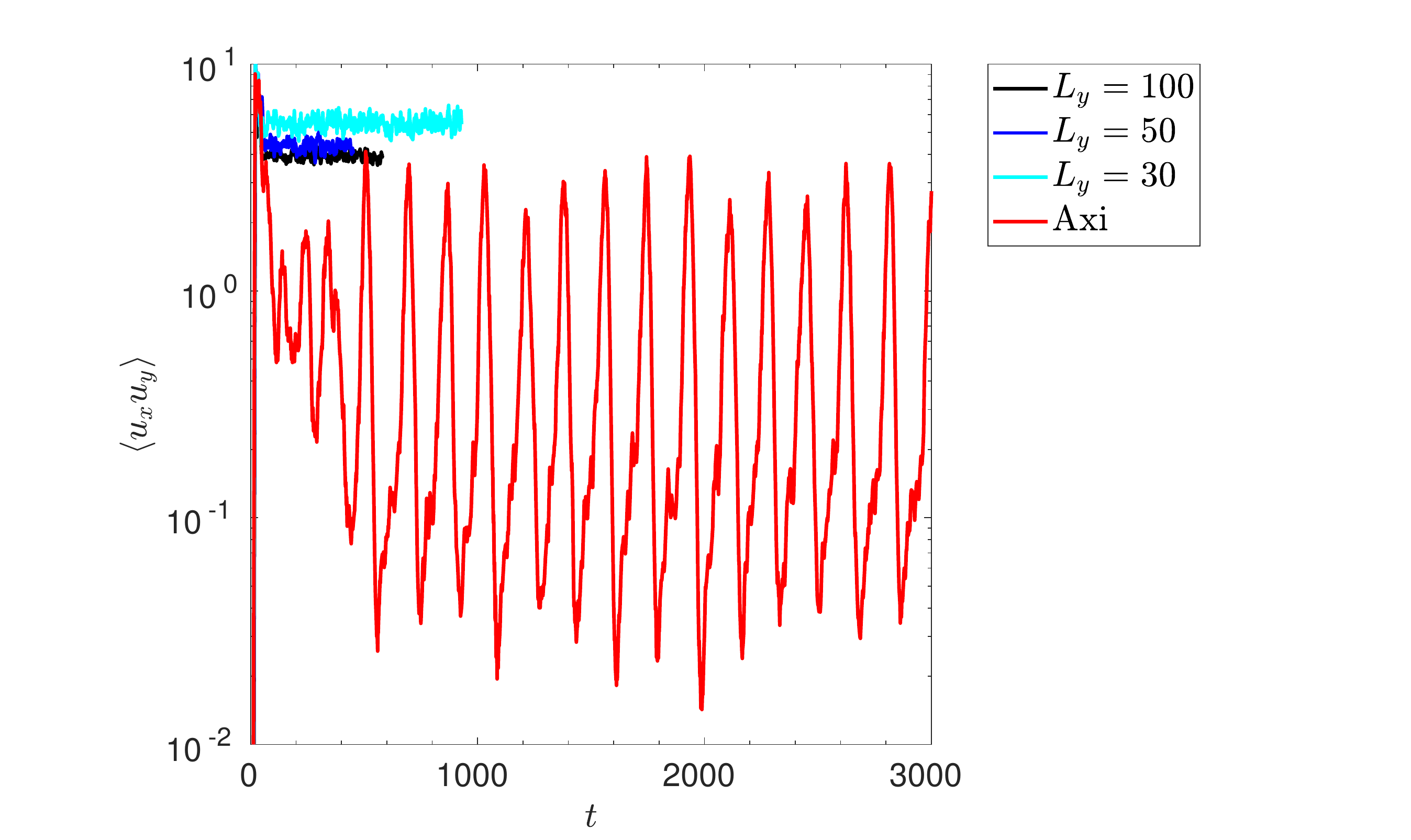}}
    \subfigure[$\sqrt{\langle u_z^2\rangle}$]{\includegraphics[trim=0.4cm 0cm 0cm 0cm, clip=true,width=0.45\textwidth]{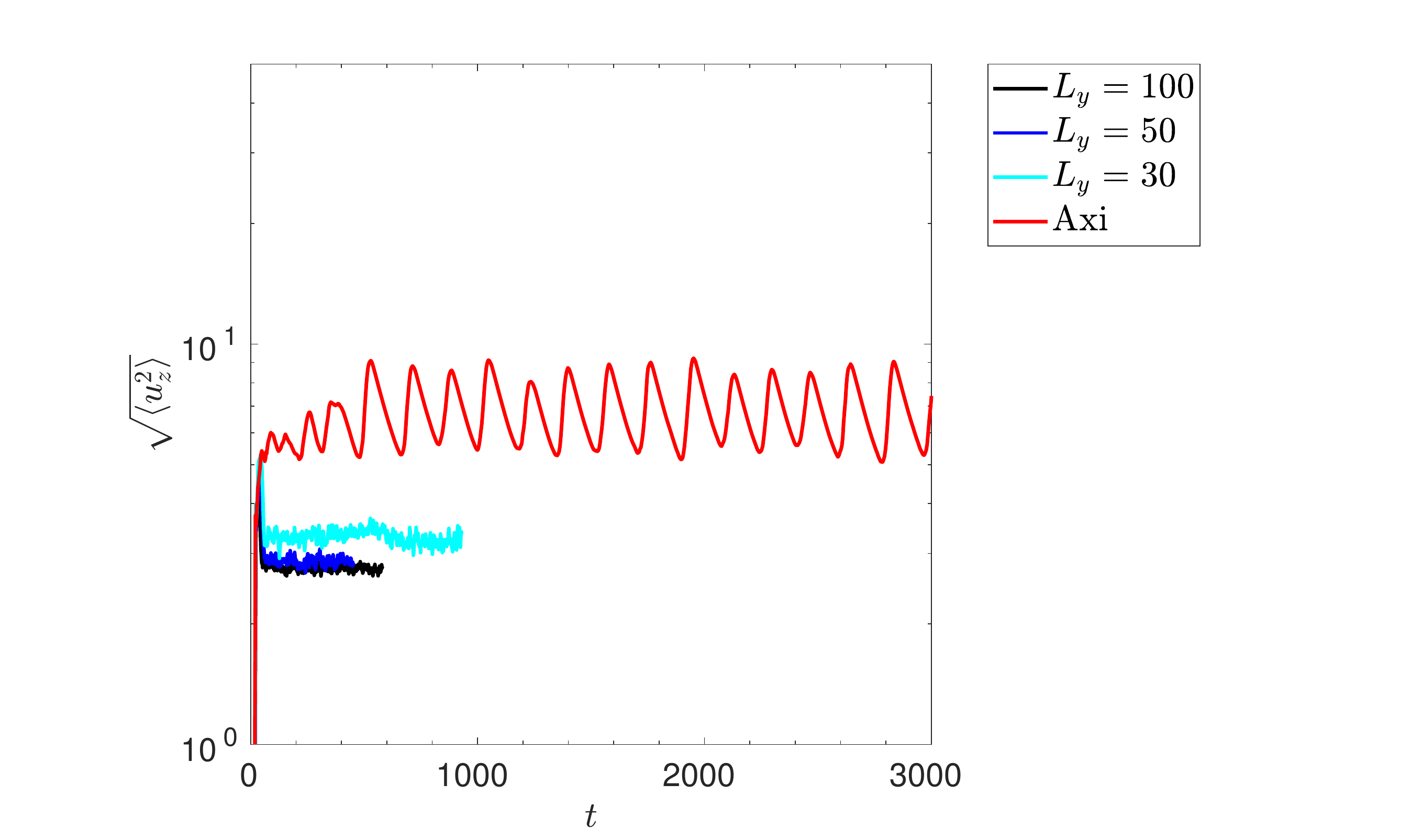}}
     \subfigure[$\langle u_x\theta\rangle$]{\includegraphics[trim=0.4cm 0cm 0cm 0cm, clip=true,width=0.45\textwidth]{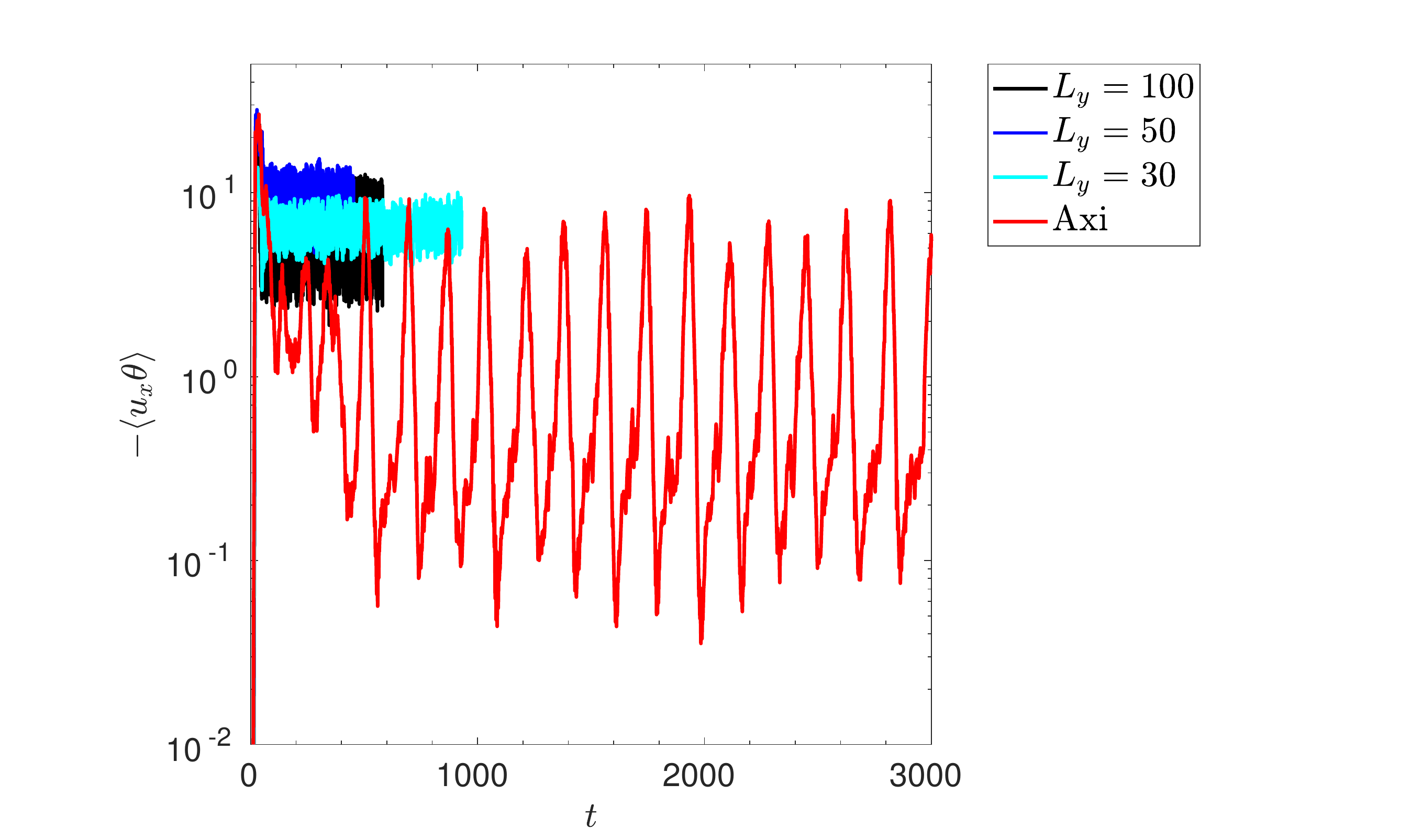}}
    \end{center}
  \caption{Temporal evolution of $K$, $\langle u_xu_y\rangle$, $v_z$, and $-\langle u_x\theta\rangle$, for a set of simulations with $S=2.5, N^2=10, \mathrm{Pr}=10^{-2}$ with various different $L_y$. This clearly illustrates the importance of three-dimensional effects on the nonlinear evolution of the GSF instability.}
  \label{S2p5_tplots}
\end{figure}

The latitudinal shear flows are even more pronounced in simulations with the stronger shear of $S=2.5$. In Fig.~\ref{S2p5_tplots} we show the evolution of the same volume-averaged quantities as in Fig.~\ref{S2p1_tplots} for these simulations (with the axisymmetric case shown in red), along with comparison 3D simulations which will be described further in \S~\ref{results}. The latitudinal flows in this simulation are not shown but are similar to those in Fig.~\ref{S2p1_2Duyplots} except that they are stronger.
Axisymmetric simulations exhibit bursty dynamics in which strong latitudinal jets inhibit instability in a cyclic manner reminiscent of predator-prey dynamics. This is similar to the effects of zonal flows driven by convection in a rotating annulus (e.g.~\citealt{RotvigJones2006,TOM2018}), and also those driven by the elliptical instability (e.g.~\citealt{Barker2016}). Rapid cyclic transitions occur between a state with strong latitudinal jets and weak momentum transport, and a state with weaker latitudinal jets and stronger momentum transport. We have also explored cases with even stronger shears ($S>2.5$), and these behave in a qualitatively similar manner, except that larger $S$ leads to even more violent bursty dynamics.

Similar evolution, including the generation of latitudinal jets, has been observed in the analogous two-dimensional salt fingering problem by \citet{GaraudBrummell2015}, and also in an asymptotically reduced model of this system by \cite{Xie2019}.

\subsection{Illustrative three-dimensional simulations}
\label{results}

\begin{figure*}
  \begin{center}
  \subfigure[$t=100$]{\includegraphics[trim=1cm 7cm 0.5cm 7cm, clip=true,width=0.35\textwidth]{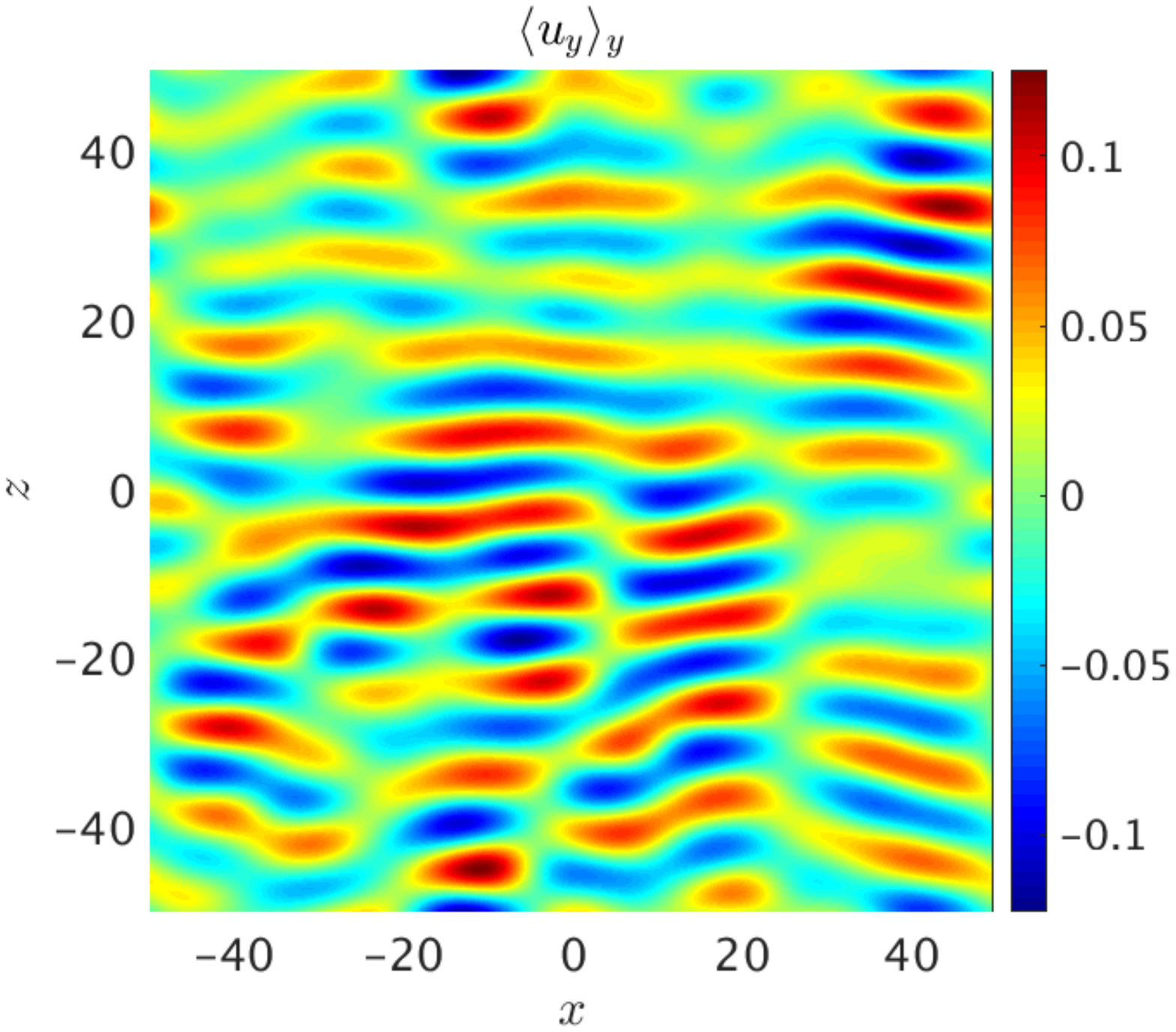}}
    \subfigure[$t=100$]{\includegraphics[trim=1cm 7cm 0.5cm 7cm, clip=true,width=0.35\textwidth]{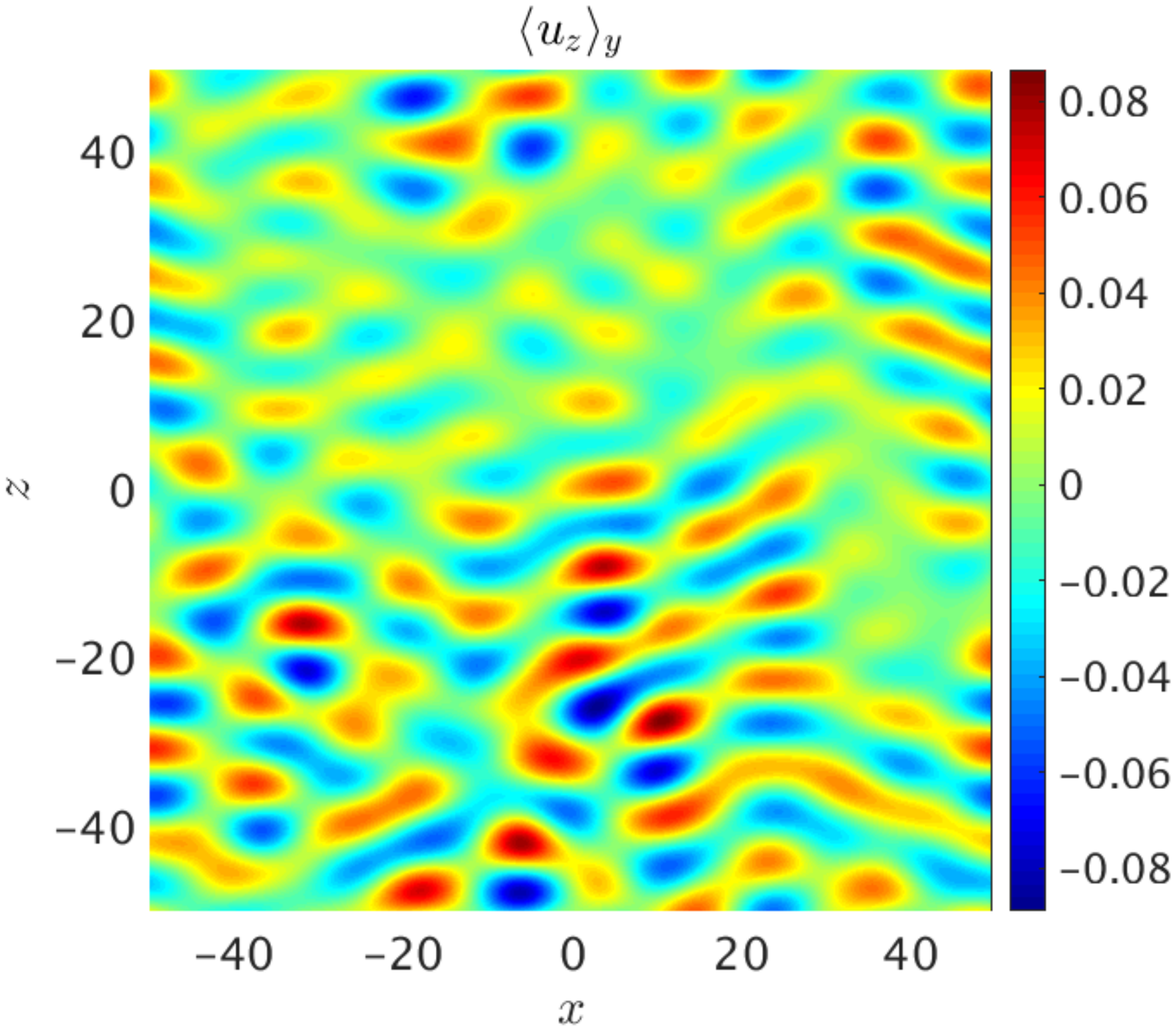}}
    \subfigure[$t=130$]{\includegraphics[trim=1cm 7cm 0.5cm 7cm, clip=true,width=0.35\textwidth]{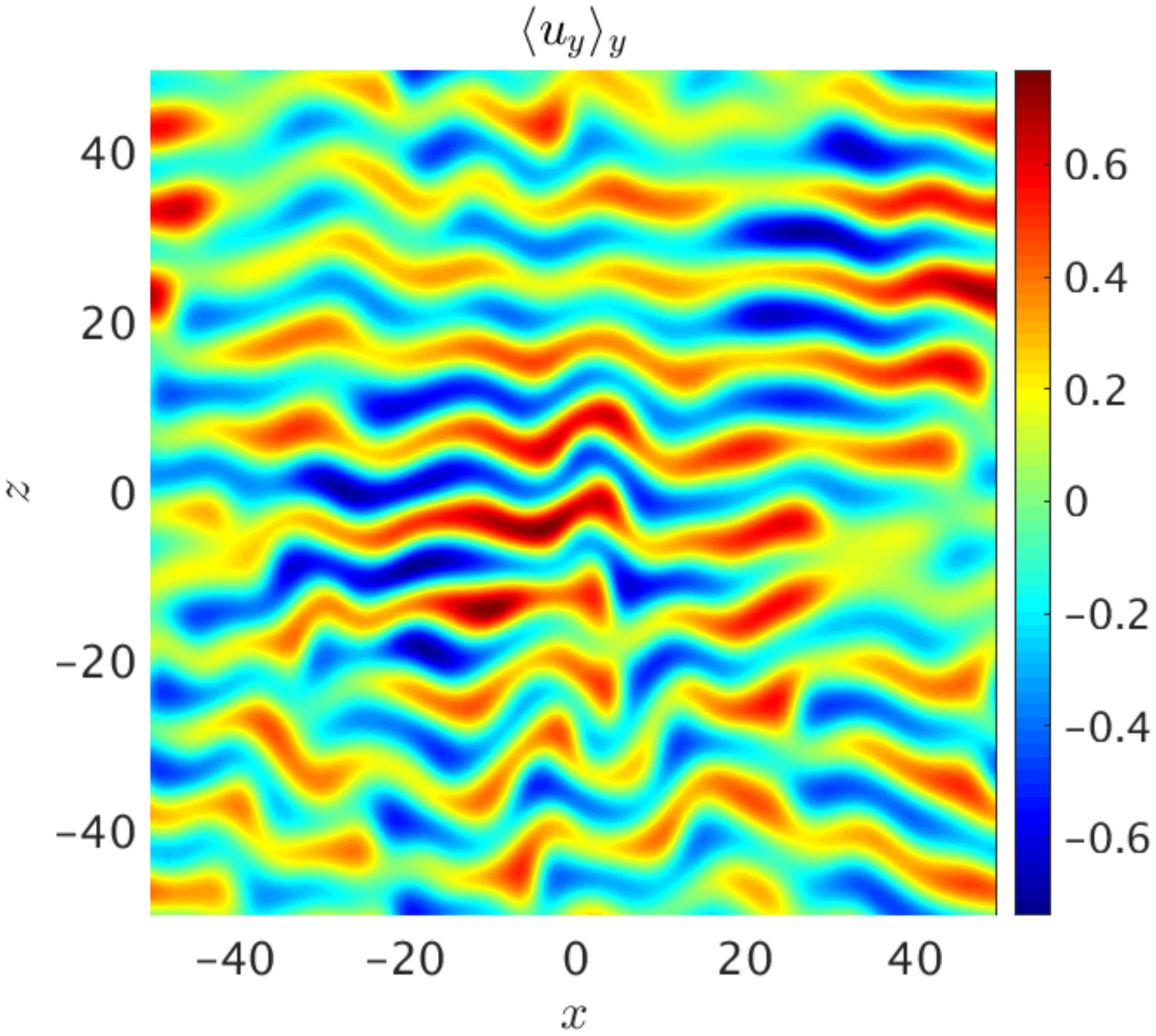}}
    \subfigure[$t=130$]{\includegraphics[trim=1cm 7cm 0.5cm 7cm, clip=true,width=0.35\textwidth]{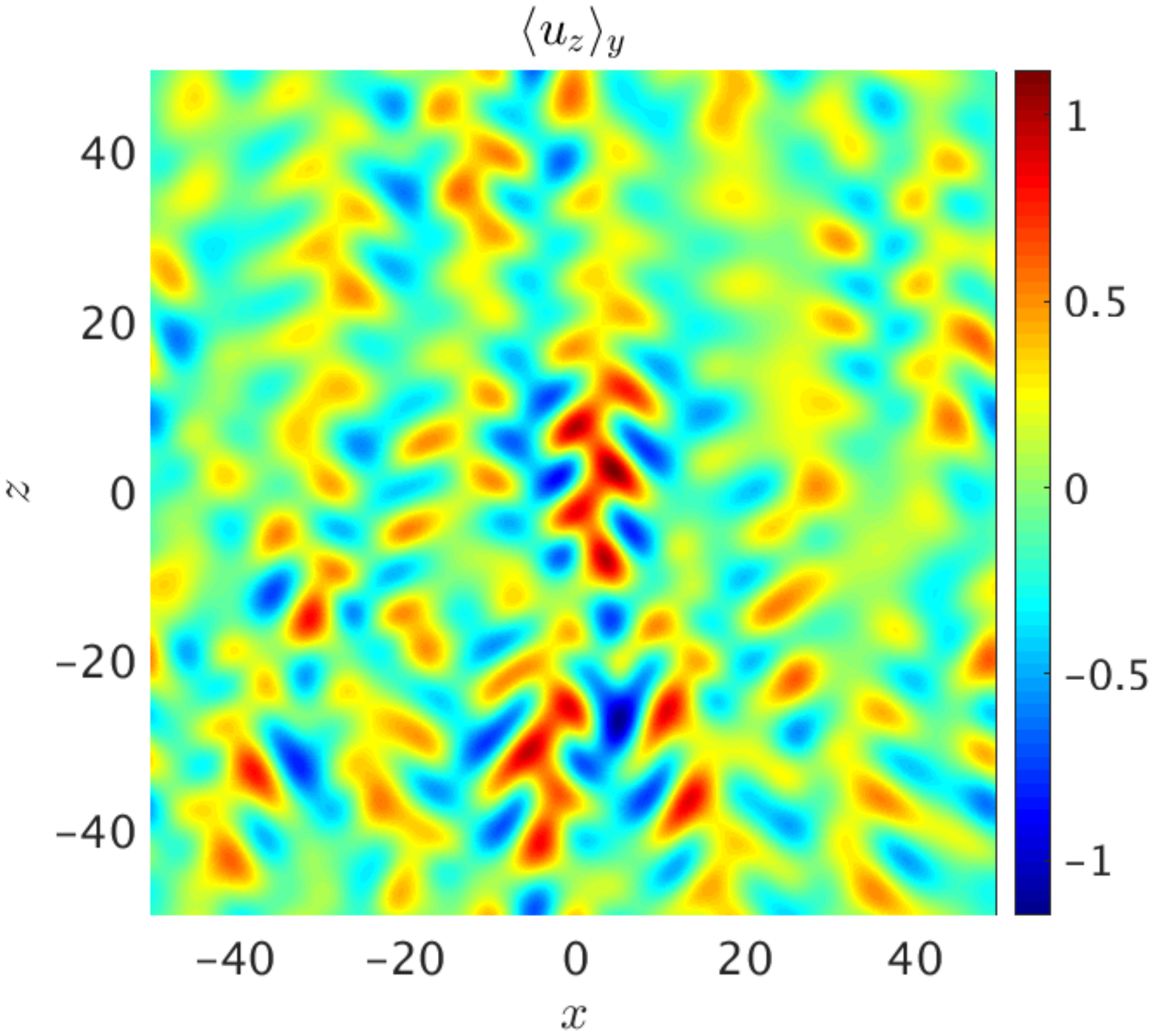}}
     \subfigure[$t=2000$]{\includegraphics[trim=1cm 7cm 0.5cm 7cm, clip=true,width=0.35\textwidth]{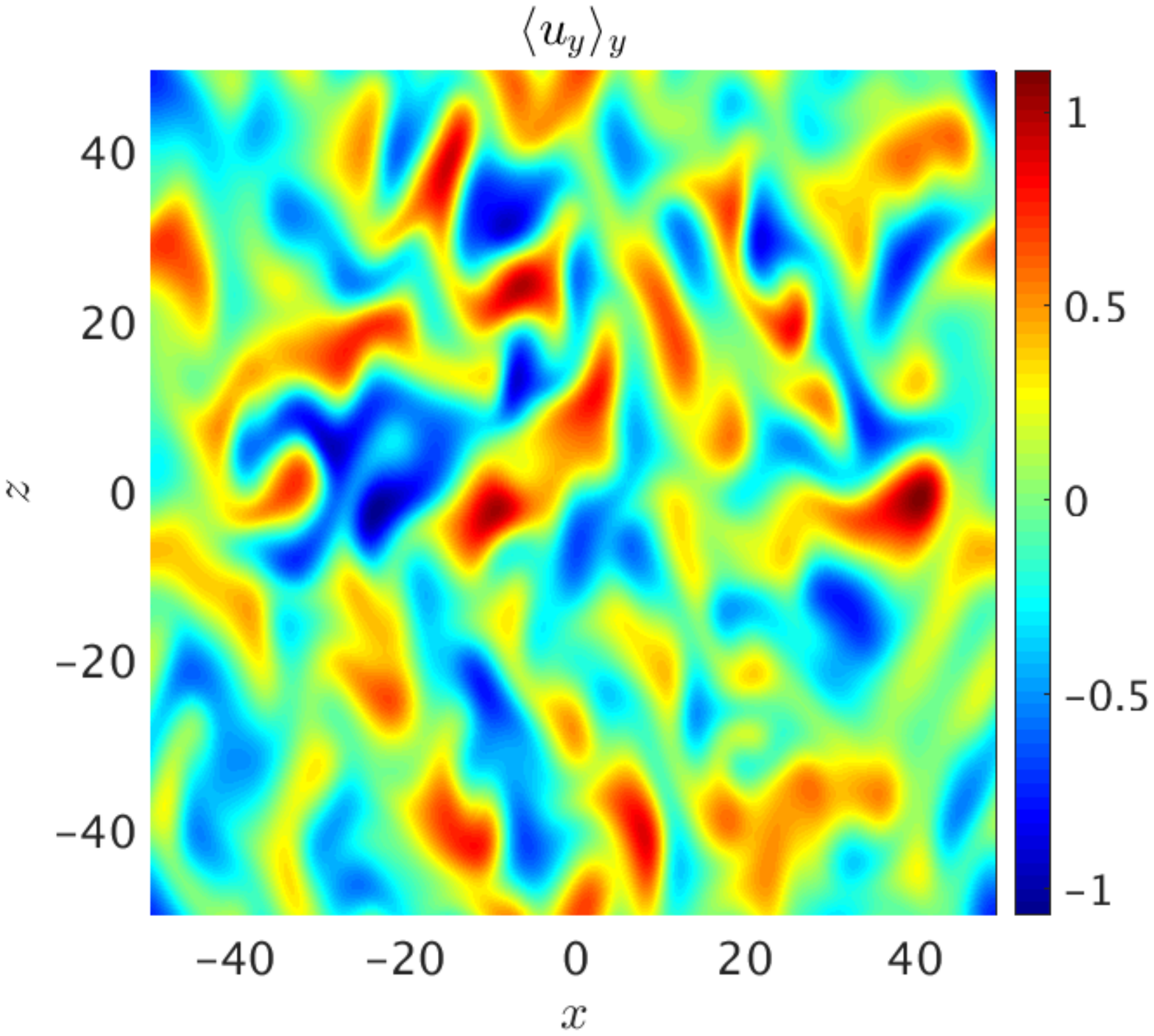}}
    \subfigure[$t=2000$]{\includegraphics[trim=1cm 7cm 0.5cm 7cm, clip=true,width=0.35\textwidth]{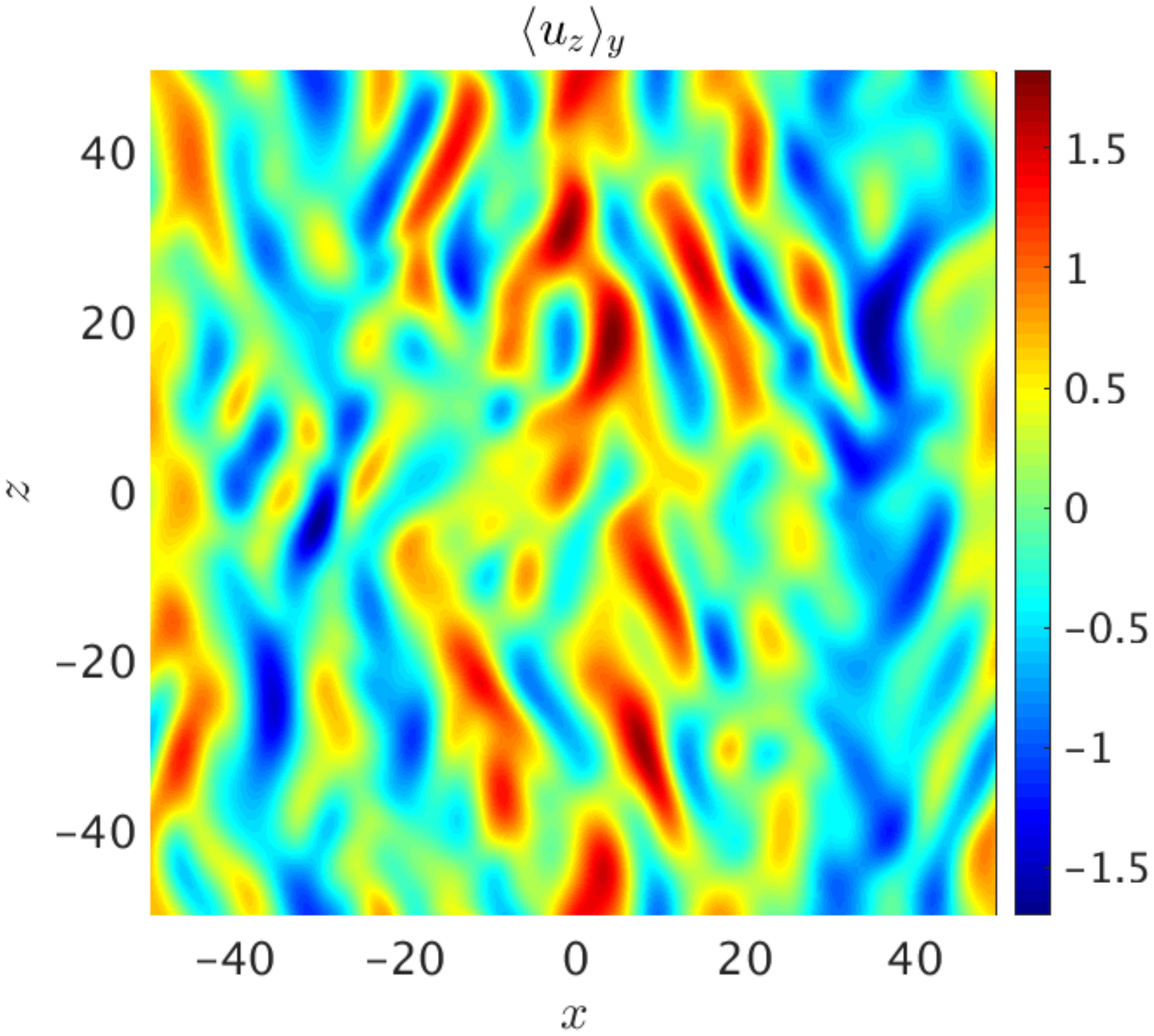}}
     \subfigure[$t=8000$]{\includegraphics[trim=1cm 7cm 0.5cm 7cm, clip=true,width=0.35\textwidth]{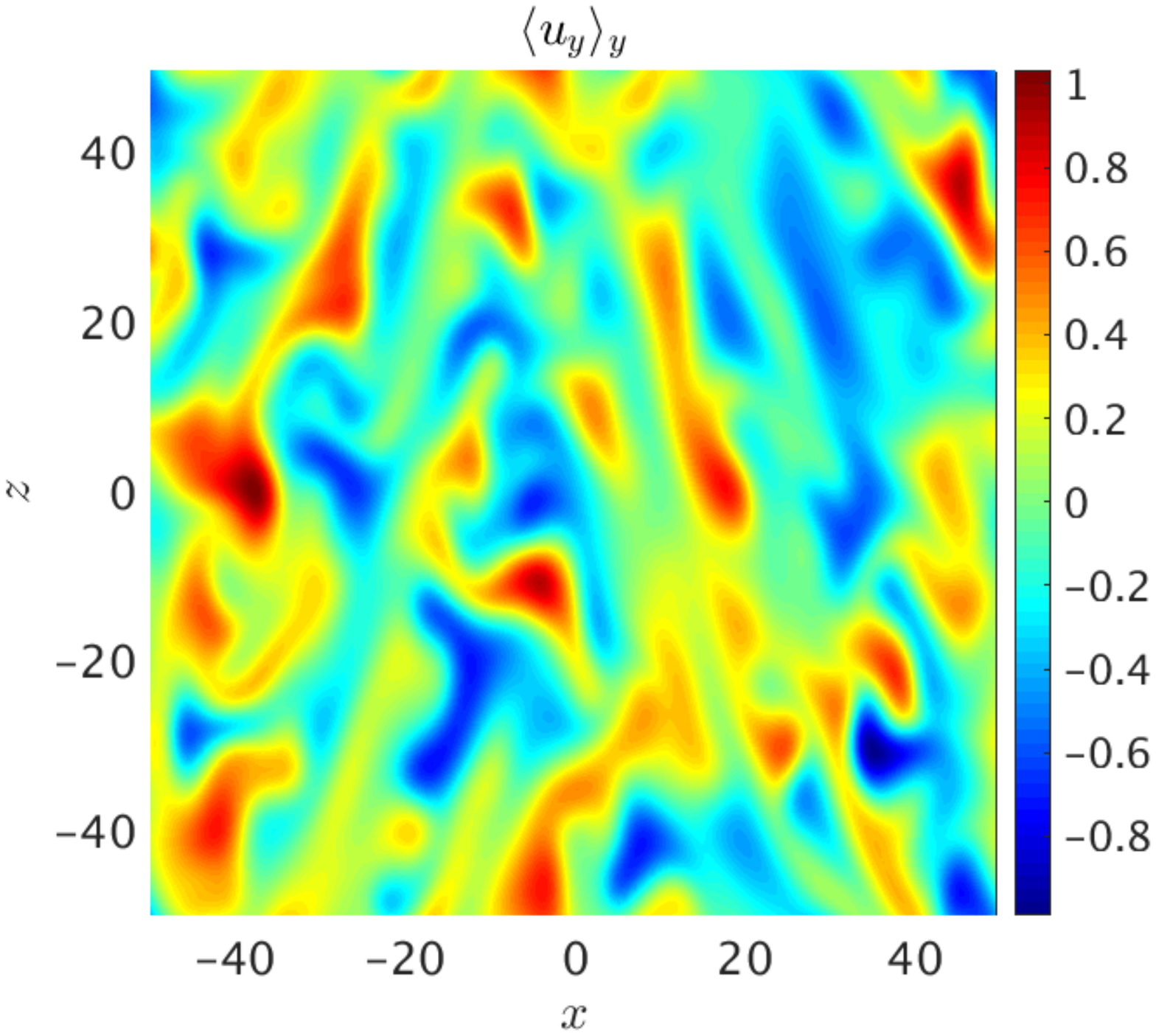}}
    \subfigure[$t=8000$]{\includegraphics[trim=1cm 7cm 0.5cm 7cm, clip=true,width=0.35\textwidth]{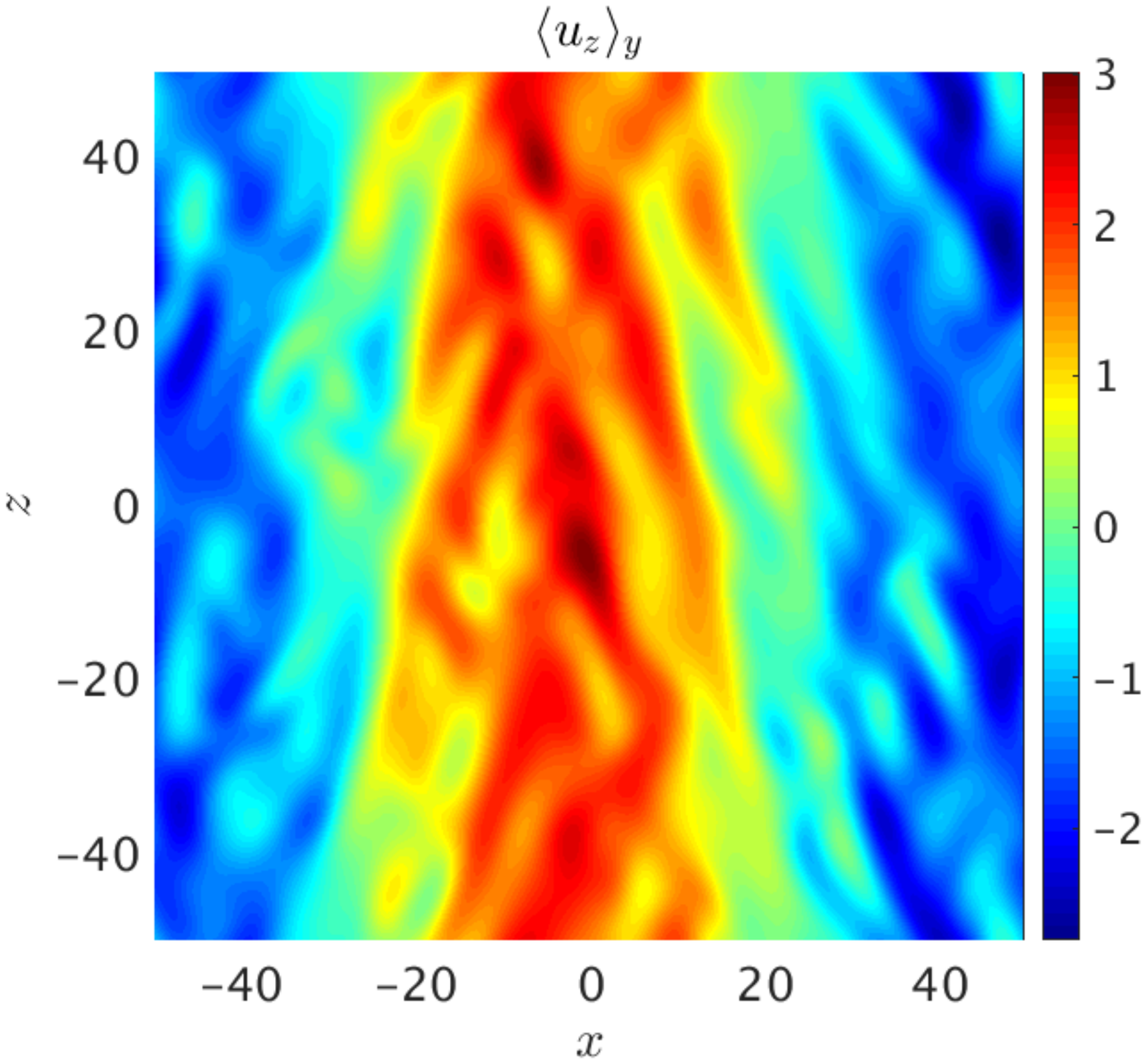}}
    \end{center}
  \caption{Snapshots of $y$-averaged $u_y$ and $u_z$ in the $(x,z)$-plane for a 3D simulation with $S=2.1, N^2=10, \mathrm{Pr}=10^{-2}$ and $L_y=100$ at various times. This can be compared with the axisymmetric simulation in Fig.~\ref{S2p1_2Duyplots} and demonstrates that three-dimensional effects can be important for the initial evolution, though in this case similar latitudinal jets eventually form.}
  \label{S2p1_3Duyplots}
\end{figure*}

We now present 3D simulations with various $L_y$ values with otherwise the same parameters as in \S~\ref{2Dresults} to explore the importance of three-dimensional effects, and to explore whether the strong latitudinal shear flows are a robust feature. The importance of three-dimensional effects can be seen in Fig.~\ref{S2p1_tplots}, which shows the evolution of volume-averaged quantities for several 3D simulations with $S=2.1$, where results with $L_y=30,50$ and 100 can be compared with the axisymmetric simulation. Snapshots of the $y$-averaged $u_y$ and $u_z$ velocity components in the $(x,z)$-plane at several times are then shown in Fig.~\ref{S2p1_3Duyplots} for the 3D simulation with $L_y=100$, which can be compared with the axisymmetric case in Fig.~\ref{S2p1_2Duyplots}.

The axisymmetric and 3D simulations behave similarly until just after the initial saturation of the instability. However, the subsequent evolution, including the generation of strong latitudinal shear flows, depends strongly on $L_y$, and hence three-dimensional effects are important. The axisymmetric simulations, and the 3D ones with $L_y=30$ and $50$, exhibit much larger kinetic energies than the case with $L_y=100$ until $t\sim 4000$, as is shown in the top panel of Fig.~\ref{S2p1_tplots}.
The time taken for latitudinal shear flows to grow is observed to depend on $L_y$ (and may also depend on the initial conditions).

The 3D simulation with $L_y=100$ behaves similarly to the axisymmetric simulation in the early nonlinear phases, but by $t\sim 2000$, it still has $u_z\sim u_y$, and strong latitudinal jets are absent at this stage. The flow is instead closer to a homogeneous turbulence state. This is shown in snapshots at various times in Fig.~\ref{S2p1_3Duyplots}. The initial absence of strong latitudinal jets that advect and stretch the unstable motions in $x$ leads to enhanced, and persistent, momentum (and buoyancy) transport relative to cases with smaller $L_y$, as is shown in the second panel of Fig.~\ref{S2p1_tplots}. However, the latitudinal jets do eventually develop in this example (and are shown in the bottom right panel of Fig.~\ref{S2p1_3Duyplots}), even if their effects on the flow are somewhat weaker than in the corresponding axisymmetric simulation.

\begin{figure}
  \begin{center}
  \subfigure{\includegraphics[trim=1cm 7cm 0.5cm 7cm, clip=true,width=0.35\textwidth]{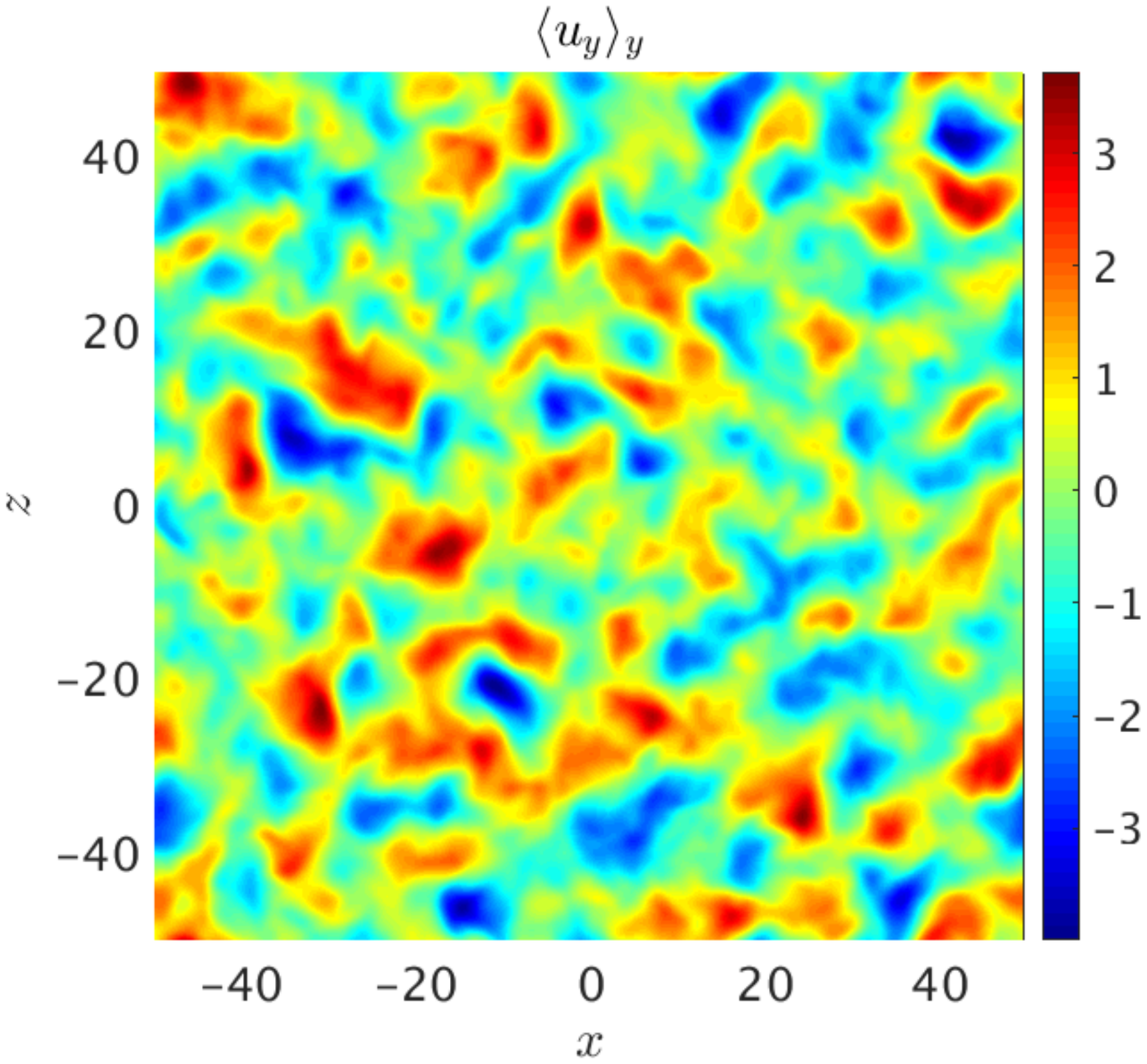}}
    \subfigure{\includegraphics[trim=1cm 7cm 0.5cm 7cm, clip=true,width=0.35\textwidth]{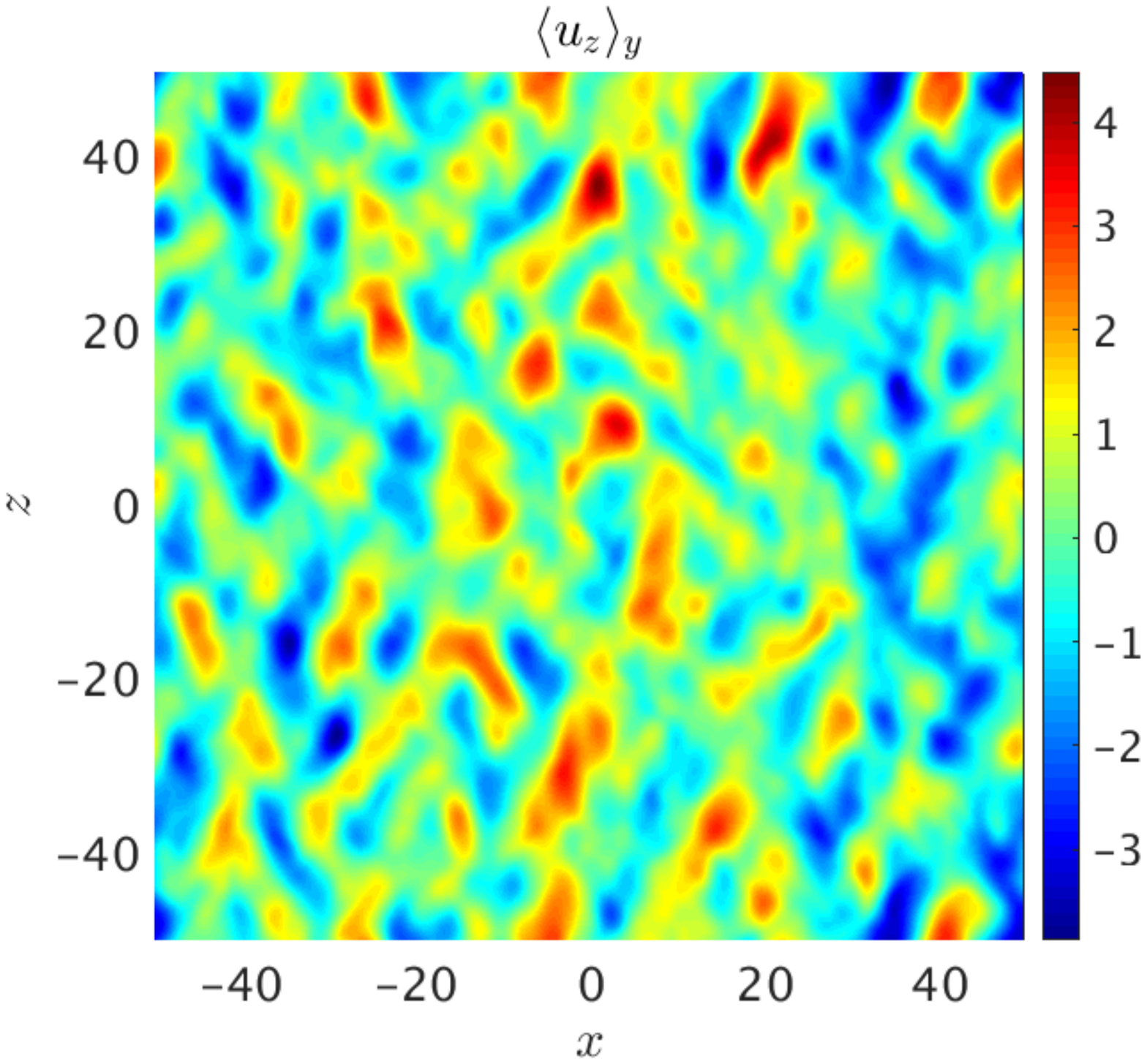}}
    \end{center}
  \caption{Snapshots of $y$-averaged $u_y$ and $u_z$ in the $(x,z)$-plane for a 3D simulation with $S=2.5, N^2=10, \mathrm{Pr}=10^{-2}$ and $L_y=100$ at $t=100$ in the saturated quasi-homogeneous turbulent state. The flow is very different from the corresponding axisymmetric simulation, which is dominated by strong latitudinal jets much like those in Fig.~\ref{S2p1_2Duyplots}.}
  \label{S2p5_3Duyplots}
\end{figure}

The differences between axisymmetric and 3D simulations are much clearer in a set of simulations with a stronger shear of $S=2.5$. In Fig.~\ref{S2p5_tplots} we show the evolution of the same volume-averaged quantities as Fig.~\ref{S2p1_tplots} for these simulations, which have $L_y=30, 50$ and 100. The $y$-averaged $u_y$ and $u_z$ velocity components on the $(x,z)$-plane are shown at $t=100$ in Fig.~\ref{S2p5_3Duyplots}. The instability now saturates in homogeneous turbulence in all 3D simulations for all $L_y$ considered. However, the energy level attained and the corresponding momentum transport does depend on $L_y$, with a trend towards convergence for $L_y\gtrsim50$.

The striking predator-prey-like dynamics observed in the axisymmetric simulation with $S=2.5$ discussed in the previous section does not occur in three-dimensions for any case with $L_y\geq30$; jets are not observed even at longer times (though they would presumably also occur for $S=2.5$ with small enough $L_y$). Presumably, strong latitudinal shears can only persist when they are stable to parasitic shear instabilities with long enough wavelengths along $y$. Naively, we might expect a requirement on $L_y\gtrsim \lambda_x$ for these strong shears to be suppressed, where $\lambda_x$ is the radial wavelength of the latitudinal shear flow.

The simulations in the previous section and this one highlight that the nonlinear evolution can significantly differ between axisymmetric and 3D simulations. Axisymmetric simulations develop strong latitudinal jets whereas 3D simulations with large enough $L_y$ prefer to saturate in homogeneous turbulence. This is reminiscent of the results of \citet{GaraudBrummell2015} for salt fingering. This is what we might have expected based on \S~\ref{saltfinger}, but does not directly follow from the formal analogy presented there.

We have also explored 3D cases with even stronger shears ($S>2.5$), and these behave in a qualitatively similar manner 
-- though see \S~\ref{largeshear} for a further discussion of cases with very large $S$. Given that the anisotropy in axisymmetric simulations, or those with small domains along $y$, is artificially imposed (rather than developing naturally from a more weakly constrained system), we advocate that 3D simulations with $L_x\sim L_z\sim L_y$ are likely to provide the most useful information regarding the nonlinear evolution of the GSF instability in stars. Indeed, in real stars, there is no enforced azimuthal periodicity on a short length-scale, so cases in which $L_y$ is large enough not to artificially constrain the flow are likely to be the most realistic. We will therefore focus on these simulations when we later consider the astrophysical consequences of our results. In addition, the formation of latitudinal jets is likely to be related to the adoption of periodic boundary conditions in the local latitudinal direction.

\subsection{Comparison with simulations with stress-free boundaries in $x$.}
\label{nekcomp}

We will now briefly consider the effects of varying the boundary conditions on the nonlinear evolution. To do this we have performed a pair of  simulations with impenetrable, stress-free, fixed temperature boundary conditions. These conditions differ from shearing-periodic boundary conditions in two crucial ways: they allow the flow to modify the background shear flow even at the boundaries, and they also disallow elevator modes. We choose $Pr=10^{-2}$, $N^2=10$, and $S=2.1$ and $S=2.5$, and we also adopt $L_y=30$, since this requires the fewest number of grid points in total to resolve the flow.

\begin{figure}
  \begin{center}
     \subfigure[$K$]{\includegraphics[trim=0.4cm 0cm 0cm 0cm, clip=true,width=0.48\textwidth]{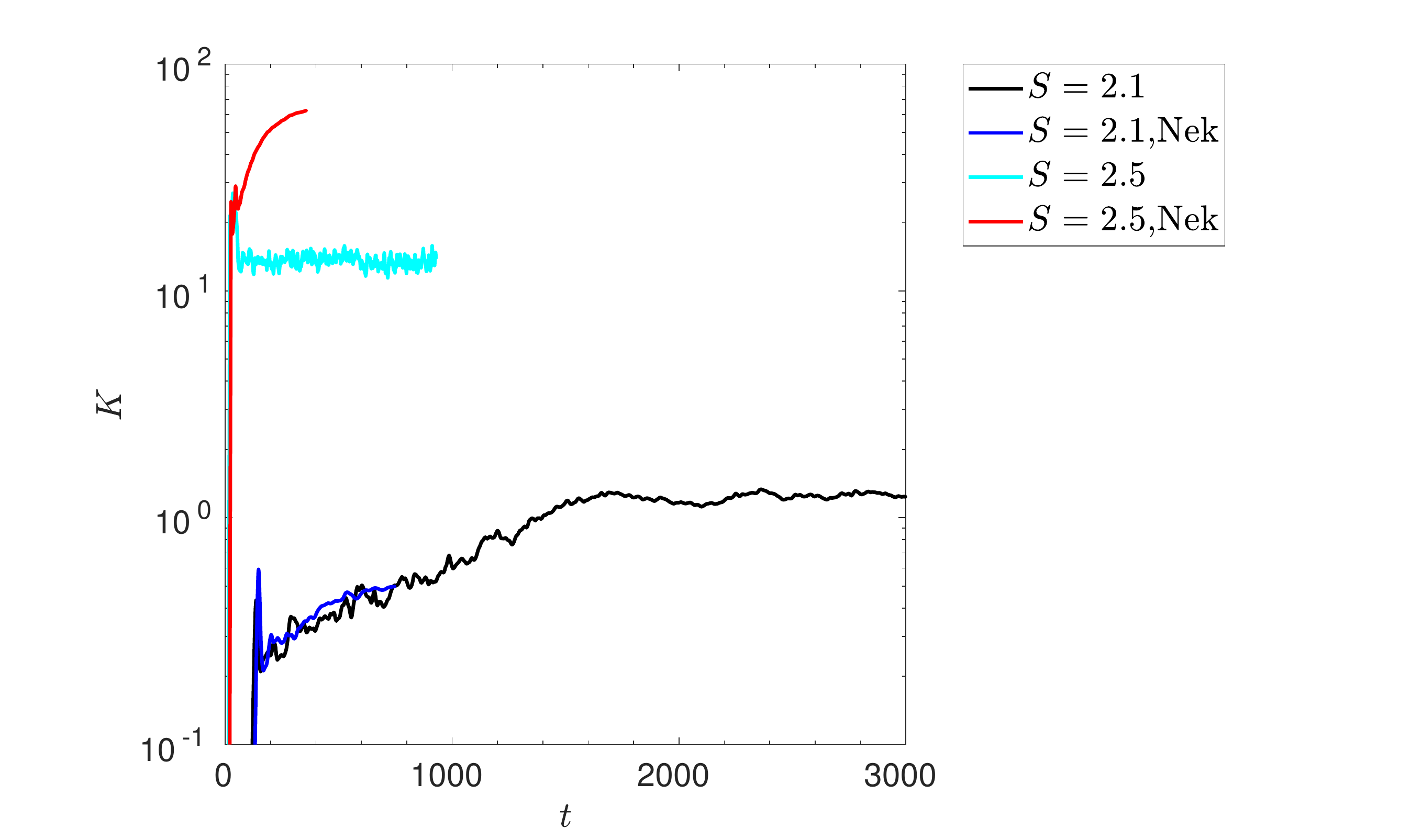}}
    \subfigure[$\langle u_xu_y\rangle$]{\includegraphics[trim=0.4cm 0cm 0cm 0cm, clip=true,width=0.48\textwidth]{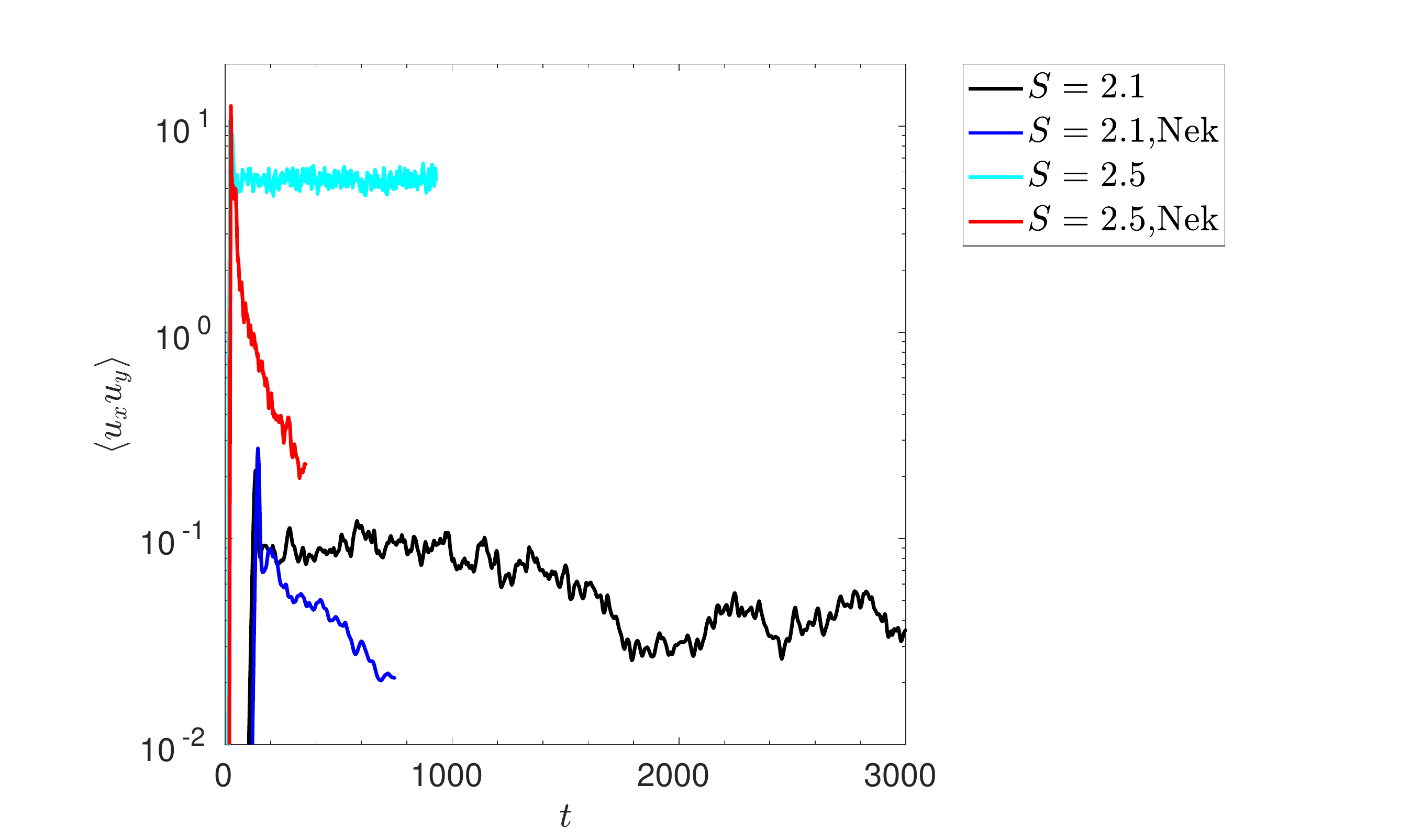}}
    \subfigure[$\sqrt{\langle u_z^2\rangle}$]{\includegraphics[trim=0.4cm 0cm 0cm 0cm, clip=true,width=0.48\textwidth]{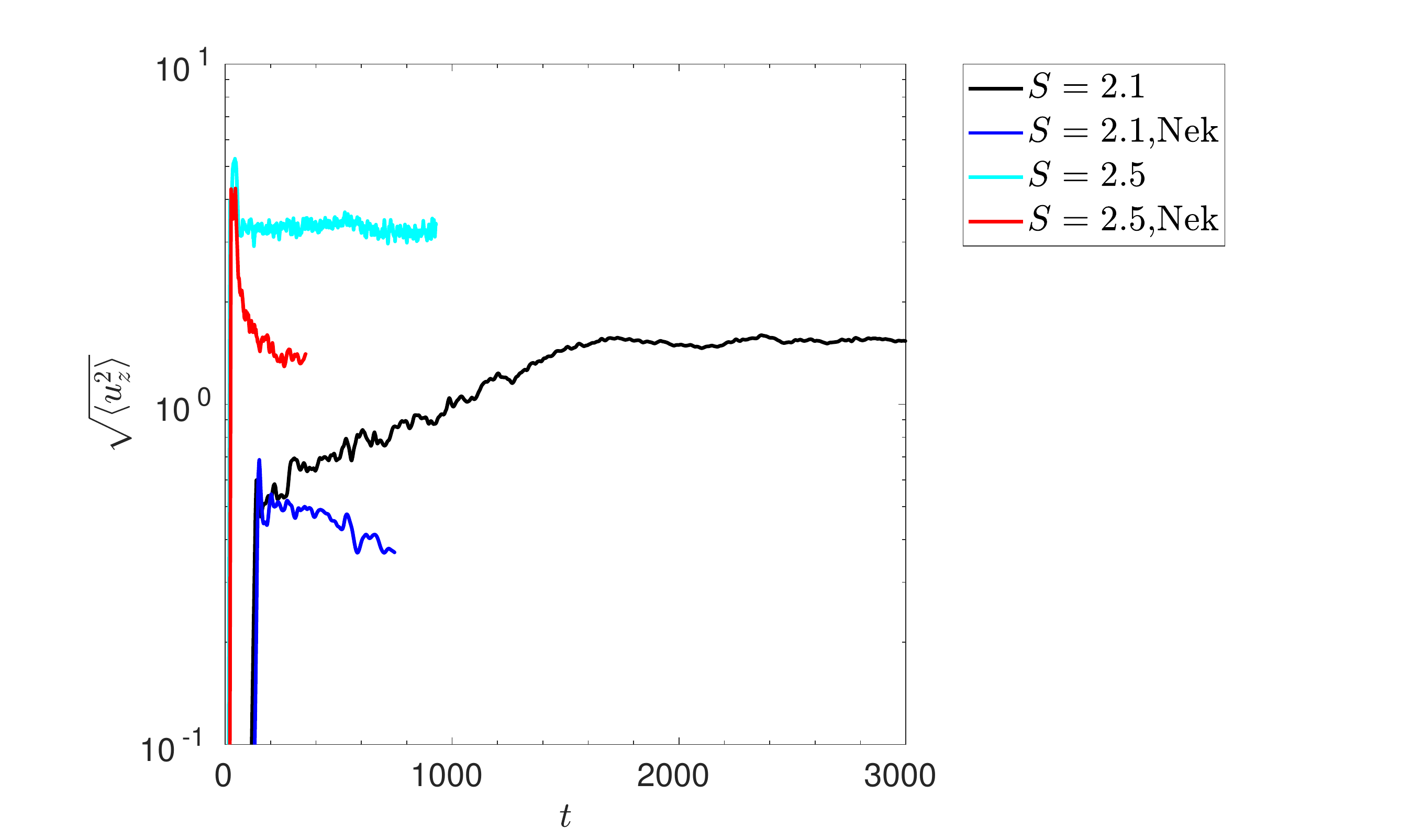}}
    \end{center}
  \caption{Comparison of the temporal evolution of $K$, $\langle u_xu_y\rangle$, and $v_z$, for a 3D simulation with stress-free boundaries (labelled ``Nek") and shearing-periodic boundaries. The parameters are $S=2.1, N^2=10, \mathrm{Pr}=10^{-2}$ and $L_y=30$. The energy with shearing-periodic boundaries grows due to the development of strong latitudinal shear flows, but in the case with stress-free impenetrable boundaries the energy instead grows due to the generation of a strong $u_y$ flow which counter-acts the imposed shear.}
  \label{S2p1_3D_CompNek5000SF}
\end{figure}

\begin{figure}
  \begin{center}
  \subfigure[$t=226$]{\includegraphics[trim=4cm 5cm 10cm 2.5cm, clip=true,width=0.35\textwidth]{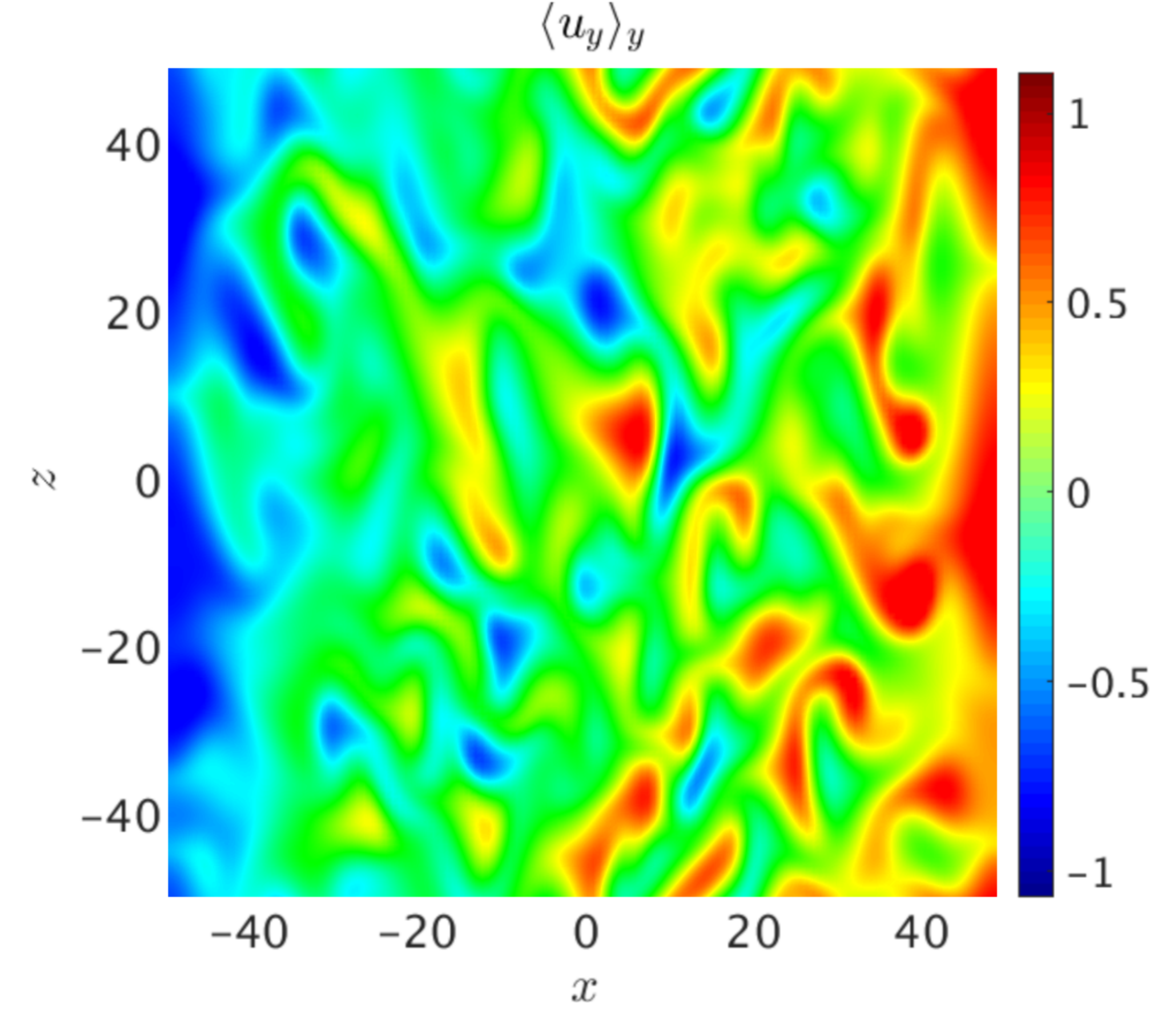}}
      \subfigure[$t=960$]{\includegraphics[trim=4cm 5cm 10cm 2.5cm, clip=true,width=0.35\textwidth]{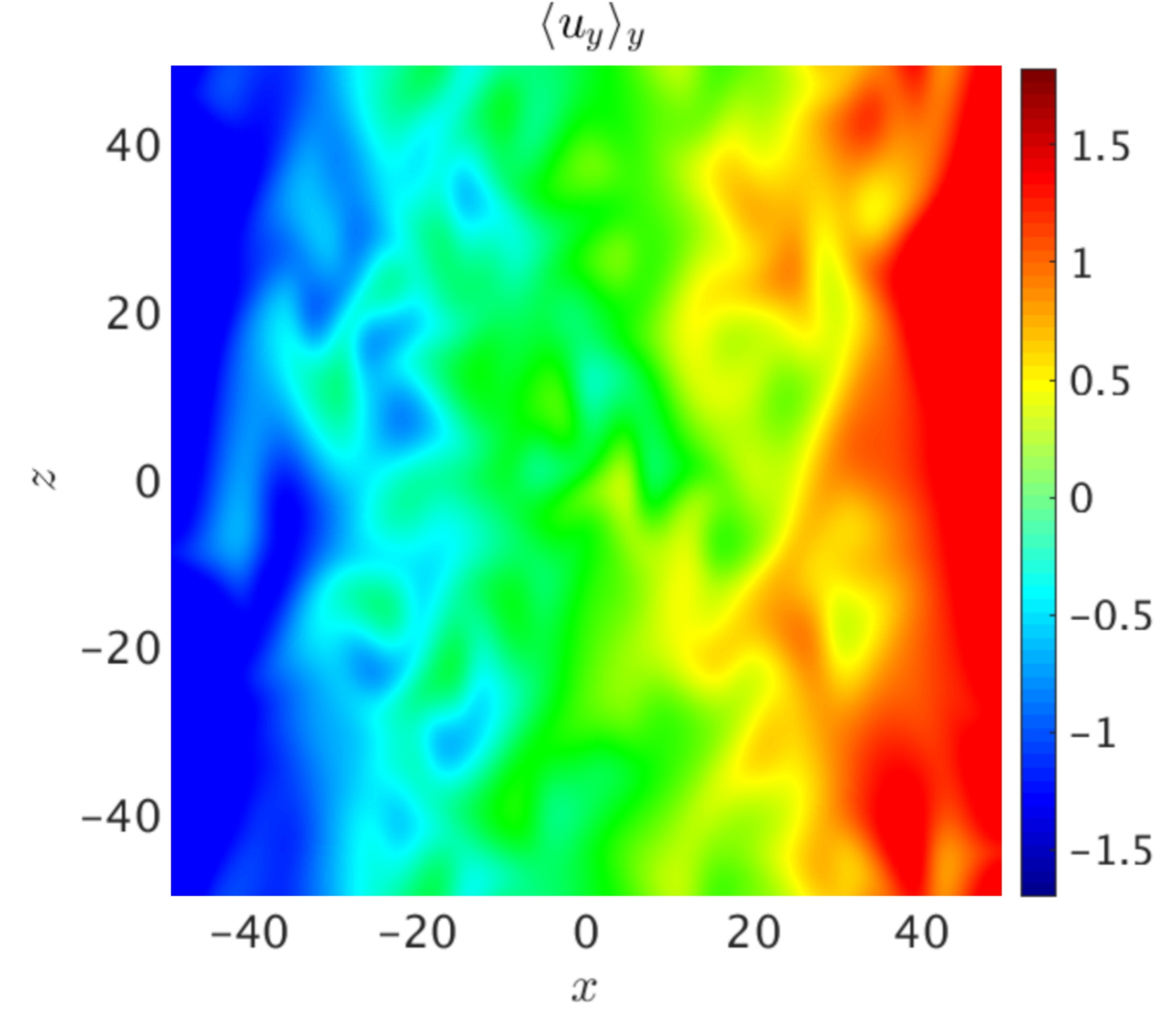}}
    \end{center}
  \caption{Snapshots of $y$-averaged $u_y$ in the $(x,z)$-plane from a 3D simulation with stress-free boundaries at two different times. The parameters are $S=2.1, N^2=10, \mathrm{Pr}=10^{-2}$ and $L_y=30$. The strong $u_y$ flow is evident, which counter-acts the imposed shear.}
  \label{S2p1_3D_CompNek5000SF1}
\end{figure}

In Figs.~\ref{S2p1_3D_CompNek5000SF} we compare the evolution of volume-averaged quantities for a case with shearing-periodic, and one with stress-free, boundary conditions for each of $S=2.1$ and $S=2.5$, respectively. Figs.~\ref{S2p1_3D_CompNek5000SF1} shows corresponding snapshots of the flow in the $(x,z)$-plane for the case with $S=2.1$. In Fig.~\ref{S2p1_3D_CompNek5000SF}, the kinetic energy with stress-free impenetrable boundaries is observed to grow due to the generation of a strong $u_y$ flow which partially counter-acts the imposed shear flow. The kinetic energy growth is just like with shearing-periodic boundaries, with the crucial difference that in this case the latitudinal flow is in fact decreasing (bottom panel). The snapshots in Fig.~\ref{S2p1_3D_CompNek5000SF1} also illustrate that a strong $u_y$ shear has developed by $t\sim200$, which gradually strengthens in time to dominate the flow by $t\sim 1000$. The shear in the total (background + perturbation) flow is now reduced by the action of the instability, relative to the initial imposed shear. However, note that this modification is still relatively weak even by $t=960$, and $\langle u_y\rangle_y$ at $x=0$ is approximately only $1.5\%$ of the initial background shear velocity ($-105$ at $x=0$). Similar behaviour is found with $S=2.5$, except that in this case the modification of the shear with these boundary conditions is then much stronger. 

These simulations clearly demonstrate that the GSF instability produces angular momentum transport that reduces the overall differential rotation. This kind of modification of the imposed shear is not permitted by shearing-periodic boundary conditions, so these simulations highlight that the long-term evolution of the instability is dependent on the boundary conditions. However, the initial saturation level is similar, even if the longer-term nonlinear reduction of the total shear acts to reduce the momentum transport over time compared with the cases with shearing-periodic boundary conditions. The initial agreement between both sets of simulations indicates that we may continue to use shearing-periodic boundary conditions to probe the momentum transport, at least during the initial phases of homogeneous turbulence. In addition, since strong latitudinal jets are absent in simulations with impenetrable radial boundaries, this suggests that we should focus on the initial phases of homogeneous turbulence in our simulations with shearing-periodic boundaries when constructing a model to apply to astrophysics.

\subsection{Simulations with very large imposed shears}
\label{largeshear}

The nonlinear behaviour described in \S~\ref{results} is typical of most of our simulations in which the GSF instability operates. However, we observe different behaviour when $S$ is very large. For our typical value of $N^2=10$, note that when $S\gtrsim 3$, $\mathrm{Ri}=N^2/S^2\lesssim 1$. For these large shears, the shear dominates over the stable stratification and we might expect different nonlinear evolution. Note also that if $S\leq 7$, the flow is linearly stable to adiabatic axisymmetric perturbations, so simulations in this regime are still probing the action of the GSF instability. Note that such large shears are probably not relevant in astrophysics, where we typically expect $\mathrm{Ri}\gg 1$ except very close to convection zones, but these simulations are nevertheless useful in allowing us to explore the nonlinear behaviour in cases where we might expect the theory that we will present in the next section to no longer apply.

\begin{figure}
  \begin{center}
   \subfigure[$K$]{\includegraphics[trim=0cm 0cm 0cm 0cm, clip=true,width=0.48\textwidth]{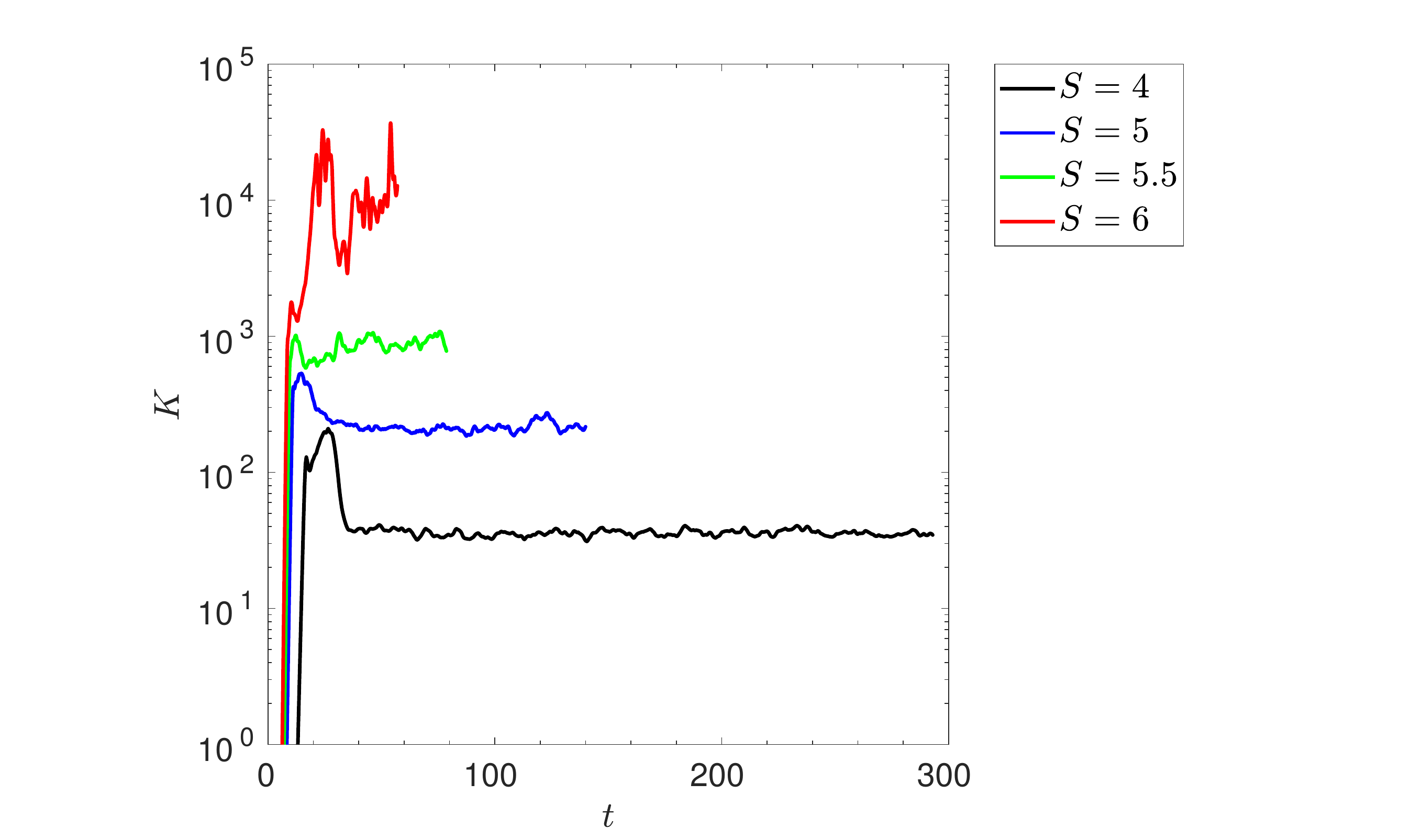}}
    \subfigure[$\langle u_xu_y\rangle$]{\includegraphics[trim=0.4cm 0cm 0cm 0cm, clip=true,width=0.48\textwidth]{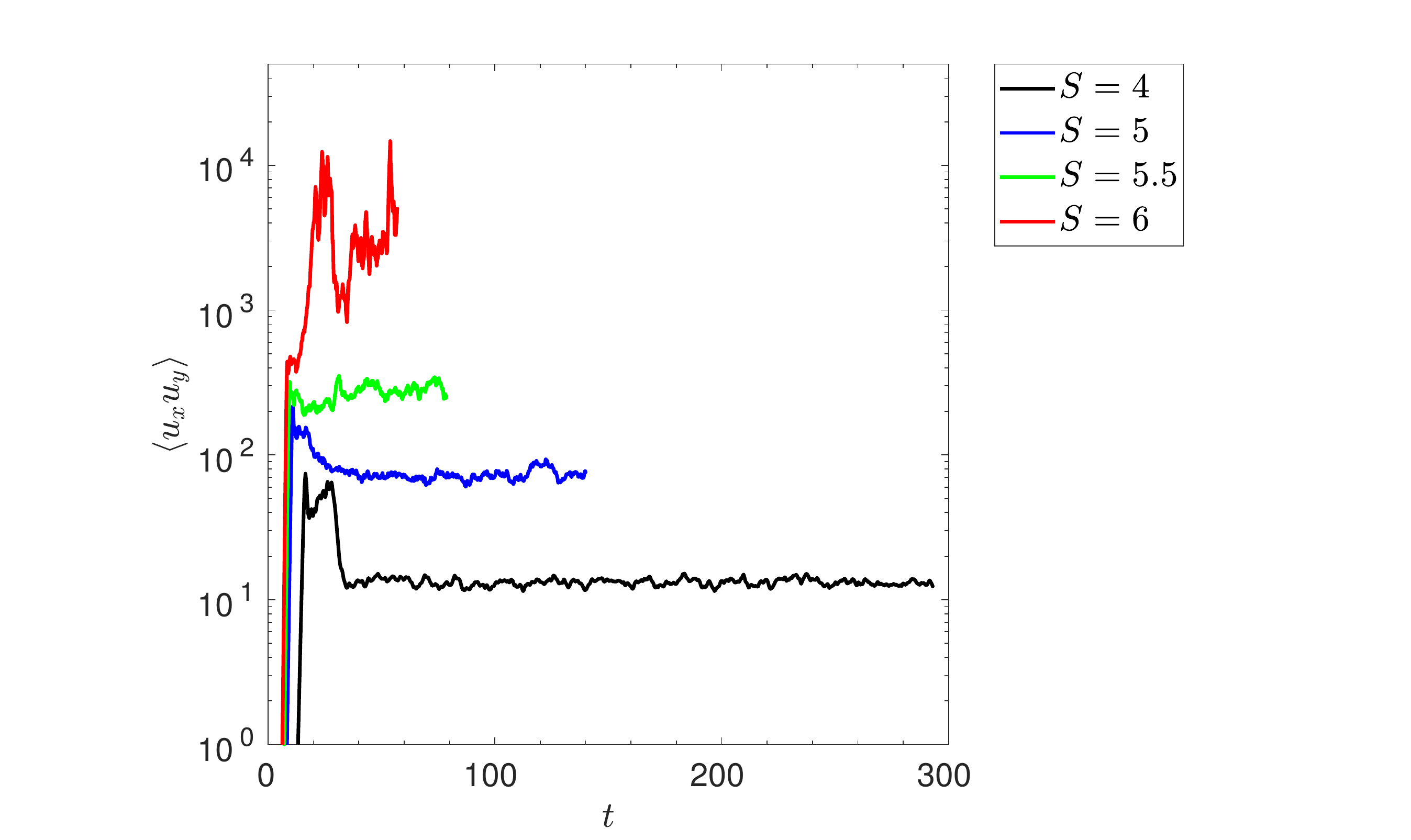}}
    \end{center}
  \caption{Temporal evolution of $K$ and $\langle u_xu_y\rangle$ for a set of simulations with $N^2=10, \mathrm{Pr}=10^{-1}, L_y=100$, for the strong shear values of $S=4,5,5.5$, and $6$.}
  \label{KElargeshear}
\end{figure}

\begin{figure*}
  \begin{center}
  \subfigure[$S=5, t=70$]{\includegraphics[trim=1cm 7cm 0.5cm 7cm,, clip=true,width=0.35\textwidth]{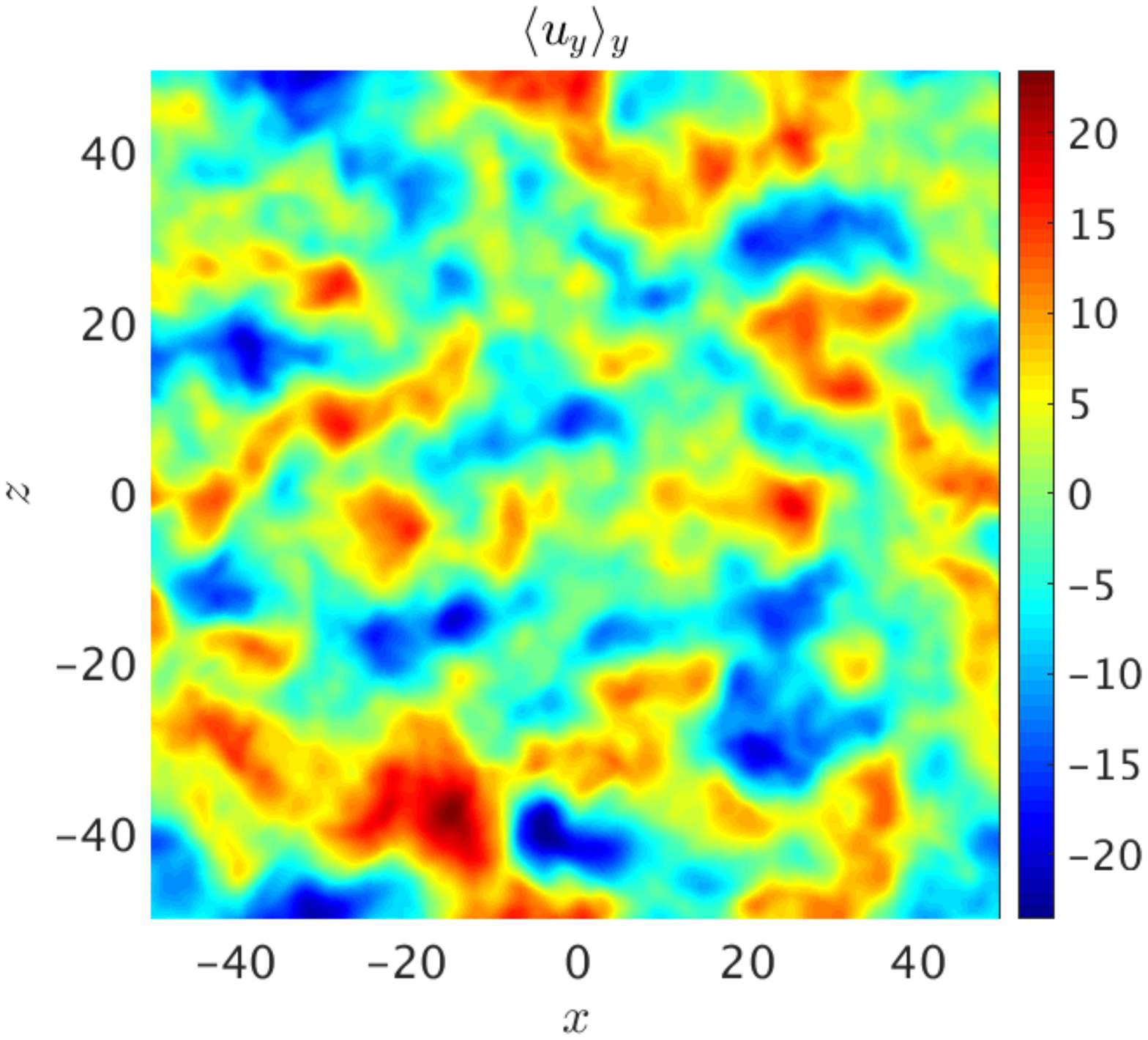}}
   \subfigure[$S=5, t=70$]{\includegraphics[trim=1cm 7cm 0.5cm 7cm,, clip=true,width=0.35\textwidth]{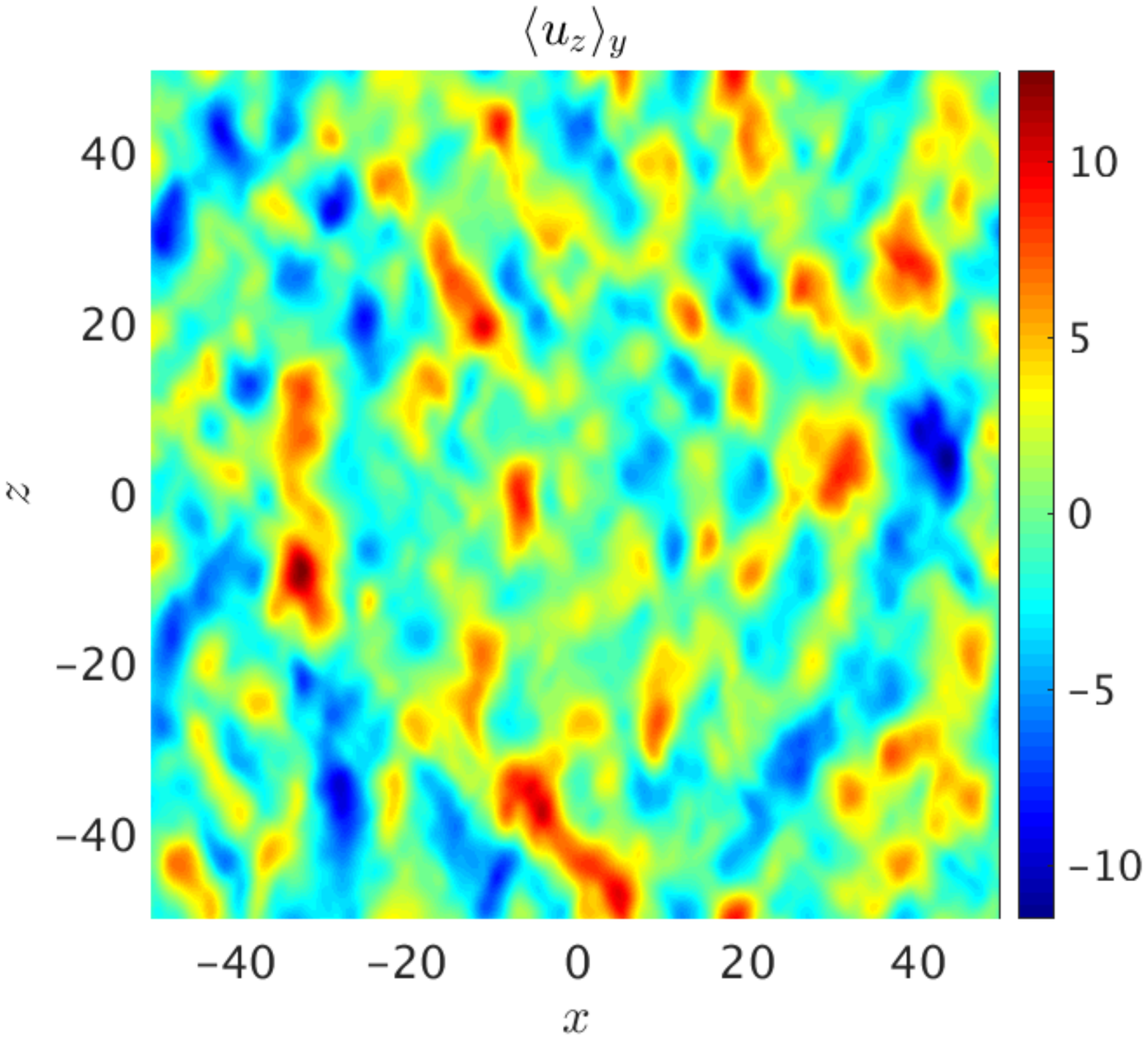}}
   \subfigure[$S=5.5, t=70$]{\includegraphics[trim=1cm 7cm 0.5cm 7cm,, clip=true,width=0.35\textwidth]{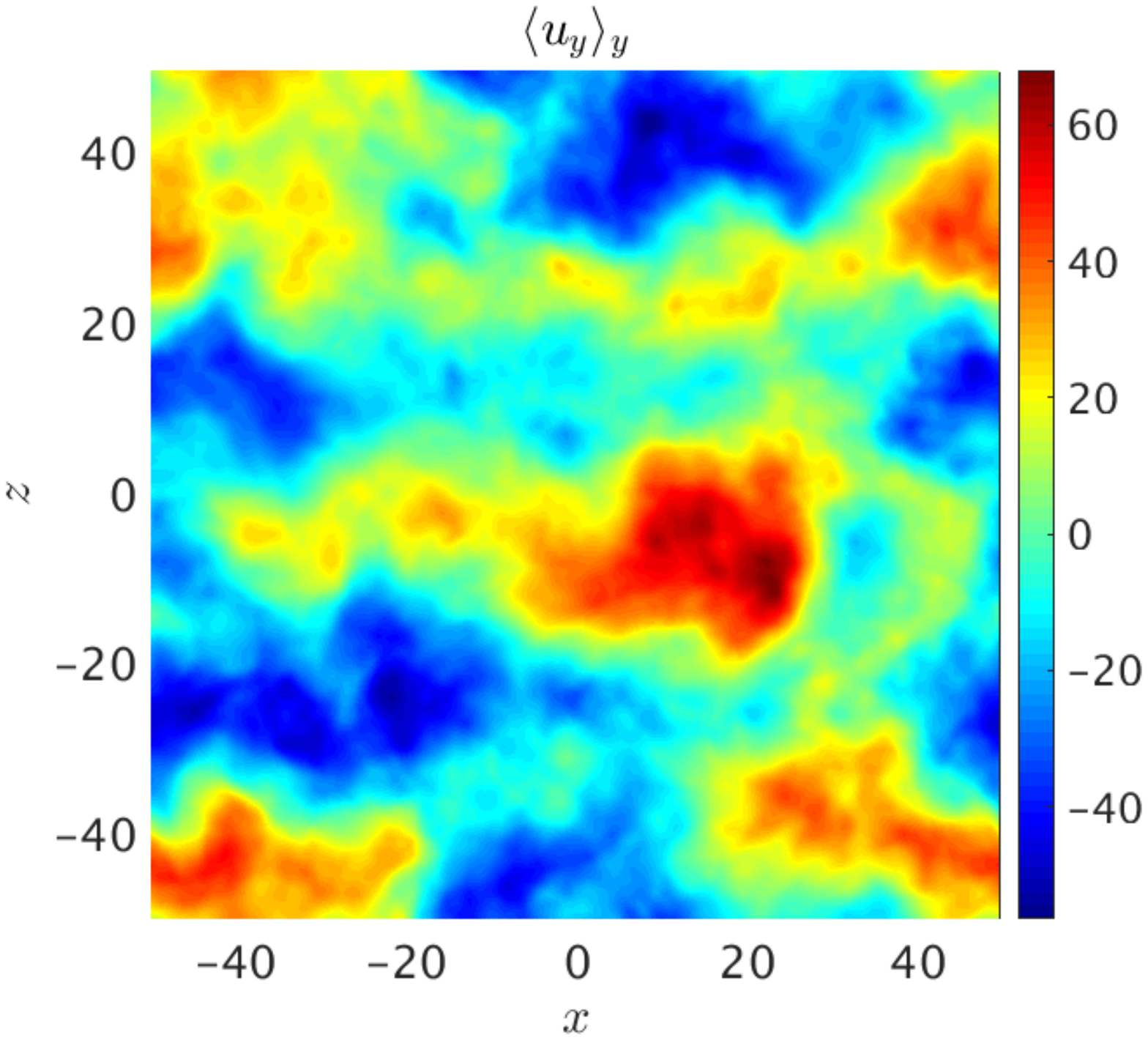}}
   \subfigure[$S=5.5, t=70$]{\includegraphics[trim=1cm 7cm 0.5cm 7cm,, clip=true,width=0.35\textwidth]{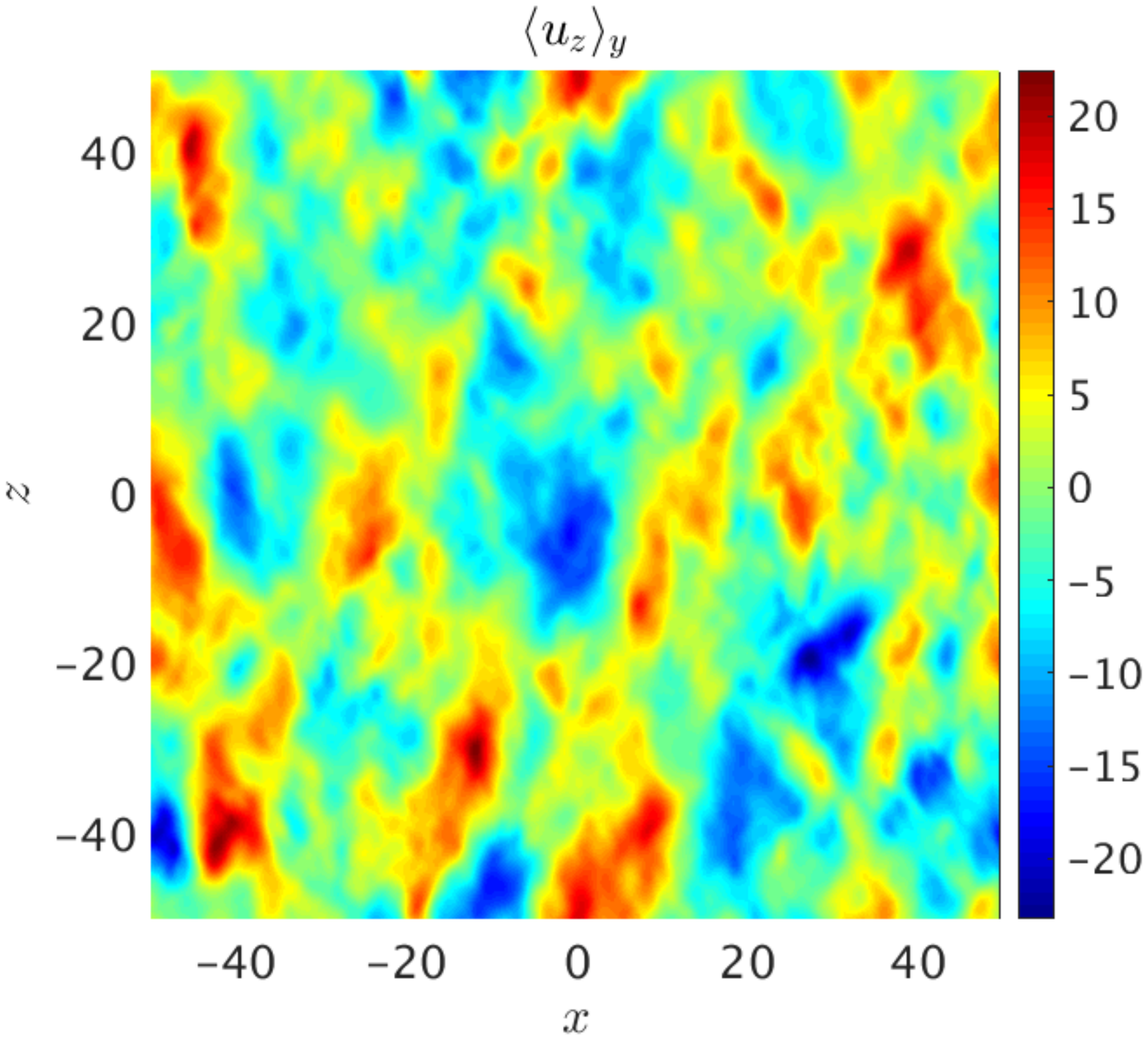}}
   \subfigure[$S=6, t=30$]{\includegraphics[trim=1cm 7cm 0.5cm 7cm,, clip=true,width=0.35\textwidth]{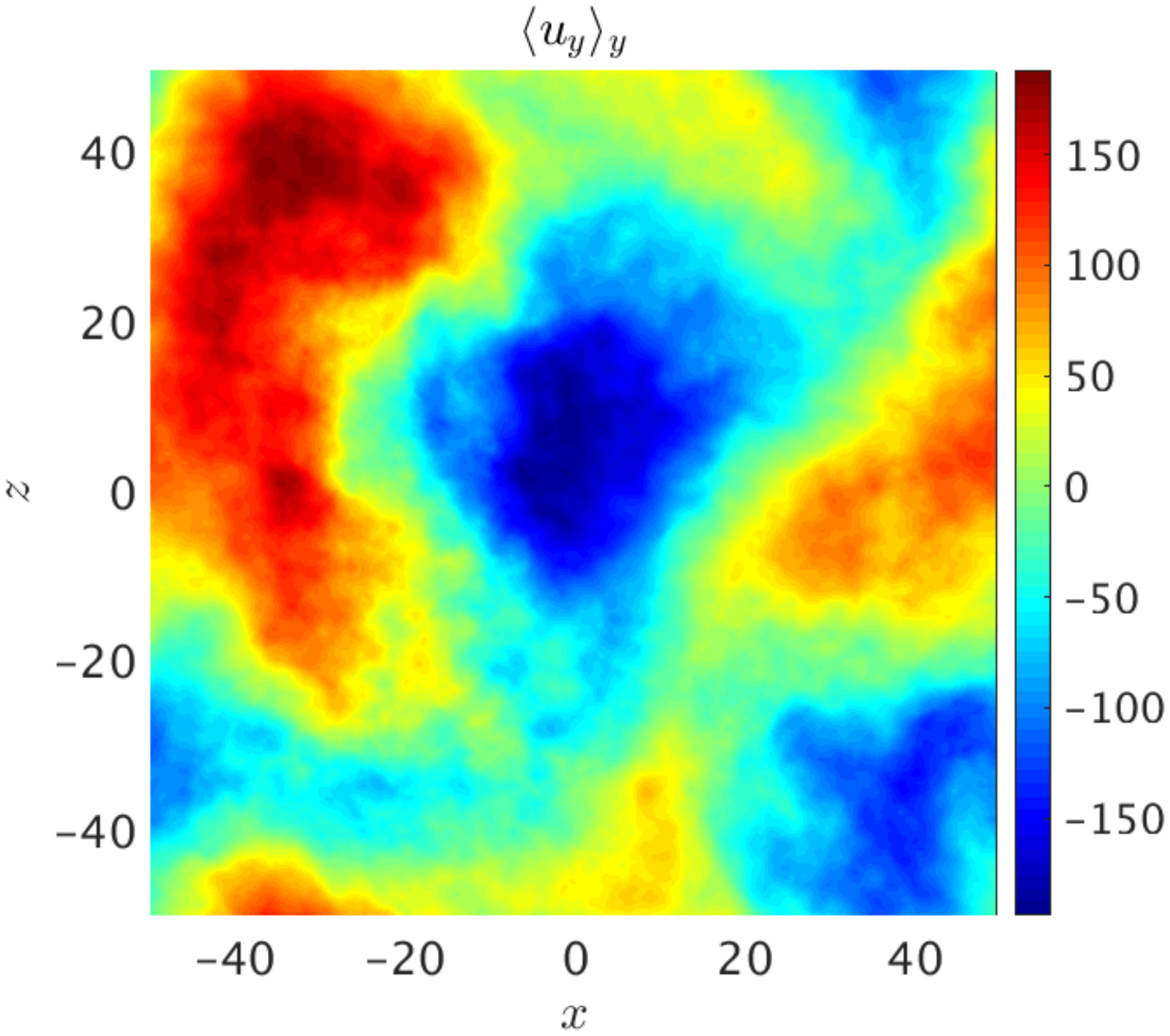}}
   \subfigure[$S=6, t=30$]{\includegraphics[trim=1cm 7cm 0.5cm 7cm,, clip=true,width=0.35\textwidth]{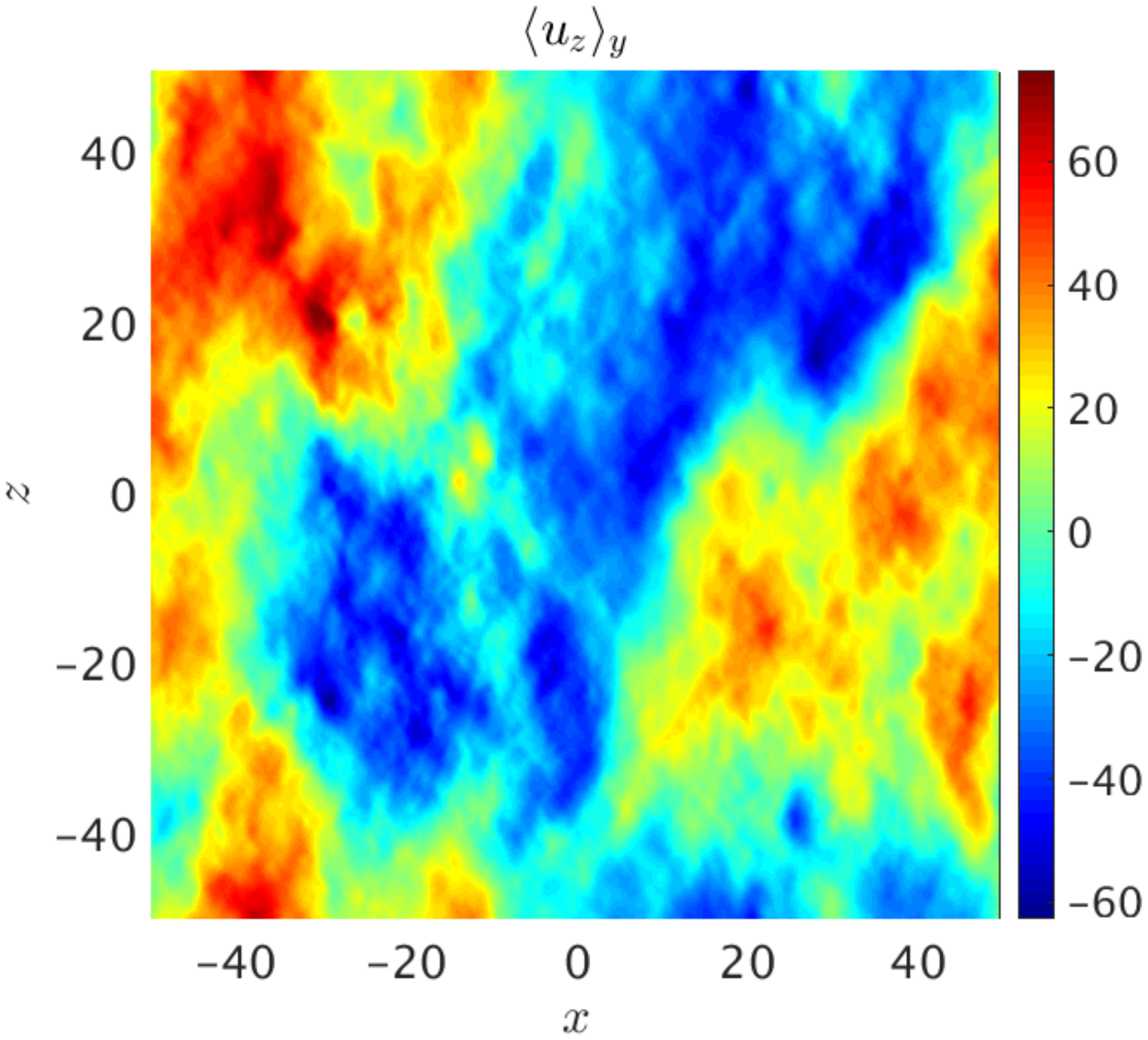}}
    \end{center}
  \caption{Snapshots of $y$-averaged $u_y$ and $u_z$ in the $(x,z)$-plane for simulations with large shears, which illustrates the large-scale flows that develop. The parameters are $N^2=10, \mathrm{Pr}=10^{-1}$, $L_y=100$, and $S=5, 5.5$, and $6$, at the times indicated in the captions.}
  \label{Plotflowlargeshear}
\end{figure*}

In this section, we use three-dimensional simulations with shearing-periodic boundary conditions with $Pr=0.1$, $N^2=10$, $L_y=100$, and $S=4, 5, 5.5$ and $6$ to illustrate the behaviour in these cases with strong shear. In Fig.~\ref{KElargeshear}, we show the mean kinetic energy and $\langle u_x u_y\rangle$ as a function of time. Then, in Fig.~\ref{Plotflowlargeshear}, we show the $y$-averaged $u_y$ and $u_z$ flow components at a given time in the turbulent state in a simulation with $S=5, 5.5$ and $6$. We observe that the simulations with $S\leq 5.5$ reach a statistically steady turbulent state with only small fluctuations about the mean kinetic energy and momentum transport. On the other hand, the simulation with $S=6$ is strongly bursty, with large fluctuations in the kinetic energy and momentum transport. The flow corresponding to this simulation is shown in the bottom two panels of Fig.~\ref{Plotflowlargeshear}, which shows the presence of large-scale flow structures in both $u_y$ and $u_z$. Fig.~\ref{Plotflowlargeshear} shows that as $S$ is increased, the flow saturates in large length-scale flows (which have also been found in salt fingering by e.g.~\citealt{Brownetal2013}, but not necessarily for the same reason).

Linear theory predicts that larger values of $S$ allow instability for increasingly larger wavelength modes, but this cannot be the sole explanation for this behaviour. Indeed, the fastest growing mode has a wavelength of $\lambda_z\approx 9.3$ (with $k_x=0$) when $S=4$, and $\lambda_z\approx 12.5$ when $S=6$, but the flow structures in the nonlinear state are larger than this by more than a factor of 3 in the latter case. Hence, the formation of large-scale flows for large shears in likely to be related to the modification of the nonlinear cascade when the shear dominates over the stratification. Due to their oscillatory nature, these large-scale flows may correspond with large-wavelength gravity waves, and they significantly enhance the transport over cases with smaller shears.

The formation of these large-scale flows is only observed for very strong shears, and such large shears are unlikely to be astrophysically relevant -- the possible exception being very close to interface between the convective and radiative regions in very early phases of stellar evolution. Hence, we will not focus on explaining these simulations with large $S$, though we will note that they do saturate differently from those with smaller shears. This means that we would not expect the nonlinear behaviour in simulations with large $S$ to be explained by a theory (such as the one that we will present in the next section) that is designed to explain simulations with smaller values of $S$.

\section{Theory for saturation of the GSF instability}
\label{theoryvalidation}

For astrophysical applications we would like to quantify the angular momentum transport produced by the equatorial GSF instability. It is simplest to focus on developing a model for the saturation where the instability produces homogeneous turbulence, rather than coherent shear flows. As a result, our primary focus here is on explaining the nonlinear behaviour in simulations with $L_x=L_z=L_y$ during the phases of homogeneous turbulence. A different (quasi-linear) theory (see e.g.~\citealt{MQT2014}) would be required to explain the behaviour in simulations where strong shear flows develop.

The simplest model of saturation of the equatorial instability is to assume that the flow is dominated by the fastest growing linear mode, and that this mode saturates when its growth rate balances its nonlinear cascade rate. The basic idea is that the fastest growing modes predominantly involve radial ($u_x$) flows with significant latitudinal (along $z$) shear, and that parasitic shear instabilities acting on these flows would be expected to grow, and draw energy from the primary mode, at a rate of order $k_z |u_x |$. Note that the theory in this section will be expressed using dimensional quantities for clarity. We may expect saturation of the primary instability when
\begin{eqnarray}
s \sim k_z u_x .
\end{eqnarray}
We define a constant of proportionality $A$, which will be chosen later to fit our simulation data, through
\begin{eqnarray}
\label{saturation}
u_x \equiv A \frac{s}{k_z}.
\end{eqnarray}
For a single linear mode, Eqs.~\ref{Eq1}--\ref{Eq4} relate the perturbations by
\begin{eqnarray}
\label{linear1}
u_y &=& \frac{\left(\mathcal{S}-2\Omega\right)}{s_\nu}u_x, \\
\label{linear2}
u_z &=& -\frac{k_x}{k_z}u_x, \\
\label{linear3}
\theta &=& \frac{-\mathcal{N}^2}{s_\kappa} u_x,
\end{eqnarray}
in terms of the radial velocity $u_x$, which is specified by Eq.~\ref{saturation}. The corresponding time- and volume-averaged rates of momentum and heat transport, and the kinetic energy, for a single such mode are given by
\begin{eqnarray}
\langle u_x u_y\rangle &=& \frac{\left(\mathcal{S}-2\Omega \right)}{2 s_{\nu}}|u_x|^2, \\
\langle u_yu_z \rangle &=& -\frac{\left(\mathcal{S}-2\Omega \right)}{2 s_{\nu}}\frac{k_x}{k_z}|u_x|^2, \\
\langle u_xu_z \rangle &=& -\frac{k_x}{2k_z}|u_x|^2, \\
\langle u_x \theta \rangle &=& \frac{-\mathcal{N}^2}{2 s_{\kappa}}|u_x|^2, \\
\langle \frac{1}{2}|\boldsymbol{u}^2|\rangle &=&\frac{1}{4}\left(1+\frac{\left(\mathcal{S}-2\Omega\right)^2}{s_{\nu}^2}+\frac{k_x^2}{k_z^2}\right)|u_x|^2.
\end{eqnarray}
The values of $s$ and $k_z$ for the fastest growing linear mode can be determined by solving Eqs.~\ref{disprel} and \ref{maxgrowthkz}.

This simple theory predicts the energy and momentum transport in terms of the properties of the linear instability and a single constant $A$, which is supposed to be independent of the parameters of our problem. However, with large values of $S$, or when strong shear flows develop, we may expect the growth and cascade rates to be modified by the shear, so this theory would no longer be expected to hold, and a different type of theory would be needed (e.g.~\citealt{Bouchet2013}). We will determine $A$ numerically by fitting this model to our data from a suite of numerical simulations. Note that the model predicts $\langle u_xu_z\rangle=\langle u_yu_z\rangle=0$ for ``elevator modes" with $k_x=0$. This theory is essentially equivalent to the ones proposed and tested for salt fingering by \cite{Den2010} and \cite{Brownetal2013}.

\begin{figure}
  \begin{center}
    \subfigure[$A=4$]{\includegraphics[trim=1.5cm 9cm 8cm 2.5cm, clip=true,width=0.48\textwidth]{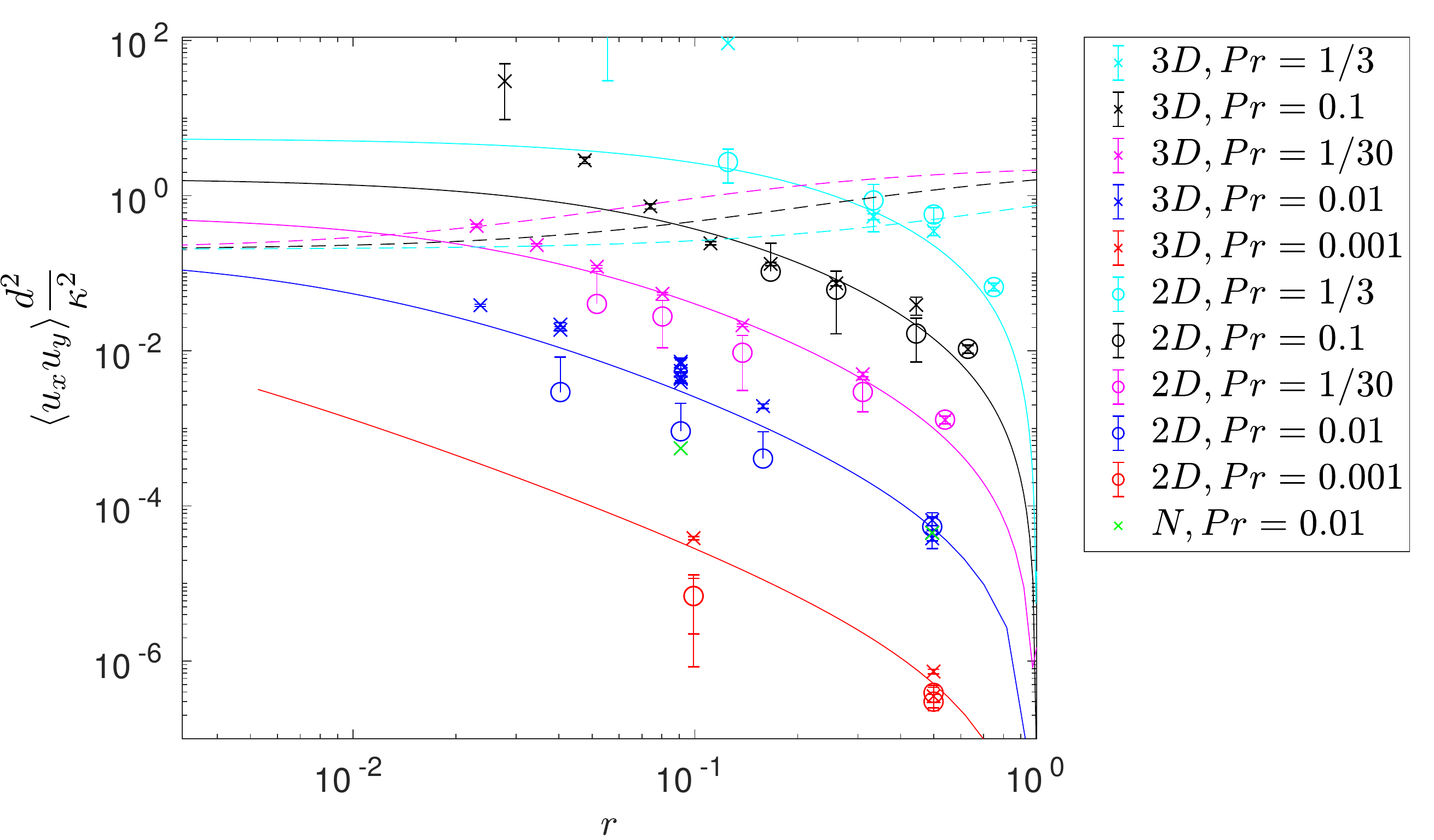}}
    \subfigure[$A=3.4$]{\includegraphics[trim=1.5cm 9cm 8cm 2.5cm, clip=true,width=0.48\textwidth]{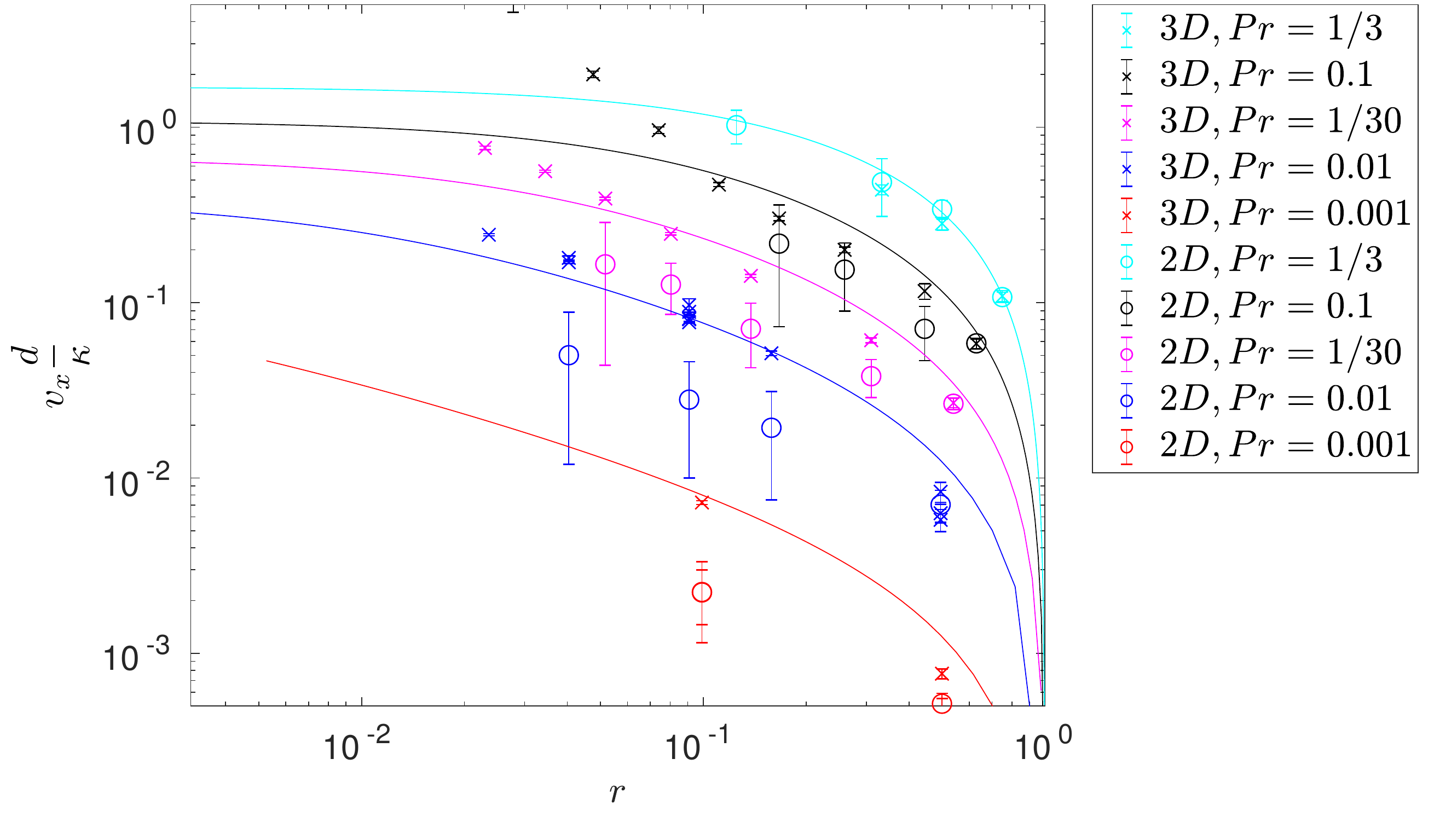}}
     \subfigure[$A=9$]{\includegraphics[trim=1.5cm 9cm 8cm 2cm, clip=true,width=0.48\textwidth]{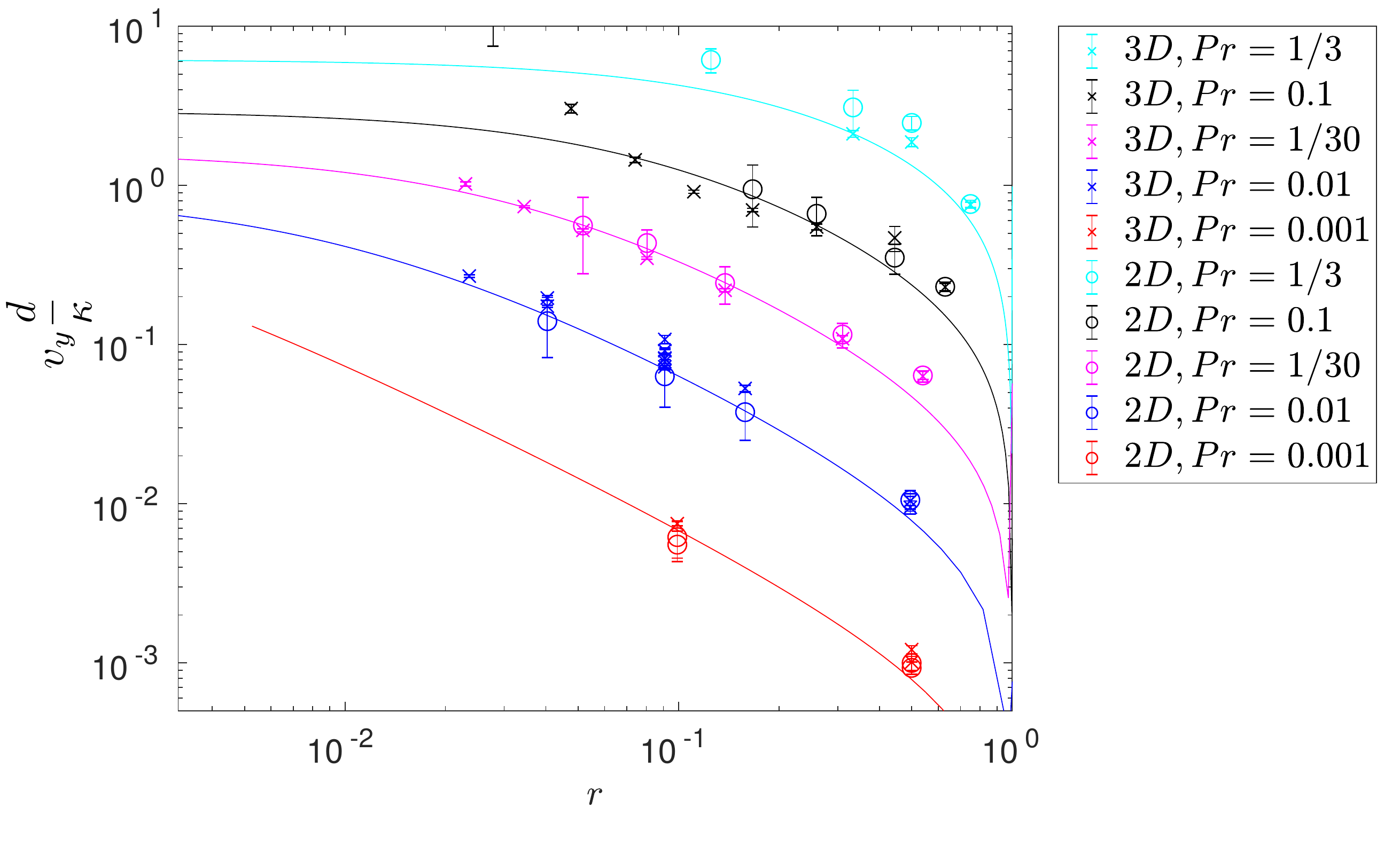}}
    \end{center}
  \caption{Comparison between the theory (solid lines) presented in \S~\ref{theoryvalidation} and simulations (symbols) for several volume- and time-averaged quantities. 3D simulations are represented as crosses and axisymmetric simulations as open circles. The colours of each solid line and the symbols represent a given set of parameters, as identified in the legend. The panel captions represent the value of $A$ used in the theory, which was varied in the bottom two panels for a better fit. The top panel also shows the line $\mathrm{Ri}=1$ as dotted lines, which indicates that the simulations that disagree with theory for small $r$ are those for which $\mathrm{Ri}\lesssim 1$.}
  \label{TheoryComparison}
\end{figure}

\begin{figure}
  \begin{center}
         \subfigure[$A=4$]{\includegraphics[trim=1.5cm 9cm 8cm 2.5cm, clip=true,width=0.49\textwidth]{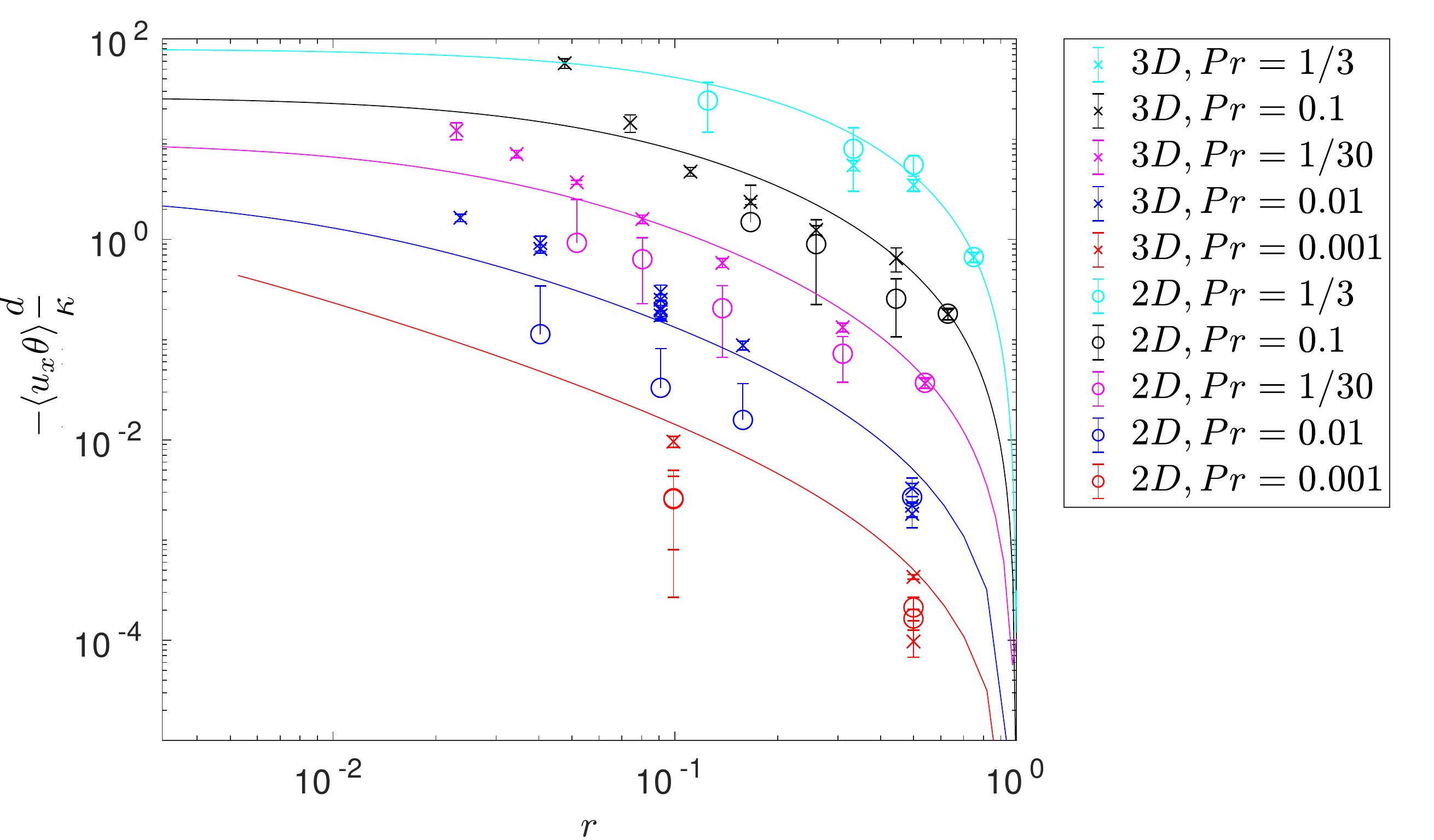}}
    \subfigure[$A=4$]{\includegraphics[trim=0cm 0cm 1.75cm 0cm, clip=true,width=0.47\textwidth]{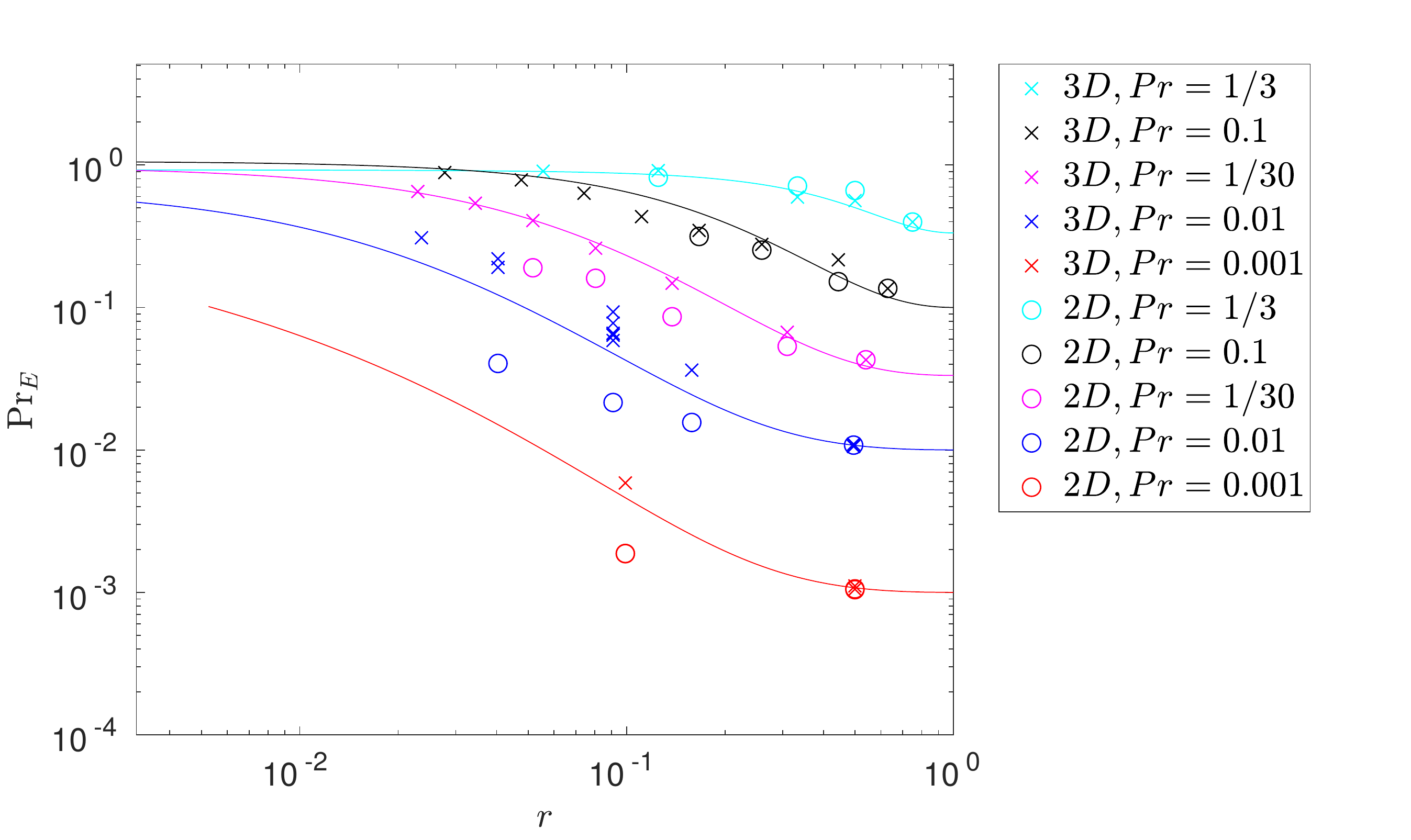}}
    \end{center}
  \caption{Buoyancy flux $-\langle u_x\theta\rangle$ and ``turbulent Prandtl number" $\mathrm{Pr}_E$, compared with theory for $A=4$. Note that since we always have stable stratification, the buoyancy flux is negative.}
  \label{TheoryComparison3}
\end{figure}

To test this theory we have performed a suite of simulations with the parameters listed in Table~\ref{Table}.  We show the comparison of the theory and simulations for $\langle u_x u_y\rangle d^2 /\kappa^2$ vs $r$, where
\begin{eqnarray}
r=\left(\frac{\mathrm{Pr}}{\mathrm{Pr}-1}\right)\left(1+\frac{\mathcal{N}^2}{\kappa_{ep}^2}\right),
\end{eqnarray}
in the top panel of Fig.~\ref{TheoryComparison}. Here all theoretical lines have been computed by using the same value of $A=4$. We follow \cite{Brownetal2013} in defining $r$ to map the region of parameter space that is unstable to the GSF instability to $r\in [0,1]$.
Note that when $r\geq 1$, the system is stable to the GSF instability, and when $r<0$, the system is unstable via the adiabatic Solberg-H{\o}iland criterion, and that the super-criticality increases as $r\rightarrow 0$. We have scaled the 
velocities in figures \ref{TheoryComparison} and \ref{TheoryComparison3} in units 
of $\kappa/d$, rather than $\Omega d$ as in the earlier figures. This separates the data with different Pr, and also aids the comparison with Fig.~5 of \cite{Brownetal2013} (which shows the heavy element transport by the salt fingering instability). The agreement is very good for most 3D simulations, indicating that this theory is essentially correct to explain the transport driven by the GSF instability. The agreement is also reasonable for 2D simulations for values of $r$ that are not too small $r<0.1$ or so. However, there is a departure from the theory for small values of $r$, which corresponds with the large shear cases described in \S~\ref{largeshear}. The top panel shows the line $\mathrm{Ri}=1$ as dotted lines for the first three values of Pr. This indicates that the theory works well below this line, but fails to apply when $\mathrm{Ri}\lesssim 1$, which lies above these lines.

In the middle and bottom panels of Fig.~\ref{TheoryComparison}, we also compare the theoretical predictions for the scaled RMS radial ($v_x d/\kappa=\sqrt{\langle u_x^2\rangle}d/\kappa$) and azimuthal velocity ($v_y d/\kappa=\sqrt{\langle u_y^2\rangle}d/\kappa$). All curves use the same value of $A$, but we have selected a different constant value to fit the data better, with $A=3.4$ for $v_x$ and $A=9$ for $v_y$. The difference in the values of $A$ required to fit the data for these quantities is presumably because the flow does not only consist of a single mode, but contains modes with several $k_x$ and $k_z$. In reality, the flow will consist of several modes, and the theory implicitly involves an integration over the domain, for which the spatial structure of the flow is important. As a result of the flow consisting of several modes, quantities that involve different products may require a different constant which accounts for this integration. Hence, we may not expect the same $A$ to be applicable for all quantities.

We note that we obtain enhanced transport and velocity amplitudes, not explained by our simple theory, for very large shears. For these simulations, $\mathrm{Ri}=N^2/S^2\lesssim 1$, and the flow is no longer strongly stratified. Those simulations are less relevant than those with larger $r$ for astrophysics, since we generally expect $\mathrm{Ri}=N^2/S^2\gg 1$ in stellar radiation zones, though as noted previously, such cases may be relevant very near convection zones.

We have also calculated $\langle u_xu_z\rangle$, $\langle u_yu_z\rangle$, and the RMS latitudinal velocity $v_z$. The theory would predict these quantities to be exactly zero because the fastest growing modes are elevator modes with $k_x=0$. We observe them to be nonzero in general, but we confirm that they fluctuate about zero, consistent with theoretical expectations.

In Fig.~\ref{TheoryComparison3}, we show the scaled buoyancy flux $-\langle u_x\theta\rangle d/\kappa$, along with the theoretical predictions assuming $A=4$. The agreement is very good apart from cases with small $r$, just as with $\langle u_x u_y\rangle$. The buoyancy flux in 3D simulations typically exceeds that in axisymmetric cases, presumably due to the presence of strong latitudinal shear flows that inhibit radial transport in the axisymmetric case.

We may also crudely define a ``turbulent Prandtl number" by:
\begin{eqnarray}
\mathrm{Pr}_E=\frac{\nu+\nu_E}{\kappa+\kappa_E} = \frac{\nu+\langle u_xu_y\rangle/\mathcal{S}}{\kappa+|\langle u_x\theta\rangle/\mathcal{N}^2|}
\end{eqnarray}
This is shown in the bottom panel of Fig.~\ref{TheoryComparison3}, and is found to depend on $r$, and is not a constant that is equal to the laminar Pr. Note that $\nu\ll \nu_E$ except when $r\sim 1$. We find that $\mathrm{Pr}_E\leq 1$ for all simulations performed. This indicates that the saturated state maintains an effective Prandtl number smaller than one, which supports further action of the GSF, rather than by saturating by increasing $\mathrm{Pr}_E>1$, which would eliminate GSF.

\subsection{The absence of ``layering" by the GSF instability}

Salt fingering in the oceans is known to produce layering of the density field, leading to the formation of density staircases, in which convective layers are separated by thin diffusive interfaces. The presence of layering is associated with a significant enhancement in the rates of turbulent transport over that of a homogeneous turbulent medium. For low Pr fluids, \cite{Brownetal2013} showed that layer formation is possible by the salt fingering instability by the ``collective instability" (which involves the excitation of large-scale gravity waves that form layers when the waves break), but not by the linear mean-field ``$\gamma$-instability" of \cite{Radko2003} and \cite{Traxler2011a}. The $\gamma$-instability involves slow-growing, horizontally-invariant but vertically-varying modes. Both of these are secondary instabilities that are driven by a positive feedback mechanism between large-scale temperature and salinity perturbations, and the turbulent fluxes induced by them, which can be derived within a mean-field framework.
Given the potential importance of such layering for turbulent transport, we wish to determine whether layering of the angular momentum field by the GSF instability may be possible. In our simulations at the equator we do not observe layering in $u_y$. Instead, we observe the formation of $u_z$ jets, which is not what we are attempting to explain here. In this section we are interested in exploring whether the absence of observed layering in $u_y$ is consistent with mean-field theories that have been tested for the related salt fingering problem. This is motivated by the analogy discussed in \S~\ref{saltfinger}.

\begin{figure}
  \begin{center}
     \subfigure{\includegraphics[trim=0cm 0cm 1.75cm 0cm, clip=true,width=0.47\textwidth]{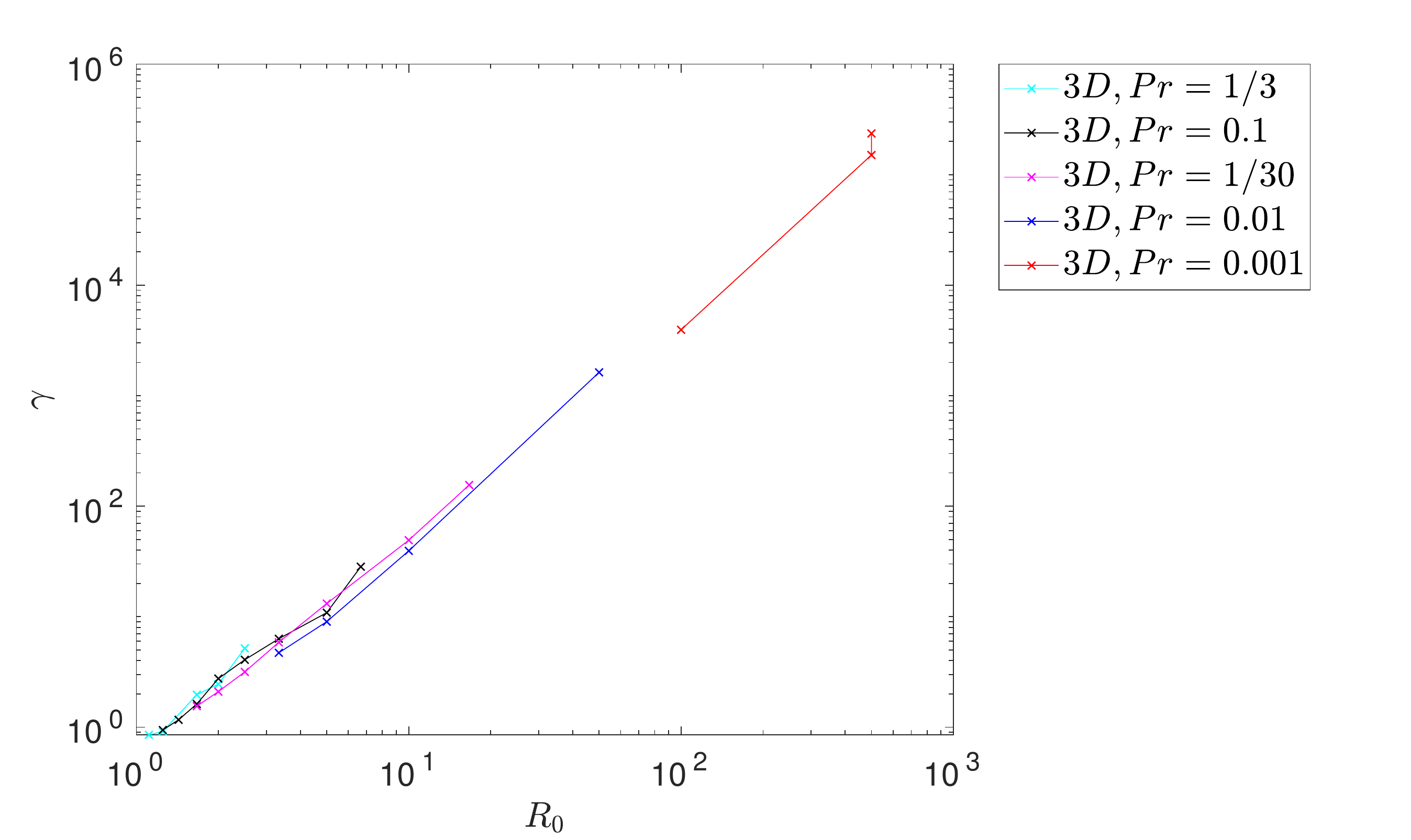}}
    \end{center}
  \caption{The flux ratio $\gamma$ as a function of the density ratio $R_0$. Since this is monotonically increasing, the mean-field $\gamma$ instability of \citet{Radko2003} and \citet{Traxler2011a} cannot produce layering in $u_y$ in these simulations. This is consistent with our observations of the flow.}
  \label{GammaInst}
\end{figure}

In order to explore whether the GSF instability would be expected to produce such layering i.e.~generate mean flows in $u_y$ that vary in $x$, we can calculate whether the mean-field $\gamma$-instability may occur. Given that the axisymmetric problem is equivalent to salt fingering, by analogy with that problem, we may define the ``density ratio"
\begin{eqnarray}
R_0=\frac{\mathcal{N}^2}{-\kappa_{ep}^2}.
\end{eqnarray}
The ratio of turbulent buoyancy flux to angular momentum flux is
\begin{eqnarray}
\gamma = \frac{(\kappa \mathcal{N}^2-\langle u_x\theta\rangle)}{-\nu\kappa_{ep}^2+2\Omega\langle u_xu_y\rangle}.
\end{eqnarray}
If this is a monotonically increasing function of $R_0$, the mean-field $\gamma$-instability is unable to produce ``layering" of $u_y$ flows along $z$ that vary in $x$. In Fig.~\ref{GammaInst}, we plot $\gamma$ as a function of $R_0$. This clearly indicates that for the parameters considered, layering cannot occur via the $\gamma$-instability for the equatorial GSF instability. The smallest values of $R_0$ for Pr$=0.1$ are slightly non-monotonic, but still increasing with $R_0$. This is consistent with the absence of layering in our simulations. These results are similar to those of \cite{Brownetal2013} for salt fingering. 

However, as we have discussed in \S~\ref{largeshear}, we do observe the excitation of large-scale flows that may correspond with gravity waves in our system for very large $S$ (very small $r$), which significantly enhance the transport over cases with smaller shears. This is qualitatively different from the layering that we may expect from a mean-field instability such as the $\gamma$-instability, and we have also not observed the formation of layers in these cases with larger shears either. It is possible that layering may occur for GSF away from the equator, and this will be investigated in the second paper in this series.

\section{Astrophysical implications}
\label{implications}

We now turn to discuss the astrophysical implications of our results, and to apply the theory presented in \S~\ref{theoryvalidation}. To do so, we must convert our dimensionless units to obtain physical quantities that describe the rates of angular momentum transport. We can use
\begin{eqnarray}
\langle u_xu_y\rangle_{\mathrm{real}}= \Omega^2d^2 \langle u_xu_y\rangle_{\mathrm{code}},
\end{eqnarray}
to relate the momentum transport in physical units (subscript ``real") with that in our dimensional units (subscript ``code"). To make the simplest estimate to quantify the importance of the GSF instability, we assume that it transports angular momentum radially in the form of an eddy diffusion with a diffusivity $\nu_E$. An effective viscosity can then be defined by 
\begin{eqnarray}
\label{nuEexp}
\nu_E = \frac{\langle u_xu_y\rangle_{\mathrm{real}}}{\mathcal{S}} = \nu 
S^{-1} N^{-1}  Pr^{-1/2}\langle u_xu_y\rangle_{\mathrm{code}},
\end{eqnarray}
indicating the extent to which $\nu_E$ is enhanced over laminar viscosity $\nu$.  The effective viscous timescale for angular momentum transport over a distance $L$ is then
\begin{eqnarray}
\nonumber
t_{\nu}&=&\frac{L^2}{\nu_E}\\ 
 &=&\frac{1}{\langle u_xu_y\rangle_{\mathrm{code}}} \frac{L^2}{d^2} S \Omega^{-1}.
 \label{tnuExp}
\end{eqnarray}
This relates the transport to the length-scales of the differential rotation.

 We should note that our simulations have been performed with accessible $\mathrm{Pr}$ values that are larger than in most astrophysical applications by several orders of magnitude. Consequently, it is possible that the values of $\langle u_xu_y\rangle_{\mathrm{code}}$ may differ in reality for much smaller Pr. However, as long as our simple theory remains applicable, we speculate that this difference is unimportant. From Eq.~\ref{saturation} we see that the velocity amplitudes depend on
 $s$ and $k_z$, and from section 3 the fastest growing modes have $k_z=k$. In the limit
 $\mathrm{Pr} \to 0$, Eqs.~\ref{disprel} and \ref{maxgrowthkz} give finite values, $s \to
 \sqrt{2S-4}$ and $k \to 2^{-1/4}$.  This suggests the resulting momentum transport shouldn't depend strongly on Pr. However, it is predicted to depend strongly on $S$.

We will briefly consider two astrophysical examples where the GSF instability may be important. We will assume that the instability operates in each case, and focus on estimating the rates of momentum transport informed by the theory that we have validated using simulations.

\subsection{Solar Tachocline}
Our first example is the solar tachocline, and to apply our results we can estimate the relevant parameters using a solar model (e.g.~\citealt{ModelS1996}). We find $\mathcal{N}\approx 0.6-6 \times 10^{-4}\mathrm{s}^{-1}$ (depending on which radial location is chosen), the local rotation period is of order 25 days, i.e. $\Omega\approx 3 \times 10^{-6}\mathrm{s}^{-1}$, with $\mathcal{S}/\Omega\approx 0.1-0.2$, $\kappa\approx 2.5\times 10^{7}\mathrm{cm}^2 \mathrm{s}^{-1}$, $\nu\approx 20 \mathrm{cm}^2 \mathrm{s}^{-1}$ (adopting some of these numbers from \citealt{Caleo2016a}). These numbers predict
\begin{eqnarray}
d\approx 100-200 \;\mathrm{km},
\end{eqnarray}
indicating that this instability occurs on short length-scales. Substituting the above numbers into Eq.~\ref{nuEexp} and \ref{tnuExp}, we obtain
\begin{eqnarray}
\nu_E\approx 5\times 10^3 \mathrm{cm}^2 \mathrm{s}^{-1} \langle u_xu_y\rangle_{\mathrm{code}},
\end{eqnarray}
and an effective viscous timescale 
\begin{eqnarray}
t_{\nu}\approx 3\mathrm{Myr} \left(\frac{L}{0.01 R_{\odot}}\right)^2\frac{1}{\langle u_xu_y\rangle_{\mathrm{code}}},
\end{eqnarray}
to transport angular momentum over the radial extent of the tachocline region, which is taken to have size $L=0.01R_{\odot}$. We note that in simulations we find $\langle u_xu_y\rangle_{\mathrm{code}}=0.02-4\times 10^{3}$, depending on Pr and $S$. Using our simple theory, we estimate that if the solar interior had a stronger differential rotation so that $S=3\Omega$, $\langle u_xu_y\rangle_{\mathrm{code}}\approx 5$.

This estimate suggests that the GSF instability could be important in producing long-term angular momentum transport in the tachocline. We should point out though that at the equator, a value $\mathcal{S}/\Omega \lesssim 0.2$ would be stable to the GSF. The differential rotation required to drive the instability is however much weaker at non-equatorial latitudes. \citet{Caleo2016a} found that the GSF is at most marginally unstable in the bulk of the solar radiative interior according to the present-day internal rotation profile from helio-seismology. \cite{Rashid2008} suggested that in the tachocline (i.e. where $\mathrm{Ri}$ is somewhat smaller, though still large), GSF instability is possible at non-equatorial latitudes presently. In addition, the GSF instability may have been important in the past in playing a role in the evolution of the solar rotation profile \citep{Menou2006}. Note that non-diffusive baroclinic instabilities may also be important in the tachocline (e.g.~\citealt{GilmanDikpati2014,Gilman2016}).

\subsection{Red Giant stars}
Our second example is to estimate the relevance of the GSF instability for angular momentum transport in red giant stars. The models of \cite{Eggenberger2017} suggest an additional viscosity of $\nu=10^3-10^4 \mathrm{cm}^2\mathrm{s}^{-1}$ is required to explain the relatively weak core-envelope differential rotations that are observed. For an estimate, we adopt $\mathcal{N}\approx 6\times 10^{-3}\mathrm{s}^{-1}$, $\Omega\approx 10^{-8}-10^{-7}\mathrm{s}^{-1}$, $\mathcal{S}\sim\Omega$, $\kappa\approx 10^{9}\mathrm{cm}^2 \mathrm{s}^{-1}$, $\nu\approx 10^2-10^3 \mathrm{cm}^2 \mathrm{s}^{-1}$ (using numbers from e.g. \citealt{Caleo2016a,Eggenberger2017}). These numbers predict
\begin{eqnarray}
d\approx 100 \;\mathrm{km},
\end{eqnarray}
and using Eq.~\ref{nuEexp}, we estimate:
\begin{eqnarray}
\nu_E\approx 5\mathrm{cm}^2 \mathrm{s}^{-1} \langle u_xu_y\rangle_{\mathrm{code}},
\end{eqnarray}
which may take a maximum value of $\nu_E\approx 10^4 \mathrm{cm}^2 \mathrm{s}^{-1}$ for the optimal rates observed in our simulations at the equator. Using our simple theory, we estimate that if the central regions of the star had a much stronger differential rotation such that $S=5\Omega$, $\langle u_xu_y\rangle_{\mathrm{code}}\approx 5\times 10^3$.

This crude estimate suggests that an additional viscosity of $\nu_E\approx 10^4 \mathrm{cm}^2\mathrm{s}^{-1}$ does appear to be possible by the GSF instability, though we do not suggest that such a large viscosity is likely near the equator. However, the GSF instability should be weakest at the equator, and indeed, our non-equatorial simulations (Barker et al., in prep) indicate a significant enhancement in the momentum transport at mid-latitudes, by several orders of magnitude over that at the equator for the same $S$. So such large rates of angular momentum transport are not implausible, even if they are not expected at the equator.

\subsection{Relation to previous work}

The axisymmetric simulations at the equator by \citet{Kory1991} obtain a turbulent state in which angular momentum is transported by the GSF instability. Very broadly, our results are in agreement. We have shown that the GSF instability can lead to a homogeneous turbulent state with enhanced transport (in 3D), but we have also identified significant differences between axisymmetric and 3D simulations. Korycansky's axisymmetric simulations may not have been run for long enough to develop the strong latitudinal shear flows that we have observed, or the boundary conditions may have prevented their formation (just as we have observed in \S~\ref{nekcomp}). 

His simulations suggest that $u_x\sim \ell s$, where $\ell$ is the radial extent of the fingers. We typically find $u_x\sim s/k_z$, since parasitic shear instabilities limit the amplitudes and radial extents of these motions. He observed the angular momentum to be transported on a timescale $t_{\nu}\sim 1/s$, where $s$ is the linear growth rate, whereas we find $t_{\nu}\sim L^2/\nu_E\sim (k_z L)^2/s$, where $L$ is the desired length-scale. The transport is weaker in general than that of \citet{Kory1991} as $k_z L$ is typically much smaller than one.

\citet{Menou2006} assume $\nu_E=s/k_r^2$, taking values for the fastest growing linear mode. This is equivalent to \citet{Kory1991}, and is broadly similar but not identical to our results.

\section{Conclusions}
\label{conclusions}

The Goldreich-Schubert-Fricke (GSF) instability \citep{GS1967,Fricke1968} has long been considered as a possible mechanism to transport angular momentum in the radiation zones of stars, but its nonlinear evolution has barely been explored until now. We have presented our study into the nonlinear evolution of the GSF instability near the equator of a differentially-rotating star using local hydrodynamical simulations in a modified shearing box. Our work significantly builds upon the pioneering axisymmetric simulations of \citet{Kory1991}. We have performed a combination of axisymmetric simulations over a much wider range of parameter values, as well as the first three-dimensional simulations of the GSF instability.

Our three-dimensional simulations at the equator exhibit homogeneous turbulence with sustained, and enhanced, rates of angular momentum transport. We have proposed, and validated against our simulations, a simple theory (motivated by \citealt{Den2010} and \citealt{Brownetal2013} for salt fingering) for the nonlinear saturation of the GSF instability and its resulting angular momentum transport. This theory is based on the idea that the linearly unstable modes are primarily radial motions with short transverse length-scales, which ought to be subject to strong parasitic shear instabilities that limit their amplitudes. This theory (see \S~\ref{theoryvalidation}) is straightforward to implement in stellar evolution codes incorporating rotation.

Our crude estimates in \S~\ref{implications} suggest that the GSF instability could provide an important contribution to the evolution of differential rotation in red giant stars, and it may also have played a role in long-term evolution of the solar tachocline. Further work should explore the nonlinear evolution of the GSF instability at non-equatorial latitudes, where we expect the transport to be significantly enhanced over that at the equator. Indeed, our preliminary simulations indicate that this is indeed the case, and that transport may be enhanced by several orders of magnitude at mid-latitudes (Barker at al., in prep). This suggests that the estimates made in this paper should be viewed as lower bounds on the efficacy of the GSF instability. 

We have also revisited the formal equivalence between the linear and nonlinear evolution of the axisymmetric GSF instability at the equator, with the much more widely-studied salt fingering instability. This is helpful to interpret our simulations, even if this formal analogy does not hold in three-dimensions.
Our axisymmetric simulations, and those in three dimensions in domains with short dimensions along the local azimuthal direction, quickly develop strong radially-varying jets along the rotation axis. These jets inhibit radial transport and produce bursty predator-prey-like temporal dynamics. On the other hand, in three-dimensional simulations with a domain that is sufficiently elongated along the local azimuthal direction, we observe a very different initial state of homogeneous turbulence, with drastically different transport properties, though in some cases jets do form on a much longer timescale. Overall, these results are similar to those obtained for the salt fingering instability \citep{GaraudBrummell2015}. We advocate that three-dimensional simulations (with approximately cubical domains) are probably required so as to not artificially constrain the turbulence, and that the phases of homogeneous turbulence are likely to be the most astrophysically relevant. This is because the strong latitudinal jets are absent in simulations with impenetrable radial boundaries, and their existence is probably also related to the imposition of periodic boundaries in the local latitudinal direction.

It should be noted that the GSF instability does not bring the fluid back to rotate as a solid body, since its onset requires significant differential rotation (e.g.~\citealt{Caleo2016a}), but it pushes the system towards marginal stability. Our crude estimates suggest that the GSF instability may contribute to angular momentum transport in stars, even if, in general, the instability is not a very efficient one. This is because
the unstable modes consist of ``finger-like" motions with very short transverse length-scales, so they cannot travel far radially before they are destroyed by parasitic shear instabilities. Nevertheless, our results indicate that this is still a promising mechanism of angular momentum transport which should be explored further.

Additional effects for future work include simulations at non-equatorial latitudes (these are underway; Barker et al., in prep), investigation of lower Prandtl numbers, the incorporation of gradients in heavy elements (e.g.~\citealt{KnoblochSpruit1983}), and the inclusion of magnetic fields (e.g.~\citealt{Menouetal2004,Menou2006}).

\section*{Acknowledgements}
We would like to thank the referee for a detailed and constructive report that helped us to improve the presentation. AJB was supported by the Leverhulme Trust through the award of an Early Career Fellowship and by STFC Grant ST/R00059X/1. CAJ was supported by STFC grant ST/N000765/1. SMT was supported by funding from the European Research Council (ERC) under the EU's Horizon 2020 research and innovation programme (grant agreement D5S-DLV-786780). This work was undertaken on ARC1 and ARC2, part of the High Performance Computing facilities at the University of Leeds, UK.

\setlength{\bibsep}{0pt}
\bibliography{GSF}
\bibliographystyle{mn2e}

\appendix
\section{Table of simulations}
\begin{table*}
\begin{center}
\begin{tabular}{ccccccccc|cccc}
\hline
Pr & $N^2$ & $S$ & Ri & RiPr & $L_x$ & $L_y$ & $N_x$ & $N_y$ & $\langle u_xu_y\rangle$ & $\sqrt{\langle u_y^2\rangle}$ & $\sqrt{\langle u_x^2\rangle}$ & $\sqrt{\langle u_z^2\rangle}$
 \\
\hline 
$10^{-2}$ & 10 & 2.1 & 2.27 & 0.0227 & 100 & 30 & 256 & 128 & $0.046\pm0.02$ & $0.32\pm0.05$ & $0.20\pm0.04$ & $1.46\pm0.23$ \\
$10^{-2}$ & 10 & 2.1 & 2.27 & 0.0227 & 100 & 30 & 200N & 60N & $0.045\pm0.02$ & $0.89\pm0.62$ & - & $0.45\pm0.19$  \\
$10^{-2}$ & 10 & 2.1 & 2.27 & 0.0227 & 100 & 50 & 256 & 256 & $0.067\pm0.02$ & $0.37\pm0.04$ & $0.25\pm0.04$ & $1.24\pm0.30$ \\
$10^{-2}$ & 10 & 2.1 & 2.27 & 0.0227 & 100 & 100 & 256 & 256 & $0.066\pm0.02$ & $0.33\pm0.04$ & $0.26\pm0.03$ & $1.14\pm0.26$ \\
$10^{-2}$ & 10 & 2.1 & 2.27 & 0.0227 & 100 & 0 & 256 & 1 & $0.053\pm0.02$ & $0.33\pm0.04$ & $0.22\pm0.04$ & $1.30\pm0.28$ \\
$10^{-2}$ & 10 & 2.3 & 1.89 & 0.0189 & 100 & 30 & 256 & 128 & $1.94\pm0.14$ & $1.70\pm0.13$ & $1.62\pm0.07$ & $2.70\pm0.21$ \\
$10^{-2}$ & 10 & 2.3 & 1.89 & 0.0189 & 100 & 0 & 256 & 1 & $0.37\pm0.44$ & $1.15\pm0.37$ & $0.58\pm0.35$ & $6.54\pm1.7$ \\
$10^{-2}$ & 10 & 2.5 & 1.6 & 0.016 & 100 & 30 & 256 & 128 & $5.50\pm0.35$ & $2.89\pm0.11$ & $2.80\pm0.08$ & $3.3\pm0.14$ \\
$10^{-2}$ & 10 & 2.5 & 1.6 & 0.016 & 100 & 30 & 200N & 60N & $0.55\pm0.32$ & $10.04\pm5.97$ & - & $1.50\pm0.78$  \\
$10^{-2}$ & 10 & 2.5 & 1.6 & 0.016 & 100 & 50 & 256 & 256 & $4.39\pm0.29$ & $2.48\pm0.10$ & $2.55\pm0.08$ & $2.89\pm0.11$ \\
$10^{-2}$ & 10 & 2.5 & 1.6 & 0.016 & 100 & 100 & 256 & 256 & $3.93\pm0.13$ & $2.29\pm0.05$ & $2.44\pm0.04$ & $2.76\pm0.05$ \\
$10^{-2}$ & 10 & 2.5 & 1.6 & 0.016 & 100 & 0 & 256 & 1 & $0.81\pm0.96$ & $1.91\pm0.64$ & $0.81\pm0.50$ & $6.95\pm1.17$ \\
$10^{-2}$ & 10 & 3 & 1.11 & 0.0111 & 100 & 30 & 256 & 128 & $21.8\pm1.31$ & $6.20\pm0.22$ & $5.69\pm0.15$ & $6.02\pm0.14$ \\
$10^{-2}$ & 10 & 3 & 1.11 & 0.0111 & 100 & 100 & 256 & 256 & $18.6\pm0.57$ & $5.51\pm0.09$ & $5.37\pm0.07$ & $5.76\pm0.07$ \\
$10^{-2}$ & 10 & 3 & 1.11 & 0.0111 & 100 & 0 & 256 & 1 & $2.67\pm4.64$ & $4.35\pm1.72$ & $1.49\pm1.11$ & $17.3\pm5.9$ \\
$10^{-2}$ & 10 & 3.5 & 1.11 & 0.0111 & 100 & 100 & 256 & 256 & $38.7\pm4.64$ & $8.52\pm0.15$ & $7.71\pm0.11$ & $8.20\pm0.1$ \\
\hline
$10^{-1}$ & $10$ & 2.75 & 1.32 & 0.132 & 100 & 100 & 256 & 256 & $1.04\pm0.12$ & $2.27\pm0.15$ & $0.58\pm0.04$ & $1.04\pm0.07$ \\
$10^{-1}$ & $10$ & 2.75 & 1.32 & 0.132 & 100 & 0 & 256 & 1 & $1.06\pm0.13$ & $2.31\pm0.15$ & $0.59\pm0.04$ & $1.10\pm0.08$ \\
$10^{-1}$ & $10$ & 3 & 1.11 & 0.111 & 100 & 100 & 256 & 256 & $4.04\pm0.99$ & $4.77\pm0.79$ & $1.18\pm0.12$ & $2.26\pm0.39$ \\
$10^{-1}$ & $10$ & 3 & 1.11 & 0.111 & 100 & 0 & 256 & 1 & $2.03\pm1.2$ & $3.77\pm0.87$ & $0.79\pm0.28$ & $5.02\pm0.90$ \\
$10^{-1}$ & $10$ & 3.5 & 0.816 & 0.0816 & 100 & 100 & 256 & 256 & $7.57\pm1.32$ & $5.61\pm0.57$ & $2.04\pm0.22$ & $2.76\pm0.41$ \\
$10^{-1}$ & $10$ & 3.5 & 0.816 & 0.0816 & 100 & 0 & 256 & 1 & $6.36\pm4.44$ & $6.73\pm1.80$ & $1.57\pm0.64$ & $7.55\pm1.1$ \\
$10^{-1}$ & $10$ & 4 & 0.625 & 0.0625 & 100 & 100 & 256 & 256 & $13.19\pm0.71$ & $7.00\pm0.19$ & $3.02\pm0.09$ & $3.67\pm0.1$ \\
$10^{-1}$ & $10$ & 4 & 0.625 & 0.0625 & 100 & 0 & 256 & 1 & $10.56\pm13.7$ & $9.5\pm3.98$ & $2.18\pm1.44$ & $15.66\pm4.6$ \\
$10^{-1}$ & $10$ & 4.5 & 0.494 & 0.0494 & 100 & 100 & 256 & 256 & $24.3\pm1.14$ & $9.12\pm0.22$ & $4.7\pm0.12$ & $5.72\pm0.14$ \\
$10^{-1}$ & $10$ & 5 & 0.4 & 0.04 & 100 & 100 & 256 & 256 & $72.5\pm5.99$ & $14.4\pm0.58$ & $9.6\pm0.44$ & $11.22\pm0.46$ \\
$10^{-1}$ & $10$ & 5.5 & 0.33 & 0.033 & 100 & 100 & 256 & 256 & $282.8\pm25.7$ & $30.2\pm1.93$ & $19.9\pm0.8$ & $21.9\pm0.9$ \\
\hline
$1/3$ & 10 & 4 & 0.625 & 0.208 & 100 & 100 & 256 & 256 & $2.03\pm0.44$ & $4.12\pm0.22$ & $0.61\pm0.08$ & $0.77\pm0.05$ \\
$1/3$ & 10 & 4 & 0.625 & 0.208 & 100 & 0 & 256 & 1 & $1.99\pm0.22$ & $4.19\pm0.23$ & $0.59\pm0.04$ & $0.78\pm0.05$ \\
$1/3$ & 10 & 4.5 & 0.49 & 0.163 & 100 & 100 & 256 & 256 & $12.35\pm3.76$ & $11.0\pm1.38$ & $1.68\pm0.26$ & $2.37\pm0.26$ \\
$1/3$ & 10 & 4.5 & 0.49 & 0.163 & 100 & 0 & 256 & 1 & $16.9\pm3.83$ & $13.5\pm1.32$ & $1.85\pm0.23$ & $4.17\pm0.7$ \\
$1/3$ & 10 & 5 & 0.4 & 0.133 & 100 & 100 & 256 & 256 & $15.75\pm2.15$ & $11.4\pm0.78$ & $2.39\pm0.22$ & $3.05\pm0.27$ \\
$1/3$ & 10 & 5 & 0.4 & 0.133 & 100 & 0 & 256 & 1 & $26.0\pm15.8$ & $16.9\pm4.78$ & $2.66\pm0.96$ & $11.2\pm2.52$ \\
$1/3$ & 10 & 6 & 0.28 & 0.09 & 100 & 100 & 256 & 256 & $5303\pm4730$ & $131.0\pm43.0$ & $81.5\pm27.0$ & $78.3\pm23.9$ \\
\hline
$1/30$ & 10 & 2.3 & 1.89 & 0.063 & 100 & 100 & 256 & 256 & $0.39\pm0.04$ & $1.09\pm0.09$ & $0.46\pm0.03$ & $1.06\pm0.09$ \\
$1/30$ & 10 & 2.3 & 1.89 & 0.063 & 100 & 0 & 256 & 1 & $0.39\pm0.04$ & $1.11\pm0.06$ & $0.45\pm0.03$ & $1.19\pm0.11$ \\
$1/30$ & 10 & 2.5 & 1.6 & 0.053 & 100 & 100 & 256 & 256 & $1.61\pm0.42$ & $2.08\pm0.39$ & $1.06\pm0.1$ & $2.07\pm0.67$ \\
$1/30$ & 10 & 2.5 & 1.6 & 0.053 & 100 & 0 & 256 & 1 & $0.89\pm0.39$ & $2.00\pm0.35$ & $0.66\pm0.16$ & $3.53\pm0.39$ \\
$1/30$ & 10 & 3 & 1.11 & 0.037 & 100 & 100 & 256 & 256 & $6.40\pm0.27$ & $3.80\pm0.09$ & $2.46\pm0.05$ & $2.96\pm0.06$ \\
$1/30$ & 10 & 3 & 1.11 & 0.037 & 100 & 0 & 256 & 1 & $2.83\pm1.91$ & $4.22\pm1.12$ & $1.23\pm0.49$ & $9.11\pm1.32$ \\
$1/30$ & 10 & 3.5 & 0.82 & 0.027 & 100 & 100 & 256 & 256 & $16.3\pm0.58$ & $6.03\pm0.12$ & $4.27\pm0.07$ & $4.97\pm0.08$ \\
$1/30$ & 10 & 3.5 & 0.82 & 0.027 & 100 & 0 & 256 & 1 & $8.31\pm5.03$ & $7.52\pm1.59$ & $2.19\pm0.71$ & $10.7\pm1.12$ \\
$1/30$ & 10 & 4 & 0.625 & 0.02 & 100 & 100 & 256 & 256 & $36.3\pm1.38$ & $9.02\pm0.17$ & $6.78\pm0.13$ & $7.74\pm0.14$ \\
$1/30$ & 10 & 4 & 0.625 & 0.02 & 100 & 0 & 256 & 1 & $11.7\pm18.9$ & $9.61\pm4.81$ & $2.82\pm2.06$ & $18.6\pm3.5$ \\
$1/30$ & 10 & 4.5 & 0.49 & 0.017 & 100 & 100 & 256 & 256 & $69.5\pm2.57$ & $12.8\pm0.17$ & $9.7\pm0.16$ & $10.9\pm0.14$ \\
$1/30$ & 10 & 5 & 0.4 & 0.013 & 100 & 100 & 256 & 256 & $122\pm6.2$ & $17.7\pm0.56$ & $13.2\pm0.31$ & $14.6\pm0.33$ \\
\hline 
$10^{-2}$ & $10^2$ & 2.6 & 14.8 & 0.148 & 100 & 50 & 256 & 256 & $0.123\pm0.04$ & $0.67\pm0.17$ & $0.22\pm0.06$ & $0.25\pm0.07$ \\
$10^{-2}$ & $10^2$ & 3 & 11.11 & 0.111 & 100 & 50 & 256 & 256 & $2.79\pm0.33$ & $3.92\pm0.03$ & $0.94\pm0.08$ & $2.38\pm0.37$ \\
$10^{-2}$ & $10^2$ & 3 & 11.11 & 0.111 & 100 & 0 & 256 & 1 & $2.67\pm4.67$ & $4.35\pm1.72$ & $1.49\pm1.11$ & $17.4\pm5.9$ \\
$10^{-2}$ & $10^2$ & 3.5 & 8.16 & 0.0816 & 100 & 50 & 256 & 256 & $7.14\pm1.0$ & $5.93\pm0.60$ & $1.67\pm0.08$ & $4.50\pm0.58$ \\
\hline
$10^{-3}$ & $10^2$ & 2.1 & 22.68 & 0.02268 & 100 & 50 & 512 & 512 & $0.073\pm0.01$ & $0.38\pm0.02$ & $0.24\pm0.02$ & $0.57\pm0.02$ \\
$10^{-3}$ & $10^2$ & 2.1 & 22.68 & 0.02268 & 100 & 0 & 512 & 1 & $0.039\pm0.01$ & $0.32\pm0.03$ & $0.16\pm0.02$ & $1.19\pm0.14$ \\
$10^{-3}$ & $10^2$ & 2.5 & 16 & 0.016 & 100 & 50 & 512 & 512 & $3.85\pm0.2$ & $2.37\pm0.07$ & $2.29\pm0.06$ & $2.82\pm0.07$ \\
$10^{-3}$ & $10^2$ & 2.5 & 16 & 0.016 & 100 & 0 & 512 & 1 & $0.69\pm0.47$ & $1.75\pm0.38$ & $0.7\pm0.24$ & $5.15\pm0.52$ \\
\hline
\end{tabular}
\caption{
Table of simulation parameters. $L_z=L_x$, and $N_x=N_z$, unless otherwise specified. The eighth and ninth column give the number of Fourier modes in each direction. The two simulations with Nek5000 have `N' in their $N_x$ and $N_y$ column entries and these numbers give the total number of grid points in each direction for $N_x$ and $N_y$, computed using an element distribution with $\mathcal{N}_p=10$ points in each element. Simulation parameters not listed in this table are given in \S~\ref{model}. The data listed to the right of the vertical lines are derived from the simulation results. Our simulation units are determined by setting $\Omega=d=1$.}
\end{center}
\label{Table}
\end{table*}

\label{lastpage}
\end{document}